\documentclass[twocolumn]{aastex63}

\usepackage{url}
\usepackage{graphicx}
\usepackage{calc}
\usepackage{etoolbox}
\usepackage{xspace}
\usepackage[T1]{fontenc} 
\usepackage{textcomp}
\usepackage{ifxetex}
\ifxetex
  \usepackage{fontspec}
  \defaultfontfeatures{Extension = .otf}
\fi
\usepackage{fontawesome}

\usepackage{ifmtarg}
\definecolor{linkcolor}{rgb}{0.1216,0.4667,0.7059}
\newcommand{\codeicon}[1]{{\color{linkcolor}\faFileCodeO}}
\newcommand{\missingcodeicon}[1]{{\color{red}\faFileCodeO}}
\makeatletter
\newcommand{\codelink}[1]%
{\href{https://github.com/ericagol/TRAPPIST1_Spitzer/blob/master/tex/figures/#1}%
  {\@ifmtarg{#1}{\missingcodeicon{#1}}{\codeicon{#1}}}\,\,}
\makeatother

\newcommand{\oscaption}[2]{\caption{#2 \codelink{#1}}}

\usepackage{graphicx}	
\usepackage{amsmath}	
\usepackage{amssymb}	
\usepackage{longtable}
\usepackage{wasysym}

\received{June 26, 2020}
\revised{July 10, 2020}
\accepted{\today}
\submitjournal{PSJ}

\shorttitle{T1 masses, radii, densities}
\graphicspath{{./}{figures/}}

\begin{document}

\title{Refining the transit timing and photometric
    analysis of TRAPPIST-1: Masses, radii, densities, dynamics, and ephemerides.}

\correspondingauthor{Eric Agol}
\email{agol@uw.edu}
\author[0000-0002-0802-9145]{Eric Agol}
\affiliation{Astronomy Department and Virtual Planetary Laboratory, University of Washington, Seattle, WA 98195 USA}
\author[0000-0001-6110-4610]{Caroline Dorn}
\affiliation{University of Zurich, Institute of Computational Sciences, Winterthurerstrasse 190, CH-8057, Zurich, Switzerland}
\author[0000-0002-0632-4407]{Simon L. Grimm}
\affiliation{Center for Space and Habitability, University of Bern, Gesellschaftsstrasse 6, CH-3012, Bern, Switzerland}
\author[0000-0003-2260-9856]{Martin Turbet}
\affiliation{Observatoire de Gen\`eve, Universit\'e de Gen\`eve, 51 Chemin des Maillettes, CH-1290 Sauverny, Switzerland}
\author[0000-0002-7008-6888]{Elsa Ducrot}
\affiliation{Astrobiology Research Unit, Universit\'e de Li\`ege, All\'ee du 6 Ao\^ut 19C, B-4000 Li\`ege, Belgium}
\author[0000-0001-6108-4808]{Laetitia Delrez}
\affiliation{Space Sciences, Technologies and Astrophysics Research (STAR) Institute, Universit\'e de Li\`ege, All\'ee du 6 Ao\^ut 19C, B-4000 Li\`ege, Belgium}
\affiliation{Observatoire de Gen\`eve, Universit\'e de Gen\`eve, 51 Chemin des Maillettes, CH-1290 Sauverny, Switzerland}
\affiliation{Astrobiology Research Unit, Universit\'e de Li\`ege, All\'ee du 6 Ao\^ut 19C, B-4000 Li\`ege, Belgium}
\author[0000-0003-1462-7739]{Micha\"el Gillon}
\affiliation{Astrobiology Research Unit, Universit\'e de Li\`ege, All\'ee du 6 Ao\^ut 19C, B-4000 Li\`ege, Belgium}
\author[0000-0002-9355-5165]{Brice-Olivier Demory}
\affiliation{Center for Space and Habitability, University of Bern, Gesellschaftsstrasse 6, CH-3012, Bern, Switzerland}
\author[0000-0001-9892-2406]{Artem Burdanov}
\affiliation{Department of Earth, Atmospheric and Planetary Science, Massachusetts Institute of Technology, 77 Massachusetts Avenue, Cambridge, MA 02139, USA}
\author[0000-0003-1464-9276]{Khalid Barkaoui}
\affiliation{Oukaimeden Observatory, High Energy Physics and Astrophysics Laboratory, Cadi Ayyad University, Marrakech, Morocco}
\affiliation{Astrobiology Research Unit, Universit\'e de Li\`ege, All\'ee du 6 Ao\^ut 19C, B-4000 Li\`ege, Belgium}
\author[0000-0001-6285-9847]{Zouhair Benkhaldoun}
\affiliation{Oukaimeden Observatory, High Energy Physics and Astrophysics Laboratory, Cadi Ayyad University, Marrakech, Morocco}
\author[0000-0001-5657-4503]{Emeline Bolmont}
\affiliation{Observatoire de Gen\`eve, Universit\'e de Gen\`eve, 51 Chemin des Maillettes, CH-1290 Sauverny, Switzerland}
\author[0000-0002-6523-9536]{Adam Burgasser}
\affiliation{Center for Astrophysics and Space Science, University of California San Diego, La Jolla, CA, 92093, USA}
\author[0000-0002-0221-6871]{Sean Carey}
\affiliation{IPAC, California Institute of Technology, 1200 E California Boulevard, Mail Code 314-6, Pasadena, California 91125, USA}
\author[0000-0003-2415-2191]{Julien de Wit}
\affiliation{Department of Earth, Atmospheric and Planetary Science, Massachusetts Institute of Technology, 77 Massachusetts Avenue, Cambridge, MA 02139, USA}
\author[0000-0003-3750-0183]{Daniel Fabrycky}
\affiliation{Department of Astronomy \& Astrophysics, University of Chicago, 5640 S. Ellis Ave., Chicago, IL 60637 USA}
\author[0000-0002-9328-5652]{Daniel Foreman-Mackey}
\affiliation{Center for Computational Astrophysics, Flatiron Institute, Simons Foundation, 162 5th Ave., New York, NY 10010, USA}
\author[0000-0003-1231-2389]{Jonas Haldemann}
\affiliation{University of Bern, Gesellschaftsstrasse 6, CH-3012, Bern, Switzerland}
\author[0000-0001-7648-0926]{David M. Hernandez}
\affiliation{Harvard-Smithsonian Center for Astrophysics, 60 Garden St., MS 51, Cambridge, MA 02138, USA}
\author[0000-0003-4714-1364]{James Ingalls}
\affiliation{IPAC, California Institute of Technology, 1200 E California Boulevard, Mail Code 314-6, Pasadena, California 91125, USA}
\author[0000-0001-8923-488X]{Emmanuel Jehin}
\affiliation{Space Sciences, Technologies and Astrophysics Research (STAR) Institute, Universit\'e de Li\`ege, All\'ee du 6 Ao\^ut 19C, B-4000 Li\`ege, Belgium}
\author[0000-0001-7574-4440]{Zachary Langford}
\affiliation{Astronomy Department and Virtual Planetary Laboratory, University of Washington, Seattle, WA 98195 USA}
\author[0000-0002-3555-480X]{J\'er\'emy Leconte}
\affiliation{Laboratoire d'astrophysique de Bordeaux, Univ. Bordeaux, CNRS, B18N, Allée Geoffroy Saint-Hilaire, F-33615 Pessac, France}
\author[0000-0003-2805-8653]{Susan M. Lederer}
\affiliation{NASA Johnson Space Center, 2101 NASA Pkwy., Houston TX, 77058, USA}
\author[0000-0002-0296-3826]{Rodrigo Luger}
\affiliation{Center for Computational Astrophysics, Flatiron Institute, Simons Foundation, 162 5th Ave., New York, NY 10010, USA}
\author[0000-0002-1226-3305]{Renu Malhotra}
\affiliation{Lunar and Planetary Laboratory, The University of Arizona, Tucson, AZ 85721 USA}
\author[0000-0002-1386-1710]{Victoria S. Meadows}
\affiliation{Astronomy Department and Virtual Planetary Laboratory, University of Washington, Seattle, WA 98195 USA}
\author[0000-0003-2528-3409]{Brett M. Morris}
\affiliation{Center for Space and Habitability, University of Bern, Gesellschaftsstrasse 6, CH-3012, Bern, Switzerland}
\author[0000-0003-1572-7707]{Francisco J. Pozuelos}
\affiliation{Space Sciences, Technologies and Astrophysics Research (STAR) Institute, Universit\'e de Li\`ege, All\'ee du 6 Ao\^ut 19C, B-4000 Li\`ege, Belgium}
\affiliation{Astrobiology Research Unit, Universit\'e de Li\`ege, All\'ee du 6 Ao\^ut 19C, B-4000 Li\`ege, Belgium}
\author[0000-0002-3012-0316]{Didier Queloz}
\affiliation{Cavendish Laboratory, JJ Thomson Avenue, Cambridge, CB3 0H3, UK}
\author[0000-0001-8974-0758]{Sean N. Raymond}
\affiliation{Laboratoire d'astrophysique de Bordeaux, Univ. Bordeaux, CNRS, B18N, Allée Geoffroy Saint-Hilaire, F-33615 Pessac, France}
\author[0000-0001-9619-5356]{Franck Selsis}
\affiliation{Laboratoire d'astrophysique de Bordeaux, Univ. Bordeaux, CNRS, B18N, Allée Geoffroy Saint-Hilaire, F-33615 Pessac, France}
\author[0000-0002-8124-8360]{Marko Sestovic}
\affiliation{Center for Space and Habitability, University of Bern, Gesellschaftsstrasse 6, CH-3012, Bern, Switzerland}
\author[0000-0002-5510-8751]{Amaury H.M.J. Triaud}
\affiliation{School of Physics \& Astronomy, University of Birmingham, Edgbaston, Birmimgham B15 2TT, UK}
\author[0000-0003-2144-4316]{Valerie Van Grootel}
\affiliation{Space Sciences, Technologies and Astrophysics Research (STAR) Institute, Universit\'e de Li\`ege, All\'ee du 6 Ao\^ut 19C, B-4000 Li\`ege, Belgium}

\begin{abstract}
    We have collected transit times for the TRAPPIST-1 system with the Spitzer {Space Telescope} over {four} years. We add to these ground-based, HST and K2 transit time measurements, and revisit an N-body dynamical analysis of the seven-planet system using our complete set of times from which we refine the mass ratios of the planets to the star. We next carry out a photodynamical analysis of the Spitzer light curves to derive the density of the host star and the planet densities. We find that all seven planets' densities may be described with a single rocky mass-radius relation which is depleted in iron relative to Earth, with Fe {21} wt\% versus 32 wt\% for Earth, and otherwise Earth-like in composition. Alternatively, the planets may have an Earth-like composition, but {enhanced in light elements}, such as {a} surface water layer or a core-free structure with oxidized iron in the mantle. We measure planet masses to a precision of 3-5\%, equivalent to a radial-velocity (RV) precision of 2.5 cm/sec, or two orders of magnitude more precise than current RV capabilities. We find the eccentricities of the planets are very small; the orbits are extremely coplanar; and the system is stable on 10 Myr timescales. We find evidence of infrequent timing outliers {which we cannot explain with an eighth planet;} we {instead account for the outliers} using a robust likelihood function. We forecast JWST timing observations, and speculate on possible implications of the planet densities for the formation, migration and evolution of the planet system.
\end{abstract}






\newcommand{\comment}[1]{}








\section{Introduction}

The TRAPPIST-1 planetary system took the exoplanet community by surprise
thanks to the high multiplicity of small transiting planets orbiting a very-low-mass star ($\sim0.09 M_\odot$; \citealt{Gillon2016, Gillon2017, Luger2017a, vanGrootel2018}).
These unexpected nature stems from the fact that this system was found in a survey
of only 50 nearby ultracool dwarf stars \citep{Jehin2011, Gillon2013}, suggesting either a high-frequency of such systems around the latest of the M-dwarfs \citep{He2016}, or perhaps this discovery was fortuitous \citep{Sestovic2020, Sagear2020}.
The proximity of the host star ($\sim$12pc) makes
it brighter in the infrared ($K=10.3$) than most ultracool dwarfs.
Its small size ($\sim 0.12 R_\odot$)
means that its planets' masses and radii are large relative to those of
the star, which enables precise characterization of the planets' properties.
The system provides the first opportunity for a detailed study of potentially
rocky, Earth-sized exoplanets with incident fluxes {spanning the
range of} the terrestrial
planets in our Solar System.  As such, it has galvanized the exoplanet
community to study this system in detail, both observationally and theoretically,
and has fueled hopes that atmospheric signatures (or even biosignatures) might
be detected with the James Webb Space Telescope \citep{Barstow2016, Morley2017,
    Batalha2018, Krissansen2018, Wunderlich2019, Fauchez2019a, Lustig-Yaeger2019}.

More conservatively, the system provides a potential laboratory
for comparative planetology of terrestrial planets, and may provide insight
and constraints on the formation and evolution of terrestrial planets around
the lowest-mass stars. In particular, transiting multi-planet systems afford
an opportunity to constrain the interior compositions of exoplanets.  Sizes
from transit depths combined  with masses from transit-timing variations yield
the densities of the planets \citep[e.g.][]{Agol2017}. In the case of rocky
planets with a thin atmosphere, the bulk density can constrain the core-mass
fraction and/or Mg/Fe mass-ratio \citep{Valencia2007}, although for any given
planet there is still a degeneracy between a larger core-mass fraction and a
volatile envelope \citep{Dorn2018}.  In a multi-planet system, the bulk density
as a function of planet orbital distance may be used to partly break the compositional
degeneracy by assuming a common refractory composition and a water composition
which increases with orbital distance \citep{Unterborn2018a,Lichtenberg2019}.

The TRAPPIST-1 system was {initially} found with a ground-based pilot survey using a 60-cm telescope,
revealing two short-period transiting planets, and two additional orphan transits
\citep{Gillon2016, Burdanov2018}.  Subsequent {ground-based} study of the system revealed several
additional orphan transits, leading to an incomplete picture of the number of planets
and the architecture of the system.  A 20-day observation campaign with the Spitzer Space
Telescope {\citep{Werner2004}} resolved the
confusion, revealing the periods of six of the seven transiting planets \citep{Gillon2017},
while only a single transit observed of the outermost planet left its orbit in question.
A subsequent observation campaign of the system with the K2 {mission} included four
additional transits of the outer planet, identifying its period, and revealed a series
of generalized three-body Laplace relations (GLRs)\footnote{This refers to the condition
$pP_1^{-1}-(p+q)P_2^{-1}+qP_3^{-1} \approx 0$, which is a generalization of the Laplace
resonance with $p=1$ and $q=2$ \citep{Papaloizou2014}. } between adjacent triplets of
planets \citep{Luger2017a}.  Additional observations with Spitzer
continued to monitor the transit times of the seven planets at higher precision than
afforded by ground-based observations. An initial analysis of the Spitzer data to determine
planetary radii and masses was presented in \citet{Delrez2018a} and \citet{Grimm2018}. In total,
Spitzer observed TRAPPIST-1 for more than 1075 hrs (nearly 45 days), and the resulting time-series 
photometry includes 188 transits \citep{Ducrot2020}. {In this paper we complement and
extend the analysis of \citet{Ducrot2020} to include a transit-timing and photodynamic
analysis of the system.}

Although the planets in the TRAPPIST-1 system have short orbital periods, ranging from
1.5 to 19 days, the dynamical interactions accumulate gradually with time, which requires
longer-timescale monitoring to accurately constrain the orbital model.
The GLRs also cause adjacent pairs of planets to reside near mean-motion resonances,
for which $j P_i^{-1} \approx (j+k) P_{i+1}^{-1}$ for integers $j$ and $k$ for the $i$th and $i+1$th planets.  This proximity causes a resonant timescale {for} $\mathbf{k=1}$
given by
\begin{equation}
    P_{TTV} = \frac{1}{j P_i^{-1} - (j+1) P_{i+1}^{-1}},
\end{equation}
\citep{Lithwick2012} which is the characteristic timescale of the transit timing variations (TTVs) {of
the outer five planets.}
The period of the resonant terms for each of these pairs of planets is $P_{TTV} \approx 491{\pm}5$
days (ranging from {485 to 500} days for each pair).  This timescale has two consequences for
measuring the masses of the planets from transit-timing variations:  1)  the transit times
for each planet need to be sampled on this timescale, preferable covering two cycles so that
the amplitude and phase of the cycles may be distinguished from the planets' orbital periods;
2) this resonant period also sets the timescale for the amplitude variability of ``chopping" (short-timescale
transit-timing variations), which can help to break a degeneracy between mass and eccentricity for the resonant terms
\citep{Lithwick2012,Deck2015}.  As a consequence, we expect the measurements of the masses
of the system to require sampling on a timescale of $t_{min} \approx 2 P_{TTV} \approx 2.7$ years.
The current paper is the first with a survey time, $t_{survey} \mathbf{= 4.114}$ {yr}, such that $t_{survey} {>} t_{min}$ for the TRAPPIST-1 system.

Prior studies used the data available at the time \citep{Delrez2018a}, with $t_{survey} {<} t_{min}$, causing ample
degeneracy in the dynamical model, and hence larger uncertainties in the masses of the planets
\citep{Gillon2017,Grimm2018}.  Even so, these papers were ground-breaking as they enable the
first density determinations of temperate, Earth-sized planets exterior to the Solar System.
Both papers indicated densities for the planets which {were} lower than the value expected for
an Earth-like composition (with the exception of planet e), indicating that these planets might
have a significant volatile content.  However, these conclusions were subject to significant
uncertainty in the planet masses, making the determination of the compositions less definitive as the uncertainties were still consistent with rocky bodies at the $1-2\sigma$ level.
In addition, the masses of all of the planets are highly correlated due to the fact that the
dynamical state of all of the planets needs to be solved together {and their masses and
radii are measured relative to the star}, so model comparisons with individual planets
are not independent.

In this paper we revisit a transit-timing and photometric analysis with the completed
Spitzer program using the more extensive transit dataset we now have in hand.  The goal of
this program is to provide a more precise understanding of the masses, radii, and
densities of the planets.  These measurements may be used for planetary science with the extrasolar planets in the TRAPPIST-1 system, whose similarity to the sizes, masses  and effective insolation range of the terrestrial planets in our Solar System is the closest match known.  In addition, we refine the dynamical
state of the system, revisiting some of the questions explored in \citet{Grimm2018}.
Our final goal is to prepare for upcoming observations with the James Webb Space
Telescope \citep[JWST;][]{Gardner2006}.  More precise constraints on the parameters of the planets will not only improve the precision with which we can schedule observations, but also provide the best possible predictions of potential environmental characteristics that could be discriminated observationally.  This work will therefore help to optimize both the acquisition and interpretation of observations of the TRAPPIST-1 system with JWST.

In \S \ref{sec:observations} we summarize the observational data which are analyzed
in this paper.  In \S \ref{sec:outliers} we discuss the nature of transit
timing outliers, and the robust likelihood function we use for characterizing the
system.  This is followed by a description of our N-body transit-timing analysis
in \S \ref{sec:transit_timing}.  With the improved N-body model, we revisit the
photometric fit to the Spitzer data {using} a photodynamical model in \S
\ref{sec:photodynamics}.  The results of these two analyses are combined to
obtain the planet bulk properties in \S \ref{sec:mass_radius_relation}.  In \S \ref{sec:stellar_params} we derive revised parameters for the host star.  In \S \ref{sec:detection} we search for an eighth planet with transit-timing.  In \S \ref{sec:theoretical_interpretation} we interpret the mass-radius measurements for the planets in terms of interior and atmospheric structure models.  Discussion and conclusions
are given in \S \ref{sec:discussion} and \S \ref{sec:conclusions}.

We provide \texttt{Julia}, \texttt{python}, and \texttt{Matlab} code for running the Markov chains,
{creating the figures,}
and creating the paper PDF in
\texttt{https://github.com/ericagol/TRAPPIST1\_Spitzer}.  
The 3.5 GB \texttt{data/} directory in the repository may be found 
at \doi{10.5281/zenodo.4060252}.
In each figure we embed links to the code (\texttt{</>}) which produced that figure.

\section{New TRAPPIST-1 observations} \label{sec:observations}

Since the work described in \citet{Grimm2018} based on 284 transits, we
have added an additional 163 transit times from a combination of Spitzer
(\S \ref{sec:spitzerobs}) and ground-based observations (\S
\ref{sec:groundbasedobs}) {for a total of 447 transits}.  
{With} preliminary transit-timing fits, we {found} evidence for outliers amongst 
the measured times (\S \ref{sec:outliers}), which we account for with a robust 
likelihood model.  Each transit time is measured as a Barycentric Julian Date 
(BJD$_\mathrm{TDB}$), correcting for the location of the Earth/spacecraft relative 
to the Solar System barycenter \citep{Eastman2010} {at the time of each
transit observation.  We next describe our data.}



\subsection{Spitzer Observations} \label{sec:spitzerobs}

The dataset used in this work includes the entire photometry database of TRAPPIST-1 observations {with Spitzer Space {Telescope}'s Infrared Array Camera \citep[IRAC;][] {Carey2004}} since the discovery of its planetary system. This represents all time series observations gathered within the DDT programs 12126 (PI: M. Gillon), 13175 (PI: L. Delrez) and 14223 (PI: E. Agol). {These cover} a total of 188 transits observed from Feb 2016 to Oct 2019 and {include} 64, 47, 23, 18, 16, 13, and 7 transits of planets b, c, d, e, f, g, and h, respectively \citep{Ducrot2020}. All {of} these data can be accessed through the online Spitzer Heritage Archive database\footnote{\url{http://sha.ipac.caltech.edu}}.  {Spitzer IRAC Channels 1 (3.6 $\mu$m, 0.75 $\mu$m wide) and
2 (4.5 $\mu$m, 1.015 $\mu$m wide) were used during the Spitzer Warm Mission {\citep{Fazio2004,StorrieLombardi2010}} with 61 and 127 transits observed in each band, respectively.}
Observations were obtained with {IRAC} in subarray mode ($32{\times}32$ pixel windowing of the detector) with an exposure time of 1.92 s and a cadence of {2.02 s}. In order to minimize the pixel-phase effect \citep{Knutson2008} the peak-up mode was used \citep{Ingalls2016} to {fine-tune} the positioning of the target on the detector following the IRAC Instrument Handbook.\footnote{\url{https://irsa.ipac.caltech.edu/data/SPITZER/docs/irac/iracinstrumenthandbook/}} Finally, calibration was performed using Spitzer pipeline S19.2.0 to output data as cubes of 64 subarray images of $32{\times}32$ pixels (the pixel scale being 1.2 arcsec). Each set of exposures was summed over  {a 2.15} minute cadence to allow for a tractable data volume for carrying out the photometric analysis, which is described in detail in \citet{Delrez2018a} and \citet{Ducrot2020}.

\subsection{Ground-based observations} \label{sec:groundbasedobs}



In addition to the new Spitzer times, 125 transits were observed by the SPECULOOS-South Observatory  at Cerro Paranal, Chile \citetext{SSO; \citealt{Burdanov2018}, \citealt{Jehin2018}, \citealt{Gillon2018}, \citealt{Delrez2018b}}, TRAPPIST-South at La Silla Chile, \citep[TS;][]{Jehin2011,Gillon2011}, and TRAPPIST-North at {Ouka\"imeden,} Morocco, \citep[TN;][]{Barkaoui2019AJ}. These observations were carried out in an I+z filter with exposure times 23s, 50s and 50s, respectively; characteristics of this filter are described in \cite{Murray2020}.
Observations were also performed with the Liverpool Telescope  \citep[LT;][]{Steele2004} and the William-Herschel Telescope (WHT), both installed at the Roque de los Muchachos {Observatory}, La Palma. Only one transit of planet b and one of d were targeted with the WHT whereas 15 transits of several planets were targeted with LT. For LT observations, the IO:O optical wide field camera was used in Sloan z' band with 20s exposure time. One transit of b was observed with the  Himalayan Chandra Telescope (HCT). Finally, a total of 26 transits were observed in the near-IR (1.2~-2.1~$\mu$m) with the WFCAM near-IR imager of the the United Kingdom Infra-Red Telescope \citep[UKIRT;][]{Casali2007}, the IRIS2IR-imager installed on the the Anglo-Australian Telescope \citep[AAT;][]{Tinney2004}, and the HAWK-I cryogenic wide-field imager installed on Unit Telescope 4 (Yepun) of the ESO Very Large Telescope \citep[VLT;][]{Siebenmorgen2011}.
These observations are summarized in Table \ref{tab:transit_time_list} {: 504 transit observations were collected with 57 duplicate (or triplicate) transits which were observed by a second (or third) observatory simultaneously, for a total of 447 unique planetary transit times which are used in our analysis.  Additional} information may be found in \citet{Gillon2016} for WHT and TRAPPIST, in  \citet{Ducrot2018} for SSO and LT, and in \citet{Gillon2017} and \citet{Burdanov2019} for AAT, UKIRT and VLT.

For all ground-based observations, a standard calibration (bias, dark and flat-field correction) was applied to each image, and fluxes were measured for the stars in the field with the DAOPHOT aperture photometry software \citep{Stetson1987}. {Differential photometry was then performed using an algorithm developed by \cite{Murray2020} to automatically choose and combine multiple comparison stars, optimized to use as many stars as possible, weighted appropriately (accounting for variability, color and distance to target star), to reduce the noise levels in the final differential lightcurves. This reduction and photometry was followed by an MCMC analysis to retrieve transit parameters.}

\begin{table*}
    \centering
    \begin{tabular}{c c c c c c c c c c c}
        \hline
        Planet & HCT & SSO/TS/TN & LT & WHT & VLT/AAT/UKIRT & HST & Spitzer & K2  & Duplicates & Total ($N_i$)\\
        \hline
        b      & 1   & 45    & 7   & 1   & 10       & 1   & 64      & 48   & {17}        & 160   \\
        c      & 0   & 28    & 8   & 0   & 7        & 1   & 47      & 30   & 14        & 107   \\
        d      & 0   & 11    & 1   & 1   & 5        & 2   & 23      & {17}   & {7}         & 53    \\
        e      & 0   & 18    & 4   & 0   & 3        & 2   & 18      & {11}   & {7}         & 49    \\
        f      & 0   & 9     & 2   & 0   & 4        & 2   & 16      & 7    & {6}         & 34    \\
        g      & 0   & 11    & 0   & 0   & 3        & 2   & 13      & {5}    & {4}         & 30    \\
        h      & 0   & 3     & 2   & 0   & 0        & 0   & 7       & 4    & 2         & 14    \\
        \hline
        Total  & 1   & 125   & 24  & 2   & 32       & 10  & 188     & {122}  & {57}        & 447   \\
        \hline
    \end{tabular}
    \caption{Number of transits from ground-based and space-based observations. Duplicates {indicates the excess planet transits observed simultaneously with two or three distinct observatories (as indicated in Table \ref{tab:transit_times_observed_and_calculated})}. Details on the corresponding observations can be found in \citet{Gillon2016}, \citet{Gillon2017}, \citet{Grimm2018}, \citet{deWit2016,deWit2018}, \citet{Delrez2018a}, \citet{Ducrot2018}, \citet{Burdanov2019}, and \citet{Ducrot2020}.}
    \label{tab:transit_time_list}
\end{table*}

\subsection{K2 and HST observations}


The K2 mission \citep{Howell2014} observed the TRAPPIST-1 system over campaigns 12 and 19 \citep{Luger2017a}, in both long- and short-cadence imaging modes. We only use the short-cadence data from campaign 12 for this analysis, with $\sim1$ minute sampling. We use our own photometric pipeline to track the star and produce a light curve from the Target Pixel Files (TPF). To model and correct TRAPPIST-1's stellar variability and K2's pointing-drift-correlated systematic noise, we use a Gaussian process with a quasi-periodic kernel, following the procedure described in \cite{Grimm2018}. The campaign 12 data contains 48, 30, 17, 11, 7, 5, and 4 transits of planets b, c, d, e, f, g, and h, respectively.

Transit times for Hubble Space Telescope observations were utilized, as described in \citet{Grimm2018}, \citet{deWit2016,deWit2018}, and \citet{Wakeford2019}.

\subsection{Transit time measurements and analysis}\label{sec:transit_time_measurements}


{Gathering together} the heterogeneous sample of transits obtained from a variety of ground- and space-based telescopes,  {we transformed the time stamps} to {the} $BJD_\mathrm{TDB}$ time standard prior to {photometric} analysis. {We} analyzed {the datasets} together with a global photometric analysis of {all single-planet} transits, as described in
\citet{Ducrot2020}, with a separate analysis of the overlapping transits once the single-transit analysis was completed.

For each planet a fixed time of transit for epoch zero ($T_{0}$) and fixed period ($P$) were used, but with timing offset (``TTV") as a fitted parameter for each transit as described by \cite{Ducrot2020}.  {To derive $T_{0}$ and $P$, a linear regression of the timings as a function of their epochs was performed for each planet to derive an updated mean transit ephemeris; their exact values can be found in Table 4 of \cite{Ducrot2020}.} The timing offsets are then added back to the ephemeris to obtain the measured transit times and uncertainties.

The final observed dataset for the transit-timing analysis
is given by: $\mathbf{y} = (\{t_{\mathrm{obs},ij},\sigma_{ij}; j = 1,...,N_i\}; i=1,...,7)$, where $i$ labels each of the seven planets, $N_i$ is
the number of transits for the $i$th planet (Table \ref{tab:transit_time_list}),
and $j$ labels each transit for the $i$th planet, so that
$t_{\mathrm{obs},ij}$ is the $j$th observation of the $i$th planet, and
$\sigma_{ij}$ is the corresponding measurement error.  The total number of transits is $N_\mathrm{trans} = \sum_{i=1}^{N_p} N_i$ = 447 {where $N_p$ is the number of transiting planets}.

Table \ref{tab:transit_times_observed_and_calculated} lists the complete set
of transit times and uncertainties which were utilized in the present analysis.

With this sample of transit times collected, we proceed to describe our
dynamical analysis, starting with the likelihood function and evidence for outliers.

\section{Excess of outliers and robust likelihood model} \label{sec:outliers}

We first carried out a preliminary 7-planet, plane-parallel N-body model fit to the transit times using a $\chi^2$ log likelihood function, i.e.\ assuming a Gaussian uncertainty for each transit time given by the derived timing uncertainty, which {we} optimized using the Levenberg-Marquardt algorithm.
We found that the residuals of the fit contain many more outliers than is probable assuming
a Gaussian distribution for the timing uncertainties.

\begin{figure*}
    \centering
    \includegraphics[width=\hsize]{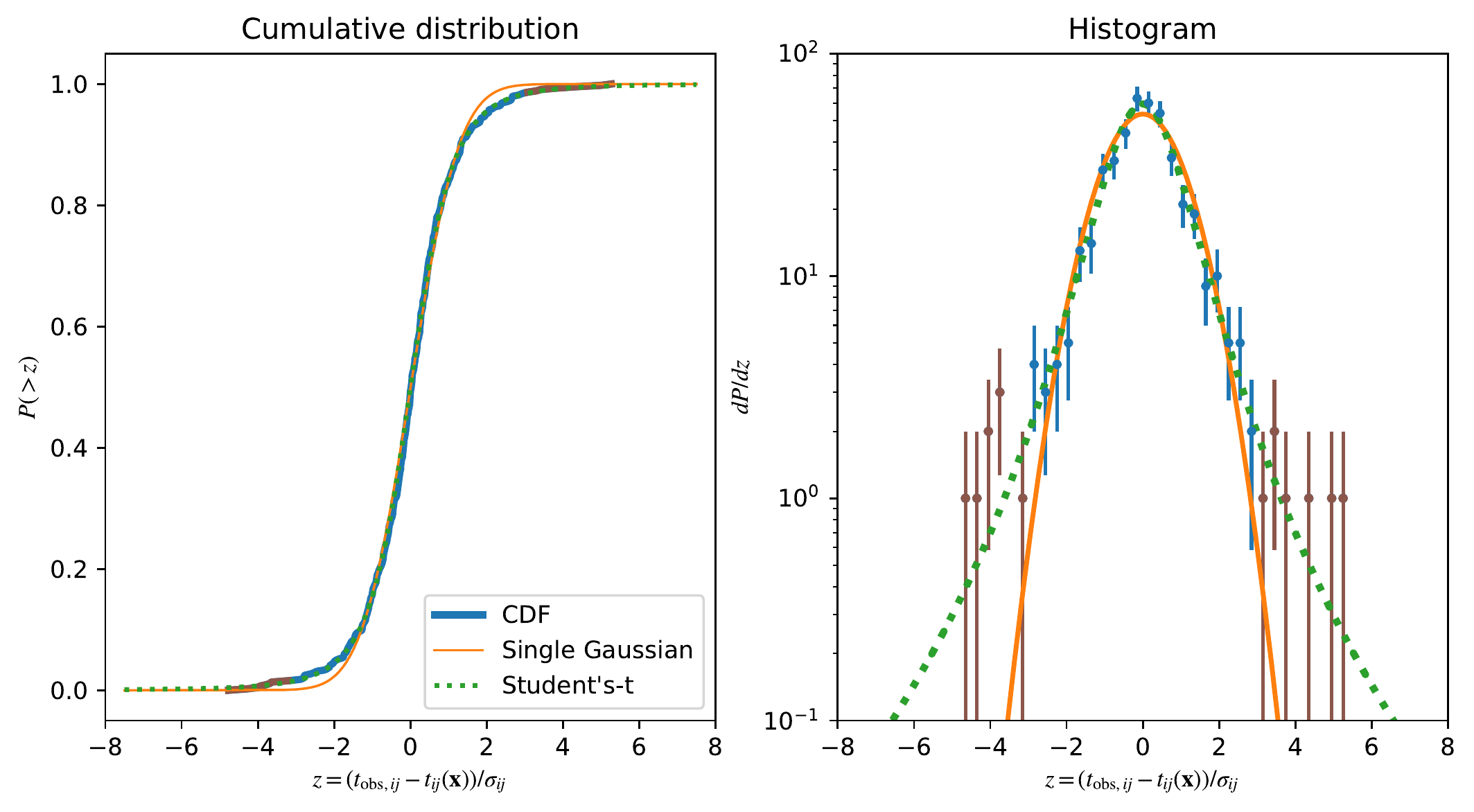}
    \oscaption{julia/plot_residuals_rescale.jl}%
    {Probability distribution of {normalized residuals}.  Left:  Cumulative distribution
        function of the normalized residuals, $z$.  Blue and brown line is a sequence of normalized residuals.
        Orange line is the CDF of a Gaussian distribution.
        Dotted green line is the CDF of a Student's t-
        distribution.  Right: Histogram of the normalized residuals.  Blue and brown data points
        are a histogram of the normalized residuals with Poisson uncertainties.  The other lines
        have the same meaning as the left panel for the probability distribution function
        (PDF),  scaled to match the histogram. {In both panels the $>3\sigma$ outliers are indicated
        in brown.}}
    \label{fig:timing_residuals}
\end{figure*}

Figure \ref{fig:timing_residuals} shows the cumulative distribution
function (CDF)
and a histogram of the normalized residuals versus a single Gaussian
probability distribution function (PDF) with unit variance (orange line).
This CDF distribution function disagrees with the Gaussian CDF
in the wings for $P({>}z) \la 0.1$ and $P({>}z) \ga 0.9$,
where $z {=} (t_{\mathrm{obs},ij} {-} t_{ij}(\mathbf{x}_\mathrm{dyn}))/\sigma_{ij}$ are the normalized residuals, with the model time, $t_{ij}(\mathbf{x}_\mathrm{dyn})$, as a function of the dynamical model parameters, $\mathbf{x}_\mathrm{dyn}$, described below.
This indicates that there is a significant excess of outliers with
large values of $\vert z\vert$
relative to a Gaussian distribution.  The histogram in Figure
\ref{fig:timing_residuals} also demonstrates this clearly:  there
are 8 data points with $z {<} -3$ and 7 with $z {>} 3$.  With 447
transit time measurements, we would only {expect} $\approx 1.2$ data points
with $\vert z\vert {>} \mathbf{3}$ if the distribution were Gaussian with accurately estimated
uncertainties.  This excess is even more apparent at
$\vert z \vert {>} 4$.

We have examined the individual transits that show these discrepancies,
and there is nothing unusual about their light curves, such as flares,
overlapping transits, or other anomalies. 
The outliers
appear for each of the planets {(save h)}, in both ground- and space-based data,
and for measurements with different sizes of uncertainties.  We do
not think that our N-body model is in error (and we have tried to fit
with an extra planet, without a significant improvement in the number
of outliers; see \S \ref{sec:detection}).  Consequently, we believe that these outliers
are due to variations in the measured times of transits which are not
associated with dynamics of the system.

We suspect instead that these outliers
are a {result} of some systematic error(s) present in the data.  There are a variety
of possibilities:  uncorrected instrumental/observational systematics;
time-correlated noise due to stellar variability;  stellar flares (which
may be too weak to be visible by eye, but might still affect the
times of transit); or stellar spots {\citep{Oshagh2013,Ioannidis2015}}.
Again, our examination of
the light curves did not point to a single culprit, so we are unable
to model and/or correct for any of these effects.  Our data are not unique in this respect:  similar outliers
have been seen in other transit-timing analyses, as described
in \citet{JontofHutter2016}.

{Our} transit-timing model will be affected by these
timing outliers, which make an excessive contribution to the
$\chi^2$ of the model, and thus can affect the inference of
the model parameters.   This can cause both the parameters {\it and}
the uncertainties to be mis-estimated.  To make progress, we have
modified the likelihood model to account for outliers.

We {use} a heavy-tailed likelihood function which
better describes the {residual distribution}: 
a Student's t-distribution \citep{JontofHutter2016}.  
{We} fit the normalized residuals to a model
in which the width of the distribution was
allowed to vary, which we parameterize with
an additional factor multiplying the variance, which we refer to below as $V_1$. 
{For the} Student's t-distribution there is only one
additional free parameter:  the number of
degrees of freedom, $\nu$, {which} we treat as a continuous parameter.

Figure \ref{fig:timing_residuals} shows a histogram of the outliers of the best-fit transit-timing model (described below),
and shows that
the Student's t-distribution gives a much higher probability
for outliers.

With the description of the dataset complete, we next describe our efforts to model the data.

\section{Transit-timing analysis} \label{sec:transit_timing}

In this section we describe our transit-timing analysis in detail, starting first with a description of our dynamical model.

\subsection{N-body integration}

We integrate the N-body dynamics in Cartesian coordinates with a novel symplectic
integrator, \textsf{NbodyGradient}, which is based on the algorithm originally described in \cite{Hernandez2015}, derived from the non-symplectic operator of \cite{GoncalvesFerrari2014}.\footnote{The code may be found at
    \texttt{https://github.com/ericagol/NbodyGradient}}  The time-evolution operator of the integrator is a succession of Kepler $2$-body problems and simple ``kick" and ``drift" operators. 
The advantage over traditional symplectic methods \citep{Wisdom1991} is that the dominant error is due to three-body interactions, while in the standard methods, the dominant error is due to two-body interactions, meaning close encounters between non-stellar bodies are treated poorly \citep{Hernandez2017}.    The Kepler problem
for each pair is solved with an efficient universal Kepler solver \citep{Wisdom2015}.
The symplectic integrator is made to be time-symmetric to yield second-order
accuracy \citep{Hernandez2015}.  Then, a simple operator is introduced to double the order of the method \citep{Dehnen2017}.
We have found that numerical cancellations occur between Kepler steps
and negative drift operators, and so we have introduced an analytic cancellation
of these terms to yield an algorithm which is numerically stable,
which converges for small time steps \citep{Agol2020}.

The initial conditions are specified with Jacobi coordinates {\citep{Hamers2016}} and
we use a set of orbital elements for each planet given by
$\mathbf{x}_{dyn} = (\{m_i,P_i,t_{0,i},e_i \cos{\omega_i},e_i\sin{\omega_i}\}; i = 1,...,N_p)$
where $N_p$ is the number of planets for a total of $5N_p$ dynamical parameters. In
addition we take the star to have a mass, $m_0 = M_*/M_\odot$, which we fix to one.
The units of time for the code are days, while the length scale of the code is taken to be $m_0^{1/3}$ AU. \footnote{Note that as we take $m_0=1$ in our simulations, we need to multiply
    the output of positions and velocities from the code by $(M_*/M_\odot)^{1/3}$ to scale to a stellar mass $M_*$.}
The initial orbital ephemeris, $(P_i,t_{0,i})$, consists of the period and initial time of
transit which each planet {\it would} have if it orbited a single
body with {a} mass of the {sum of the} star and the interior planets, 
{unperturbed} by the exterior planets.  We use these variables (in lieu of initial semi-major axis and mean longitude) as
they are well constrained by the {observed} times of transits.  We convert
these analytically to the time of periastron passage, once the
Kepler equation is solved{,} to yield the initial eccentric anomaly
for each initial Keplerian.  Finally, the eccentricity, $e_i$, and
longitude of periastron, $\omega_i$, for each Keplerian we
parameterize in terms of $e_i\cos{\omega_i}$ and $e_i\sin{\omega_i}$
to avoid the wrapping of the angle $\omega_i$.   We transform from Jacobi coordinates to Cartesian coordinates to complete the initial conditions.

For our transit-timing analysis, we assume that the planets are
plane-parallel and edge-on in their orbits, allowing us to neglect
the inclination and longitude of nodes for each planet.

A symplectic integration time step, $h$, is selected to be small,
${<} 5$\%, compared with the orbital period of the innermost planet
\citep{Wisdom1991}. For most of our integrations we use a time step
of $h=0.06$ days, or about 4\% of the orbital period of planet b.

The {model} transit times are found by tracking the positions of each
planet relative to the star across a time step.  Then, when the
dot product of the relative velocity of the planet and star
with their relative position goes from negative to positive, {and
the planet is between the star and observer},
we flag a routine which iterates with Newton's method to find
the {model} transit time, which is taken to be when this dot product
equals zero \citep{Fabrycky2010}, corresponding to the mid-point
of the transit if acceleration is negligible over the duration of the
transit.  The resulting model we obtain
is for the $j$th transit of the $i$th planet, giving each {model}
transit time as a function of the initial conditions, $t_{ij}(\mathbf{x}_\mathrm{dyn})$, which can then be compared to the observed times, $t_{\mathrm{obs},ij}$.

Once the {model} transit times have been found for every planet over the
duration of the time integration, these are then matched with the observed
transit times to compute the likelihood using the Student's t
probability distribution.
The log likelihood function for each data point is given by
\begin{equation}
    \begin{split}
        &\log{\mathcal{L}_{ij}}(\mathbf{x}_\mathrm{dyn},\nu,V_1) \\
        &= -\tfrac{\nu+1}{2}\log\left(1+\frac{(t_{\mathrm{obs},ij}-t_{ij}(\mathbf{x}_\mathrm{dyn}))^2}{\nu \sigma_{ij}^2 V_1}\right) \\
        &-\tfrac{1}{2} \log{\left(\pi \nu V_1\right)} + \log{\Gamma\left(\frac{\nu+1}{2}\right)} - \log{\Gamma\left(\frac{\nu}{2}\right)},\cr
    \end{split}
\end{equation}
where $\Gamma(x)$ is the Gamma function \citep{Fisher1925}.

The total log likelihood function {which we
optimize} is given by
\begin{equation}
    \log{\mathcal{L}}(\mathbf{x}_\mathrm{dyn},\nu,V_1) = \sum_{i=1}^{N_p} \sum_{j=1}^{N_i} \log{\mathcal{L}_{ij}}(\mathbf{x}_\mathrm{dyn},\nu,V_1),
\end{equation}
where $N_p$ is the number of planets; we use $N_p = 7$ for most of our analysis.

Note that we assume that the timing errors are uncorrelated.  Most transits
are well separated in time, and thus this is an accurate assumption as the noise
should be uncorrelated on these timescales.  There are a small number of transits
(about 6\%) that overlap in time, and thus may have correlated uncertainties; we do not account for this in the likelihood function.

\subsection{Uncertainty analysis}

We carried out the uncertainty analysis on the model parameters with
three different approaches:

\begin{enumerate}
    \item Laplace approximation.
    \item Likelihood profile.
    \item Markov-chain Monte Carlo.
\end{enumerate}

First, in our Laplace approximation analysis, we assume a uniform prior on the model parameters and  expand the likelihood as a multi-dimensional normal distribution. 
We maximize the likelihood model using
the Levenberg-Marquardt algorithm, which requires the gradient and Hessian
of the {negative} log likelihood.  Once the maximum likelihood is found, we compute an approximate Hessian at the maximum likelihood (see Appendix \ref{sec:hessian}). The inverse of the Hessian matrix yields an estimate of the covariance amongst the
parameters at the maximum likelihood, whose diagonal components
provide an initial estimate for the parameter uncertainties; we will also use the Hessian for more efficient sampling of the
Markov chain.

The second approach we use is to compute the likelihood
profile for each model parameter.  In this case each parameter
is varied over a grid of values over a range given
by ${\pm}3\sigma_{x_i}$, where $\sigma_{x_i}$ equals the
square root of the diagonal component for the $i$th model
parameter from the covariance matrix.  At each value along
the grid for each parameter we optimize the likelihood with
a constraint which keeps the parameter pinned at the grid
point.  This results in a profile of the maximum likelihood of
each parameter, optimized with respect to all other parameters, which yields
a second estimate for the uncertainties on the parameters.
The likelihood profile approach does not assume a normal distribution and is useful for checking for
a multi-modal probability distribution which can trip up Markov-chain analysis.

However, both of these error estimates are incomplete as they
do not account for non-linear correlations between parameters, for the non-Gaussian shape of the posterior probability,
nor for the prior probability distribution.\footnote{In principle we could include a prior in the Laplace and likelihood profile analyses.}
Nevertheless, the agreement between the two estimates gives
a starting point for evaluating our Markov chain analysis, and
for gauging the convergence of the chains, which we describe below.

In our initial Markov chain sampling, we found that the parameters of the Student's t-distribution, $\nu$ and $V_1$, were strongly non-linearly correlated and displayed a likelihood profile which was non-Gaussian.   After experimenting with reparameterization, we found that $\log{\nu}$ and $V_1 e^{1/(2\nu)}$ gave a parameterization which showed a nearly Gaussian likelihood profile in each parameter, and also showed more linear correlations between these two parameters.  Accordingly we chose to sample in these transformed parameters
{so that our set of model parameters is $\mathbf{x} = (\mathbf{x}_{dyn},\log \nu, V_1 e^{1/2\nu})$.} 

{In appendix \ref{sec:prior} we define the prior function $\Pi(\mathbf{x})$
which multiplies the likelihood to give}
the posterior probability distribution,
\begin{equation}
    P(\mathbf{x}) \propto \Pi(\mathbf{x}){\times}\mathcal{L}(\mathbf{x}),
\end{equation}
so that we can proceed to discussing the Markov chain sampling of the posterior probability of the model parameters given the data.

\subsection{Markov chain sampler}

We sample our posterior probability, $P(\mathbf{x})$, with a Markov chain sampler.
{There are 37 free parameters - four orbital elements and one mass-ratio for
each planet, and two parameters for the Student's t-distribution.}
Given the high dimensionality of our model we chose to use
a Markov chain sampler which efficiently samples in high dimensions:  Hamiltonian
Monte Carlo \citep[HMC; ][]{Duane1987,Neal2011,Betancourt2017,Monnahan2016}.\footnote{aka ``Hybrid Monte Carlo." Note
    that the ``Hamiltonian" referred to in HMC is not a physical Hamiltonian, but an
    artificial one used for treating the negative log probability as a potential
    energy function, and adding a kinetic energy term, with an artificial momentum conjugate
    to each model parameter (``coordinate"). {For a description of HMC and a discussion of
    applications to cosmology, including N-body, see \citet{Leclercq2014} and \citet{Jasche2010} and references therein.}}  This
sampler requires the {\it gradient} of the likelihood function with respect to
the model parameters.  The gradient of the likelihood requires the gradient
of each {model} transit time with respect to the initial conditions of the N-body
integrator.

We have written a module for our N-body integrator which computes the
gradient of each {model} transit time by propagating a Jacobian for the positions
and velocities of all bodies across every time step throughout the N-body
integration \citep{Agol2020}, which is multiplied by the Jacobian of the coordinates at the initial timestep computed with respect to the initial Keplerian elements and masses, which specify the initial conditions and comprise the N-body model parameters.

When a transit time is found during the N-body integration with \textsf{NbodyGradient}, we compute the derivative of each transit time with respect to
the coordinates at the preceding time step, which we multiply times the Jacobian
at that step to obtain the gradient of each transit time with respect to the
initial conditions.
The gradient of the prior with respect to the model parameters, and the gradient of the likelihood with respect to the model times and the Student's t-distribution parameters, are each computed with automatic
differentiation, using forward-mode derivatives  \citep{Revels2016}.  The gradient of the likelihood with
respect to the dynamical model parameters is found by applying the chain rule to the automatic derivatives of the likelihood with respect to the model times with the derivatives computed in the N-body model (from \textsf{NbodyGradient}).

For our HMC analysis, we augment the simulation parameters with a set of conjugate momenta, $\mathbf{p}$, with the same dimension. We sample from the probability distribution, $e^{-H(\mathbf{x},\mathbf{p})}$, where $H$ is a Hamiltonian given by the negative log posterior,
\begin{equation}
    H(\mathbf{p},\mathbf{x}) = \tfrac{1}{2} \mathbf{p}^T \mathbf{M}^{-1} \mathbf{p}
    - \log{\mathcal{L}}(\mathbf{x}) - \log{\Pi(\mathbf{x})},
\end{equation}
where $\mathbf{p}$ is defined from Hamilton's equations,
\begin{equation}
    \dot{\mathbf{p}} = -\frac{\partial H}{\partial \mathbf{x}}.
\end{equation}
We take the mass matrix, $\mathbf{M}$, to be the approximate Hessian matrix evaluated at the maximum likelihood, $\mathbf{M} = \mathbf{\mathcal{H}(\mathbf{x}_0)}$ (eqn.\ \ref{eqn:hessian}).
Similarly, the Hamiltonian can be used to compute the evolution of the parameter ``coordinates,"
\begin{equation}
    \dot{\mathbf{x}} = +\frac{\partial H}{\partial \mathbf{p}}.
\end{equation}
The dot represents the derivative with respect to an artificial ``time" coordinate which can be used to find a trajectory through the $(\mathbf{x},\mathbf{p})$ phase space which conserves the ``energy" defined by this Hamiltonian.

We carry out a Markov chain using the standard approach for HMC. First, we draw the initial momentum from the multi-variate Gaussian distribution defined by the kinetic energy term in the Hamiltonian,
\begin{equation}
    \mathbf{p} = \mathbf{M}^{1/2} \mathbf{Z},
\end{equation}
where $Z_n \sim N(0,1)$ is an element of a vector of random normal deviates for $n=1,...,N_\mathrm{param}$.
We then carry out a leapfrog integration of Hamilton's equations for $N_\mathrm{leap}$ steps from the starting point with a ``time" step $\epsilon$ to obtain a proposal set of parameters $(\mathbf{x}_\mathrm{prop},\mathbf{p}_\mathrm{prop})$.  Since energy is not conserved precisely due to the finite differencing of the leapfrog integration, we then apply a Metropolis rejection step to accept the proposal step with probability
\begin{equation}
    p_\mathrm{accept} = \mathrm{min}(\exp(-(H(\mathbf{x}_\mathrm{prop},\mathbf{p}_\mathrm{prop}) - H(\mathbf{x},\mathbf{p}))),1),
\end{equation}
to determine whether to accept the proposed step and add it to the Markov chain, or to reject it and copy the prior step to the chain.

We carried out some trial integrations to tune two free parameters:  $\epsilon_0$ and $N_{\mathrm{leap},0}$.  We draw the ``time"-step, $\epsilon$, for each integration from the absolute value of a Normal distribution with width $\epsilon_0$, i.e. $\epsilon \sim \vert N(0,\epsilon_0) \vert$.
The number of leapfrog steps for each integration we draw from a uniform probability,
$N_\mathrm{leap} \sim \mathrm{round}(N_{\mathrm{leap},0} \mathcal{U}(0.8,1.0))$.  We found that a choice of $\epsilon_0 = 0.1$
and $N_{\mathrm{leap},0} = 20$ results in a proposal for which the Metropolis rejection gives a high average acceptance rate of 70\%.

We ran 112 HMC chains for 2000 steps each (i.e.\ 2000 leapfrog integrations).   Each leapfrog integration averaged about 7 minutes, and so the chains took nine
days and four hours to complete.\footnote{These were run on four Broadwell Xeon Processors with 28 cores and  128 GB of memory, where each processor is a node in the Hyak Mox cluster at the University of Washington.}
We found a minimum mean effective sample size of 57 over all chains, for a
total number of independent samples of 6409.

\subsection{Results}

\begin{figure*}
    \includegraphics[width=0.95\hsize]{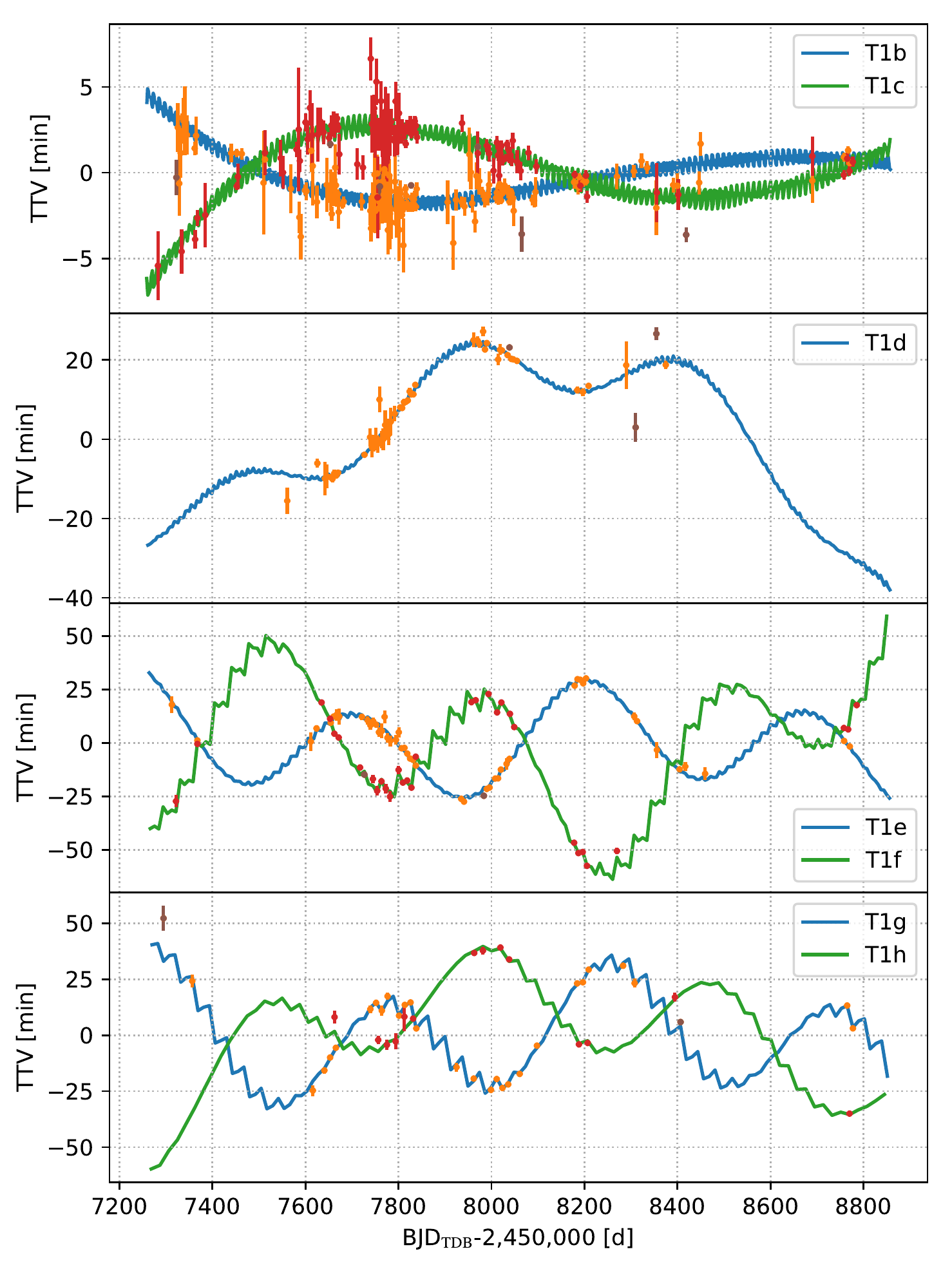}
    \oscaption{julia/plot_ttv_4panel_stacked.jl}%
    {Transit time variation measurements ({orange/red} error bars) and best-fit
        transit-time model ({blue/green} lines) for a subset of our
        Spitzer/K2/ground-based data set.  The TTVs are the transit times for each planet
        with a best-fit linear ephemeris removed.  Brown error bars indicate $>3\sigma$ outliers. 
}
    \label{fig:T1_TTVs}
\end{figure*}

\begin{figure*}
    \centering
    \includegraphics[width=\hsize]{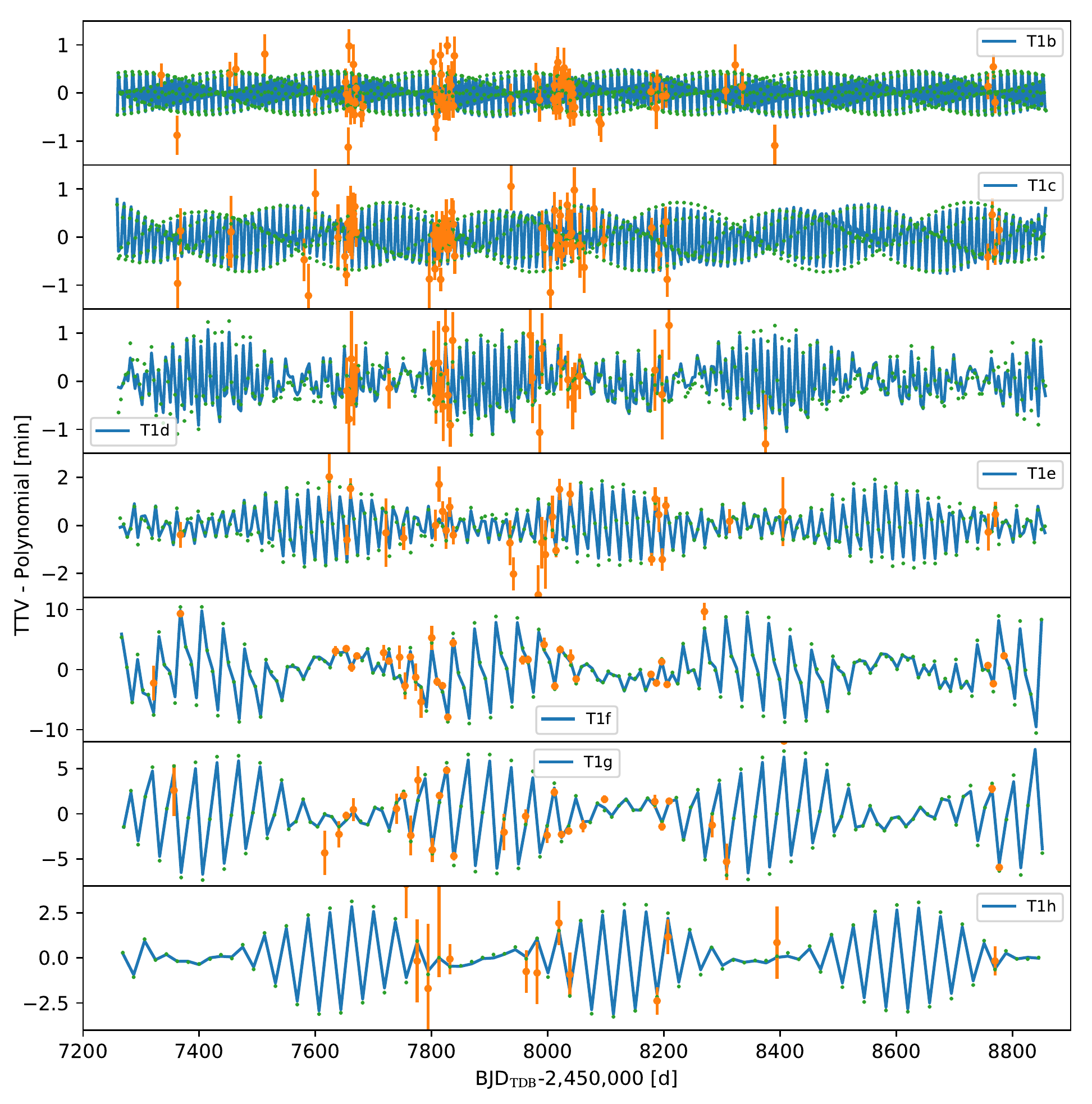}
    \oscaption{julia/plot_ttv_diff.jl}%
    {{Observed} transit times with a polynomial subtracted (orange error bars) {compared with the} short-timescale chopping variations {of the best-fit model} (blue model; same polynomial removed).  Green dots show the analytic chopping relation from \citet{Deck2015} due to adjacent planets, also with a low-order polynomial removed. For the inner four planets we have
        only plotted data with
        uncertainties smaller than the chopping semi-amplitude  (many observations have large uncertainties which would obscure the plot).}
    \label{fig:chopping}
\end{figure*}

The transit-timing variations are shown in Figure \ref{fig:T1_TTVs}, along
with our best-fit model.  The model is a very good description of the data,
although {a few} outliers are clearly visible by eye.  {As
advertised, the outer five planets show large-amplitude oscillations with
the timescale $P_{TTV}$.  We} 
have created a second figure in which a polynomial with an order between 5-30 
is fit {and removed from} the data, {and} the resulting differences
are shown in Fig.\ \ref{fig:chopping}.  The result shows high-frequency variations which are
associated with the synodic periods of pairs of adjacent planets, typically
referred to as ``chopping."  The chopping TTVs encode the mass-ratios of
the companion planets to the star without the influence of the eccentricities,
and thus provide a constraint on the planet-star mass ratios which is less
influenced by degeneracies with the orbital elements \citep{Deck2015}.  The
chopping variations are clearly detected for each planet
(except planet d), which contributes to the higher
precision of the measurements of the planet
masses in this paper.

\begin{table*}
    \centering
    \caption{Parameters of the TRAPPIST-1 system from transit-timing analysis and their $1\sigma$ uncertainties.  Note that the mass ratios, $\mu \mathbf{=M_p/M_*}$, of the planets are computed relative to the star, which is assumed to have a mass of 0.09 $M_\odot$ (this is later combined with the estimate of stellar mass to give our estimates of the planet masses).  We also report $\mu$ in units of $10^{-5}$, and the fractional precision on the measurement of $\mu$, $\sigma_\mu/\mu$.
        The parameters $P$, $t_0$, $e\cos{\omega}$, and
        $e\sin{\omega}$ describe the osculating Jacobi elements at the start
        of the simulation, on date BJD$_\mathrm{TDB}$ $-2,450,000 = 7257.93115525$ days.}
    \label{tab:TTV_parameters}
    \begin{tabular}{lccccccc}
        \hline
          & $\mu \mathbf{\left[\frac{M_\oplus}{0.09 M_\odot}\right]}$                           & $M_p$                & $\frac{\sigma_\mu}{\mu}$ & $P$                       & $t_0$                     & $ e\cos{\omega}$        & $ e\sin{\omega}$  \cr
          & $\mathbf{=\frac{M_p}{M_\oplus}\left(\frac{0.09 M_\odot}{M_*}\right)}$ & [$10^{-5} M_*$]      & \%                       & [day]                     & [BJD$_\mathrm{TDB}$-2,450,000]           &                         & \cr
        \hline
        b & $1.3771{\pm} 0.0593$                              & $ 4.596{\pm}  0.198$ & $4.3$                    & $1.510826{\pm} 0.000006$  & $7257.55044{\pm} 0.00015$ & $-0.00215{\pm} 0.00332$ & $0.00217{\pm} 0.00244$ \cr
        c & $1.3105{\pm} 0.0453$                              & $ 4.374{\pm}  0.151$ & $3.5$                    & $2.421937{\pm} 0.000018$  & $7258.58728{\pm} 0.00027$ & $0.00055{\pm} 0.00232$  & $0.00001{\pm} 0.00171$ \cr
        d & $0.3885{\pm} 0.0074$                              & $ 1.297{\pm}  0.025$ & $1.9$                    & $4.049219{\pm} 0.000026$  & $7257.06768{\pm} 0.00067$ & $-0.00496{\pm} 0.00186$ & $0.00267{\pm} 0.00112$ \cr
        e & $0.6932{\pm} 0.0128$                              & $ 2.313{\pm}  0.043$ & $1.8$                    & $6.101013{\pm} 0.000035$  & $7257.82771{\pm} 0.00041$ & $0.00433{\pm} 0.00149$  & $-0.00461{\pm} 0.00087$ \cr
        f & $1.0411{\pm} 0.0155$                              & $ 3.475{\pm}  0.052$ & $1.5$                    & $9.207540{\pm} 0.000032$  & $7257.07426{\pm} 0.00085$ & $-0.00840{\pm} 0.00130$ & $-0.00051{\pm} 0.00087$ \cr
        g & $1.3238{\pm} 0.0171$                              & $ 4.418{\pm}  0.057$ & $1.3$                    & $12.352446{\pm} 0.000054$ & $7257.71462{\pm} 0.00103$ & $0.00380{\pm} 0.00112$  & $0.00128{\pm} 0.00070$ \cr
        h & $0.3261{\pm} 0.0186$                              & $ 1.088{\pm}  0.062$ & $5.7$                    & $18.772866{\pm} 0.000214$ & $7249.60676{\pm} 0.00272$ & $-0.00365{\pm} 0.00077$ & $-0.00002{\pm} 0.00044$ \cr
        \hline
    \end{tabular}
\end{table*}

\begin{figure}
    \centering
    \includegraphics[width=\columnwidth]{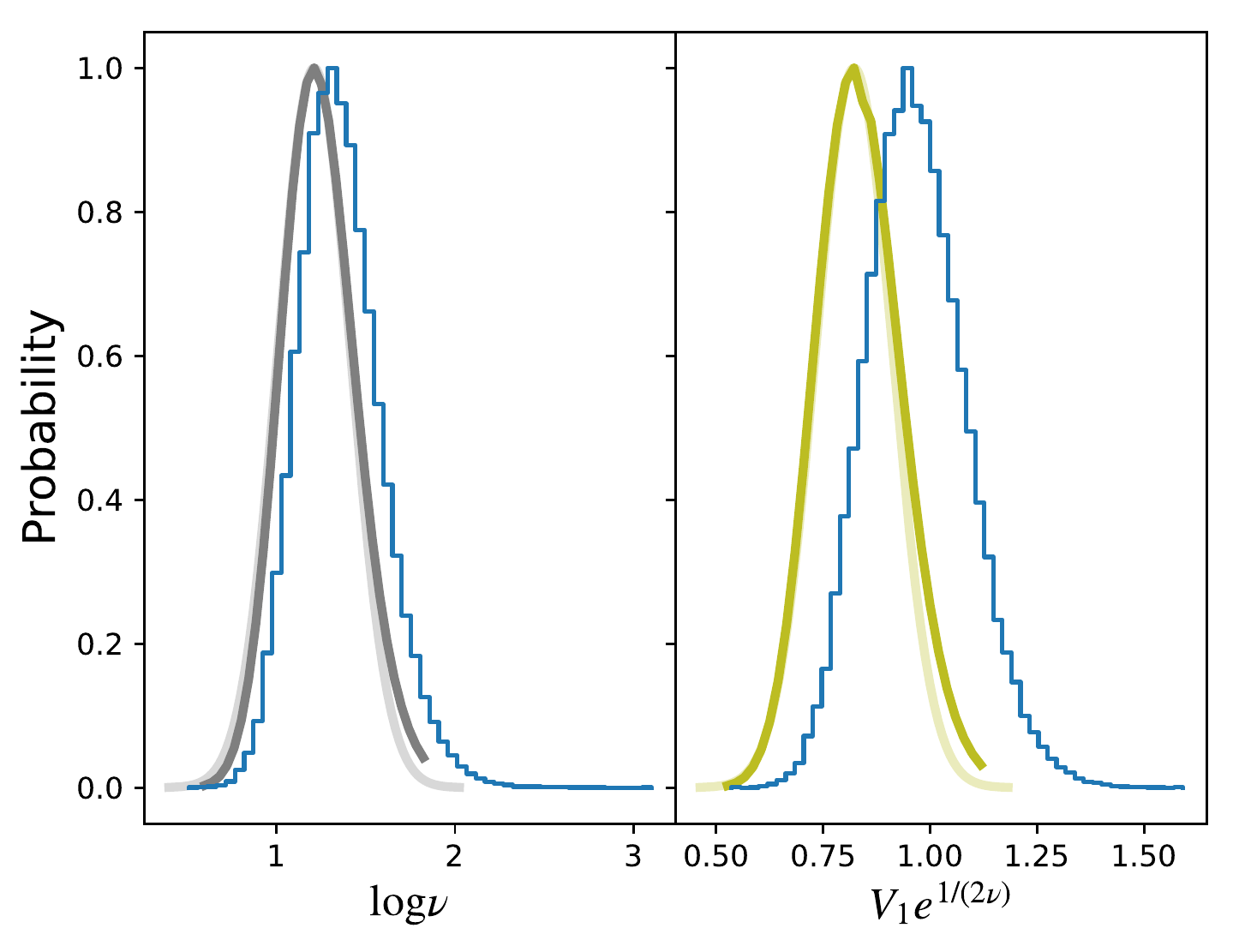}
    \oscaption{julia/plot_likelihood_profile.jl}%
    {Likelihood profile (dark line) and Gaussian distribution
        with Laplace approximation uncertainty (light line) for $\log{\nu}$ (left) and
        $V_1 e^{1/(2\nu)}$ (right).  The posterior probability distributions are shown with blue histograms.}
    \label{fig:student_param_likelihood_profile}
\end{figure}

\begin{figure*}
    \centering
    \includegraphics[width=\hsize]{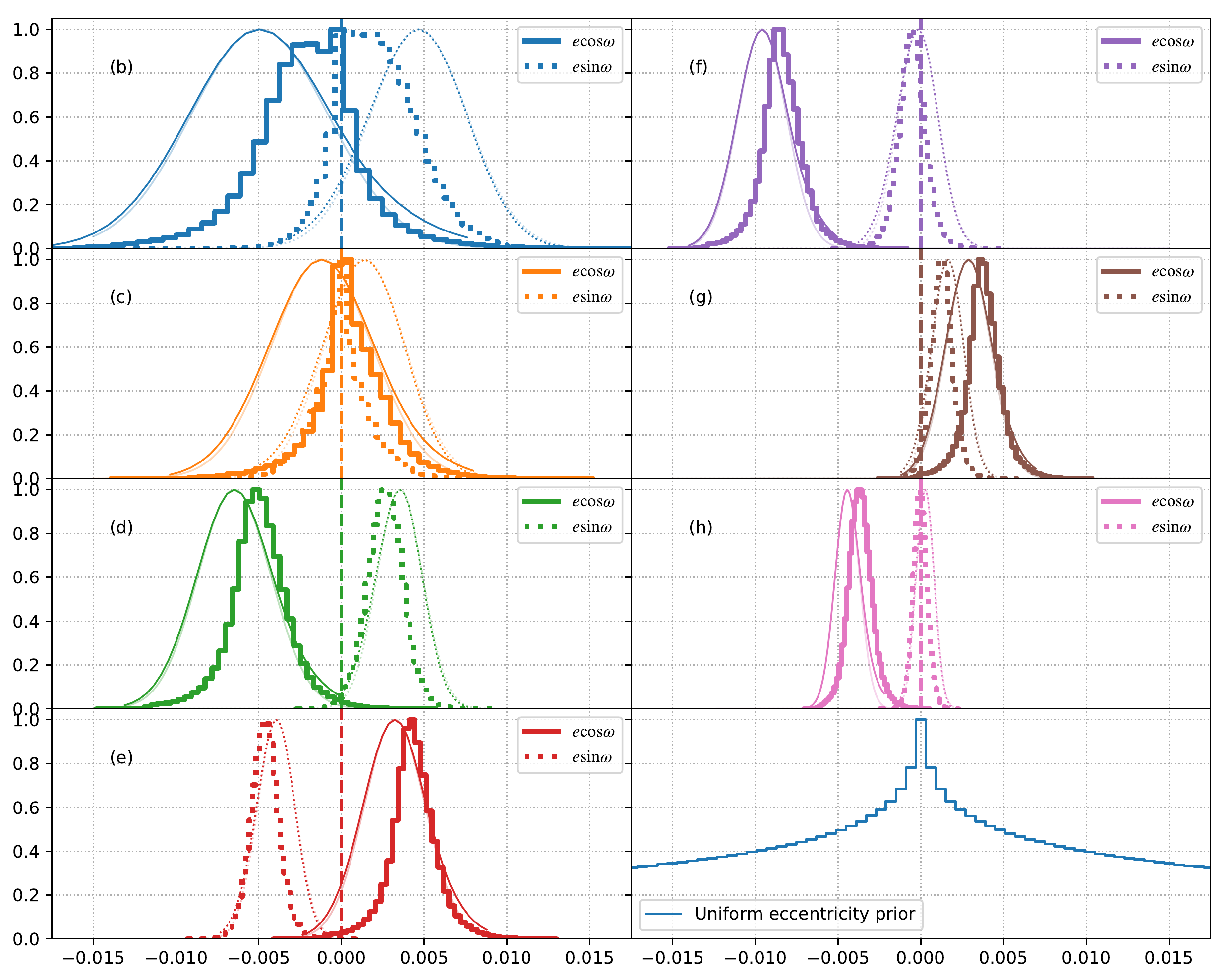}
    \oscaption{julia/plot_likelihood_profile.jl}%
    {Eccentricity vector probability distribution for each planet (y-axes are relative probability).  Thick histograms are the {marginalized} posterior distributions from the Markov chain analysis.  Light, thin lines are the Laplace approximation.  Thin dark lines are the likelihood profiles.  The lower right panel shows the distribution of
        $e\cos{\omega}$ or $e\sin{\omega}$ for a uniform prior on
        $e \in \mathcal{U}(0,0.1)$ and $\omega \in \mathcal{U}(0,2\pi)$.}
    \label{fig:ecc_likelihood_profile}
\end{figure*}

The results of the posterior distribution of our transit-timing analysis are summarized
in Table \ref{tab:TTV_parameters} {with the mean and ${\pm}34.1$\% confidence intervals (1$\sigma)$
computed from the standard deviation of the Markov chains.  The correlations between
parameters are depicted in Figure \ref{fig:corner_ttv}.} There are 35 parameters
which describe the planets, in addition to two parameters
for the Student's t-distribution, $ \log{\nu} = 1.3609{\pm}0.2337$
and $ V_1 e^{1/(2\nu)} = 0.9688{\pm}0.1166$ (Figure \ref{fig:student_param_likelihood_profile}).  The posterior mass-ratios
and ephemerides are consistent with nearly Gaussian distributions. {The}
eccentricity vectors show deviations from a Gaussian distribution for the inner
two planets b and c, as shown in Figure \ref{fig:ecc_likelihood_profile}. 
{The} Laplace approximation covariance uncertainty estimates are overplotted as
Gaussian distributions very closely match the likelihood profile
for each parameter.  This agreement is reassuring:  it indicates
that the likelihood distribution is closely approximated by
a multi-dimensional normal distribution near the maximum likelihood.
In the eccentricity-vector coordinates,
the {\it prior} probability distribution is peaked
at zero to ensure that the volume of phase-space at larger
eccentricities does not dominate the probability distribution,
as shown in the lower right panel of Figure \ref{fig:ecc_likelihood_profile}.
For the planets which have
a likelihood distribution which overlaps strongly with zero,
the prior distribution causes the Markov chain posterior to
have a significantly different distribution from the likelihood profile.  This is not
due to the prior favoring small eccentricities;  rather it is simply
a correction for the bias which results by using $e_i\cos{\omega_i}$
and $e_i\sin{\omega_i}$ as Markov chain parameters which favors
higher eccentricities \citep{Ford2006}.

\begin{figure*}
    \centering
    \includegraphics[width=\hsize]{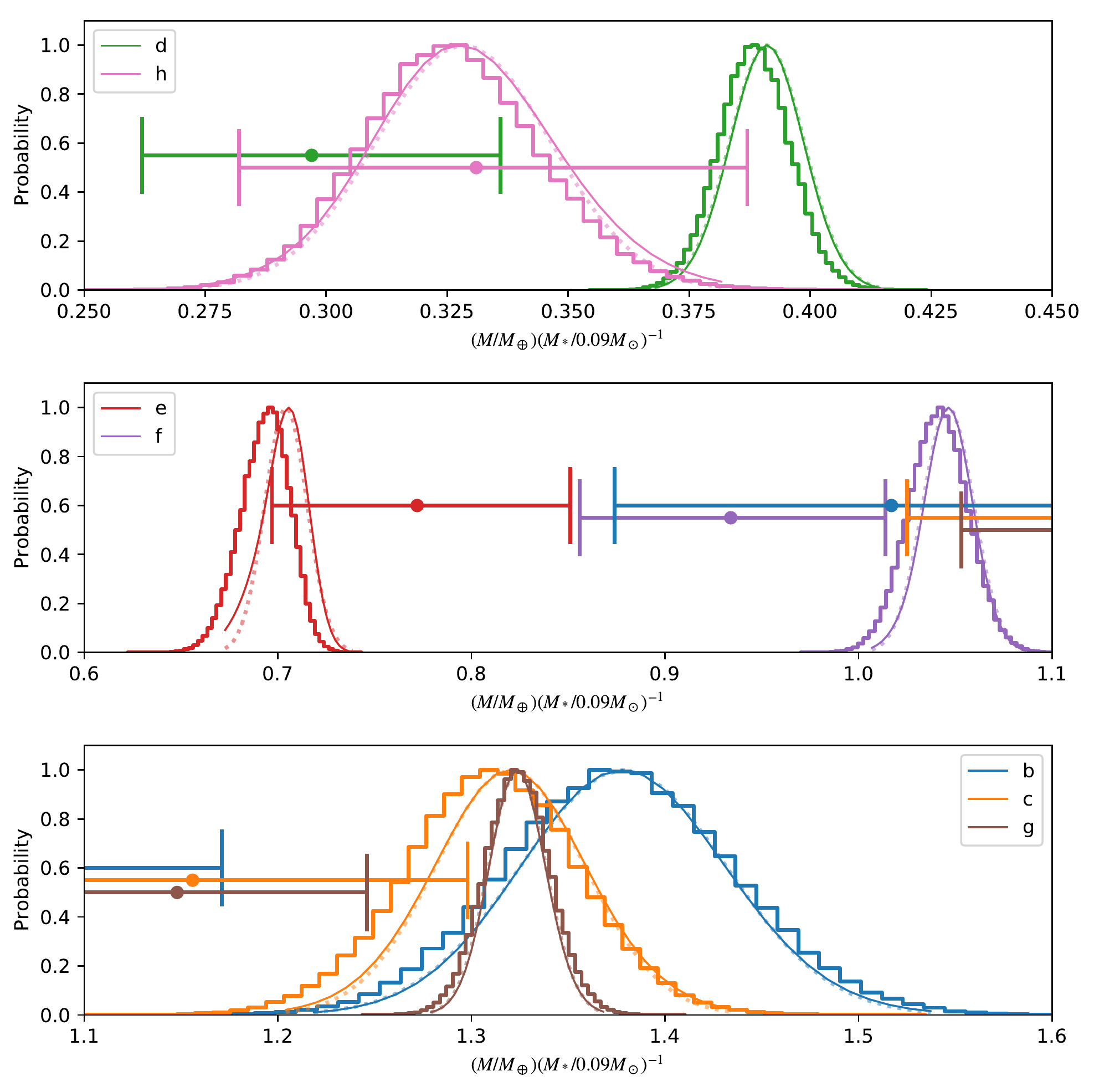}
    \oscaption{julia/plot_masses.jl}%
    {Probability distribution of the {planet-to-star mass ratios}, scaled to a stellar mass of $M_*  = 0.09 M_\odot$; panels range from small masses to large.   Thick
        histograms show the posterior probability distribution of the Markov chain analysis.
        {Horizontal error bars show the mean and $1\sigma$ intervals for the} mass-ratios from \citet{Grimm2018}.
        Dark solid {bell} curves are the likelihood profiles; light, dotted {bell} curves are the Laplace approximation.}
    \label{fig:mass_likelihood_profile}
\end{figure*}

The {marginalized} posterior distributions of the {\it ratio} of the planet masses to the star, scaled to a stellar mass of 0.09 $M_\odot$, are given in Table \ref{tab:TTV_parameters} and shown in Figure \ref{fig:mass_likelihood_profile}.  The likelihood profile of
the {planet-to-star mass ratios} is also plotted in
Figure \ref{fig:mass_likelihood_profile} and appears to be well-behaved. These
likelihood profiles are also approximately Gaussian in shape, and track the inverse Hessian evaluated at the maximum likelihood to estimate the covariance (also plotted).  Compared with the {mass
estimates from} \citet{Grimm2018}, the masses of each planet have increased
with the exception of planet e which has decreased and planet h which remains the same (Table \ref{tab:grimm_vs_agol}).   The mass ratios of the posterior distribution from the Markov chain are slightly shifted to smaller values than the likelihood profile and Laplace approximation probabilities for all planets save b and g.

\begin{table*}
    \centering
    \scriptsize
    \caption{{Planet-to-star mass ratios} in units of
        $M_\oplus/(0.09 M_\odot)$ from \citet{Grimm2018} 
       {and planet-to-star radius ratios $R_p/R_*$ from \citet{Delrez2018a}} compared with the results from this paper.}
    \begin{tabular}{l|c|c|c|c|c|c|c|c}
        Source & Quantity      & b                          & c                           & d                           & e                           & f                           & g                           & h                           \\
        Grimm & $\frac{M_p}{M_\oplus}\frac{0.09M_\odot}{M_*}$& $1.017_{-0.143}^{+ 0.154}$ & $1.156_{- 0.131 }^{+0.142}$ & $0.297_{- 0.035}^{+ 0.039}$ & $0.772_{- 0.075}^{+ 0.079}$ & $0.934_{- 0.078 }^{+0.080}$ & $1.148_{- 0.095}^{+ 0.098}$ & $0.331_{- 0.049}^{+ 0.056}$ \\
        This paper          &$\frac{M_p}{M_\oplus}\frac{0.09M_\odot}{M_*}$ & $1.3771{\pm} 0.0593$       & $1.3105{\pm}0.0453$         & $0.3885{\pm}0.0074$         & $0.6932{\pm}0.0128$         & $1.0411{\pm}0.0155$         & $1.3238{\pm}0.0171$         & $0.3261{\pm}0.0186$         \\
        \hline
        Delrez & $R_p/R_*$       & $0.0853{\pm} 0.0004$           & $0.0833{\pm} 0.0004$          & $0.0597{\pm} 0.0006$          & $0.0693{\pm} 0.0007$          & $0.0796{\pm} 0.0006$          & $0.0874{\pm} 0.0006$          & $0.0588{\pm} 0.0012$ \cr
        This paper & $R_p/R_*$       & $0.0859{\pm} 0.0004$           & $0.0844{\pm} 0.0004$          & $0.0606{\pm} 0.0005$          & $0.0708{\pm} 0.0006$          & $0.0804{\pm} 0.0005$          & $0.0869{\pm} 0.0005$          & $0.0581{\pm} 0.0009$ \cr
    \end{tabular}
    \label{tab:grimm_vs_agol}
\end{table*}

The Student's t-distribution parameters show a posterior distribution which is shifted from the likelihood profile/Laplace probability distribution (Fig.\ \ref{fig:student_param_likelihood_profile}).  This bias is due to the fact that the likelihood distribution of these parameters shifts upwards whenever the transit-timing model parameters deviate from their maximum-likelihood values.  The peak of  the  posterior distribution of these parameters corresponds to
$\nu = 3.9$ and $V_1^{1/2} = 0.87$, which indicates that the core of the distribution is narrower than the transit-timing uncertainties indicate, while the wings of the distribution are close to $\nu = 4$, which was the value used by
\citet{JontofHutter2016}.

\subsection{Independent N-body TTV analysis}

In addition to the N-body code described above, we use the GPU hybrid symplectic N-body code GENGA \citep{Grimm2014} with a Differential Evolution Markov Chain Monte Carlo Method  \citep[DEMCMC;][]{terBraak2006} as described in \citet{Grimm2018} to perform an independent TTV analysis. The parameters for the MCMC analysis are $\mathbf{x} = (\{m_i, P_i,t_{0,i},e_i,\omega_i\};i=1,...,N_p)$. The mass of the star is taken to be $M_{\star} = 0.09 M_{\odot}$, and the time step of the N-body integration is set to $h=0.05$ days.  The
likelihood is assumed to be a normal distribution with the timing errors derived from
the timing analyses.  For comparison, we have re-run the likelihood-profile
computation described above using a normal distribution in place of a Student's t-distribution.
The derived masses from the two different analyses agree well with a maximal deviation of the median masses of better than 0.4\%, {while the mass-ratio uncertainties agree to better than} 13\%.  The eccentricities and longitudes of periastron at the initial time agree as well.  We interpret this as a validation of the numerical techniques being employed in this paper.

With the transit-timing analysis completed, we now use the N-body model to improve the estimate of the stellar density and the {planet-to-star radius ratios}.  To do so we create a photodynamic model, described next.

\section{Photodynamical analysis}\label{sec:photodynamics}

With the mass-ratios and orbital parameters derived from the transit-timing
analysis, we wish to improve our derivation of the planet and stellar parameters
from the Spitzer photometry.   The transit depth, transit duration, and ingress/egress
duration combined with  orbital period constrain the impact parameters and density
of the star \citep{Seager2003}.  Combining these constraints for
each of the planets  enables a more precise constraint upon the density of the
star \citep{Kipping2012a}.
The transit durations are affected by the (small) eccentricities, but to a lesser extent.  We account
for the dynamical constraints on the transit-timing model to improve the
photometric constraints upon these parameters, albeit with the dynamical parameters fixed
at the maximum likelihood.

We fit a ``photo-dynamical" model \citep{Carter2012} to the data with the following procedure.
From the best-fit plane-parallel, edge-on transit time model,
we compute the sky velocity at each of the mid-transit times, $t_0$,
from the model (in N-body code units).  We then convert the code units to physical
units using the density of the star, obtaining the sky velocity, $v_\mathrm{sky}$, in units
of $R_*/\mathrm{day}$.
We account for quadratic limb-darkening of the star with parameters ($q_\mathrm{1,Ch1}, q_\mathrm{2,Ch1},q_\mathrm{1,Ch2},q_\mathrm{2,Ch2}$) in the two Spitzer channels, and for each
planet we specify a {planet-to-star} radius ratio ($R_p/R_*$) and we assume
mid-transit impact parameter ($b_0$), which is constant for all transits of
a given planet.  We assume that the limb-darkening parameters are a function
of wavelength for the two Spitzer channels, while we treat the planet radius
ratios as identical in both wavebands {based on their consistency across
all planets in \citet{Ducrot2020}}, giving a total of 19 free parameters
for the photodynamical model.

We ignore acceleration during the transits, treating the impact parameters as
a function of time as
\begin{equation}
    b(t) = \sqrt{(v_\mathrm{sky} (t-t_0))^2 + b_0^2},
\end{equation}
in units of the stellar radius, $R_*$.
Although this expression ignores the curvature and inclination of the orbits, as well
as the acceleration of the planet, the star is so small compared with the orbital
radii that this approximation is extremely accurate.  The transit model is integrated
with an adaptive Simpson rule over each Spitzer exposure time (which has a uniform
duration binned to 2.15 minutes), as described in \citet{Agol2019}, yielding
a light curve computed with a precision of better than $10^{-7}$ for all cadences.

We compute a photometric model for all seven
planets for all of the Spitzer data in selected windows around each of the observed
transits.  Starting with Spitzer photometric data, which were already corrected for systematic
variations based on the analysis by \citet{Ducrot2020}, we fit  each transit window
with the transit model multiplied by a cubic polynomial, whose coefficients are solved
for via regression at each step in the Markov chain.
We
transform the $q_1, q_2$ limb-darkening
parameters to $u_1,u_2$ {in each band} using the formalism of \citet{Kipping2013} for computing the transit model from \citet{Agol2019}.  After carrying out an initial
optimization of the model, we take the photometric error to be the
scatter in each observation window to yield a reduced chi-square of unity in
that  window.  With this photometric
scatter, we compute a $\chi^2$ of the model with respect to the Spitzer
photometric data, and we optimize the model using a Nelder-Mead algorithm.

\subsection{{Photodynamic Results}}

To compute the uncertainties on the photodynamical model parameters, we
use an affine-invariant Markov chain Monte Carlo algorithm \citep{Goodman2010}.\footnote{As
    implemented in the package \texttt{https://github.com/madsjulia/AffineInvariantMCMC.jl}}
We used a uniform prior with bounds on each parameter given in Table \ref{tab:pd_bounds}.
The posterior distribution of the results of the fit are given in Table \ref{tab:photodynamic_params},
{while the correlations between parameters are shown in Figure \ref{fig:corner_photdyn}.}
We utilized 100 walkers run for 50,000 generations, discarding the first 1500 generations
for burn-in.  We computed the effective sample size using the integrated
auto-correlation length, finding a minimum effective sample size of 6000 over all
19 parameters\footnote{Using \texttt{https://github.com/tpapp/MCMCDiagnostics.jl}}.

\begin{figure*}
    \centering
    \includegraphics[width=\hsize]{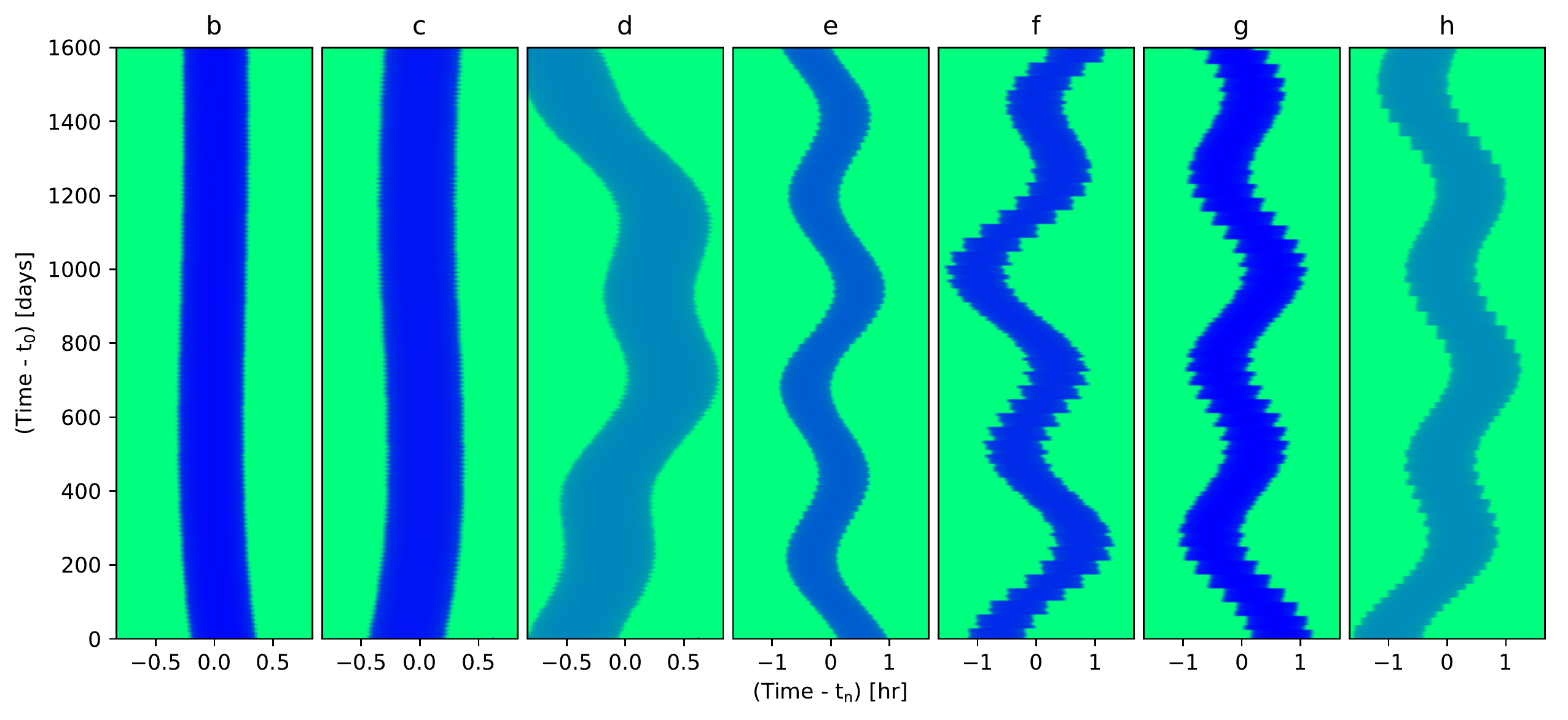}
    \oscaption{julia/T1_photodynamics_riverplot.jl}%
    {River plots showing every transit over 1600 days for one planet per panel
        (left to right are b-h, as labelled; the
        transits of companion planets are omitted from each panel).
        The x-axis ranges over 200/400${\times}$30-second exposures centered on the mean ephemeris for the $n$th transit
        for b-d/e-h respectively (note the 30 sec exposures are higher resolution than the {binned} Spitzer time-resolution).
        Each row contains a transit model, with green being the out-of-transit, and blue in
        transit.  There are (1059,661,395,262,173,129,85) transits of planets b-h, respectively.
        Planets d and h have the smallest sizes, and hence shallowest depths, causing a lighter
        color during transit.}
    \label{fig:riverplot}
\end{figure*}

\begin{table}
    \centering
    \begin{tabular}{l| c|l}
        Parameter & Units        & Prior \cr
        \hline
        $b_0$     & $R_*$        & $\mathcal{U}(0,1)$\cr
        $R_p/R_*$     & -        & $\mathcal{U}(0,0.2)$\cr
        $\rho_*$  & $\rho_\odot$ & $\mathcal{U}(0,100)$\cr
        $(q_{1,\mathrm{Ch 1}},q_{2,\mathrm{Ch 1}})$
                  & ---          & $\mathcal{U}(0,1)$\cr
        $(q_{1,\mathrm{Ch 2}},q_{2,\mathrm{Ch 2}})$
                  & ---          & $\mathcal{U}(0,1)$\cr
    \end{tabular}
    \caption{Prior bounds on photodynamic parameters.  Note that the same bounds on impact parameter, $b_0$, and radius ratio, $R_p/R_*$, are placed on all seven planets.}
    \label{tab:pd_bounds}
\end{table}
To help visualize the model, a photodynamical model with the best-fit parameters is
shown in Figure \ref{fig:riverplot} computed over 1600 days.  Planets b and c have
short periods, and are far from a $j{:}j{+}1$ period ratio.
Hence both of these planets show weak TTVs,
and straighter, but still slightly meandering, riverplots.  The outer five planets are pairwise
close to a series of $j{:}j{+}1$ resonances, showing strong transit-timing variations on
the timescale of the TTV period of $\approx 490$ days.  The other prominent feature
for the outer 4 planets is the slight zig-zag of transits due to chopping (shown in Figure \ref{fig:chopping}). 

{Table \ref{tab:grimm_vs_agol} shows the radius-ratios from \citet{Delrez2018a}
alongside those from the present analysis.  The precision of the measurements did
not improve significantly, while the radius-ratios shifted by 1-2$\sigma$.}
Figure \ref{fig:impact_parameter} shows the posterior probability distribution of impact
parameters in units of the stellar radius, $b_0$, derived from the photodynamical
model.  Figure \ref{fig:stellar_density} shows the probability distribution of stellar
density.  The density correlates with the impact
parameters of each planet, reaching a tail of lower values
for higher impact parameters of each planet.  The tail of the density probability
distribution has an approximately exponential scaling with the density below the
peak, and cuts off as a normal distribution above.  In table \ref{tab:photodynamic_params}
we report the median and 68.3\% confidence interval
of the stellar density.  The inferred density is both slightly larger and more precise
than prior analyses \citep{Delrez2018a}, which we discuss below.

Combining the measured density with the measured orbital periods of the planets,
we derive the semi-major axis of each planet in units of the stellar radius,
\begin{equation}
    \frac{a}{R_*} = \left(\frac{P^2 G M_\odot}{4\pi^2 R_\odot^3} \frac{\rho_*}{\rho_\odot}\right)^{1/3}.
\end{equation}

With the measured impact parameters, we compute the inclinations of the planets
from {\citep{Winn2010}}
\begin{equation}
    I = \cos^{-1}\left[ b_0
        \left(\frac{a}{R_*}\right)^{-1}\right],
\end{equation}
where we have neglected the eccentricity in this formula due to the extremely
small values of the eccentricities of the planets from the transit-timing
analysis (cf Table \ref{tab:TTV_parameters}).  The resulting inclination
posterior distribution is displayed in Figure \ref{fig:inclination_distribution}.
Although the inclination is derived from the impact parameters, which we constrain
to be positive, in practice the photodynamical model cannot distinguish between
inclinations of $I$ and $180-I$ (Fig.\ \ref{fig:inclination_distribution}), and so we created a histogram of these two options
with equal probability.

\subsection{Mutual inclinations and stellar density}

\begin{figure}
    \centering
    \includegraphics[width=\hsize]{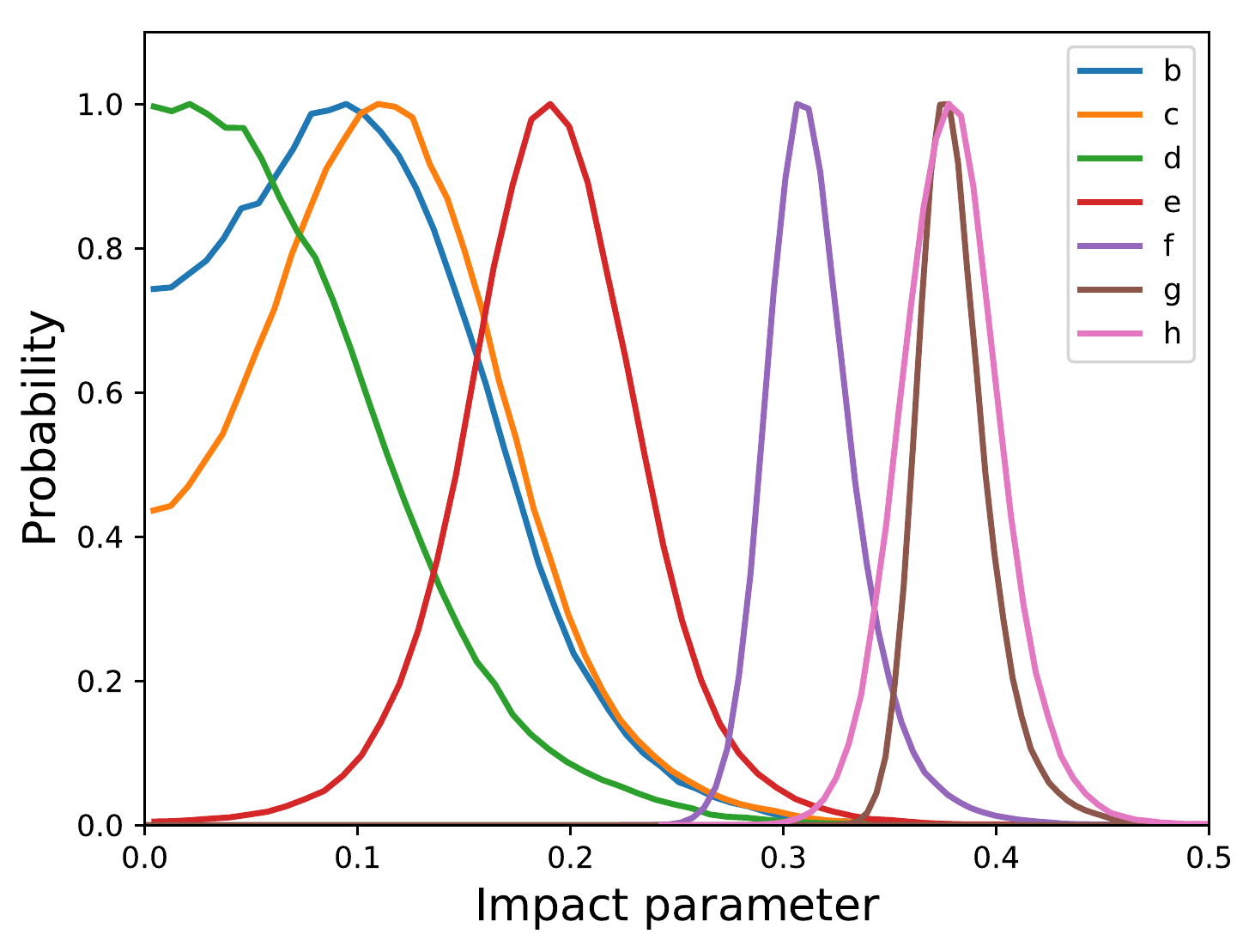}
    \oscaption{julia/photodynamic_table_noprior.jl}%
    {Probability of planet impact parameters using the photodynamic model described in the text.}
    \label{fig:impact_parameter}
\end{figure}

The outer four planets, $e$ through $h$, have inclinations which are more
precisely determined, and, remarkably, {their peak probabilities} are aligned
very closely, to less than 0.1$^\circ$, save for the degeneracy of $I$ vs.
$180-I$.  The inner three planets have poorer
constraints upon their inclinations due to the larger uncertainty of their
impact parameters (as seen in Figure \ref{fig:impact_parameter}).  Yet, their inclination posteriors have significant overlap
with the outer four planets.

As just mentioned, since each inclination may only be inferred relative to the center of
the star, the derived distribution is reflected through $180-I$.
However, if some of the planets orbited above and some below the plane of disk
of the star, it would be {\it very} improbable for the outer four planets to show
such a precise alignment.  We conclude that it {may be} likely that all of the planets {transit}
the same {hemisphere} of the star as shown in \citet{Luger2017b}: the planets'
3D orbital inclinations are likely precisely aligned.   This also implies
that their longitudes of ascending node are likely aligned as well, and
so in principle we can place a prior on the scatter of the mutual inclinations
of the planets.  We have re-run a photodynamic Markov chain with an inclination prior such
that the planets' inclinations are drawn from a Gaussian about their mean value,
with a standard deviation of $\sigma_\theta$ which is allowed to freely vary in
the chain.  We find a very tightly aligned distribution of inclinations under
this assumption, shown in Figure \ref{fig:inclination_prior}.  We also find that
very small values of $\sigma_\theta$ are preferred, with $ \sigma_\theta = {0.041^\circ}_{-0.016^\circ}^{+0.031^\circ} $.
If the outer and inner planets are in fact derived from a common inclination
distribution, this implies that the TRAPPIST-1 planetary orbits are {\it extremely}
flat, even flatter than the Galilean moons which have a dispersion in inclination of
$0.25^\circ$.

\begin{figure}
    \centering
    \includegraphics[width=\hsize]{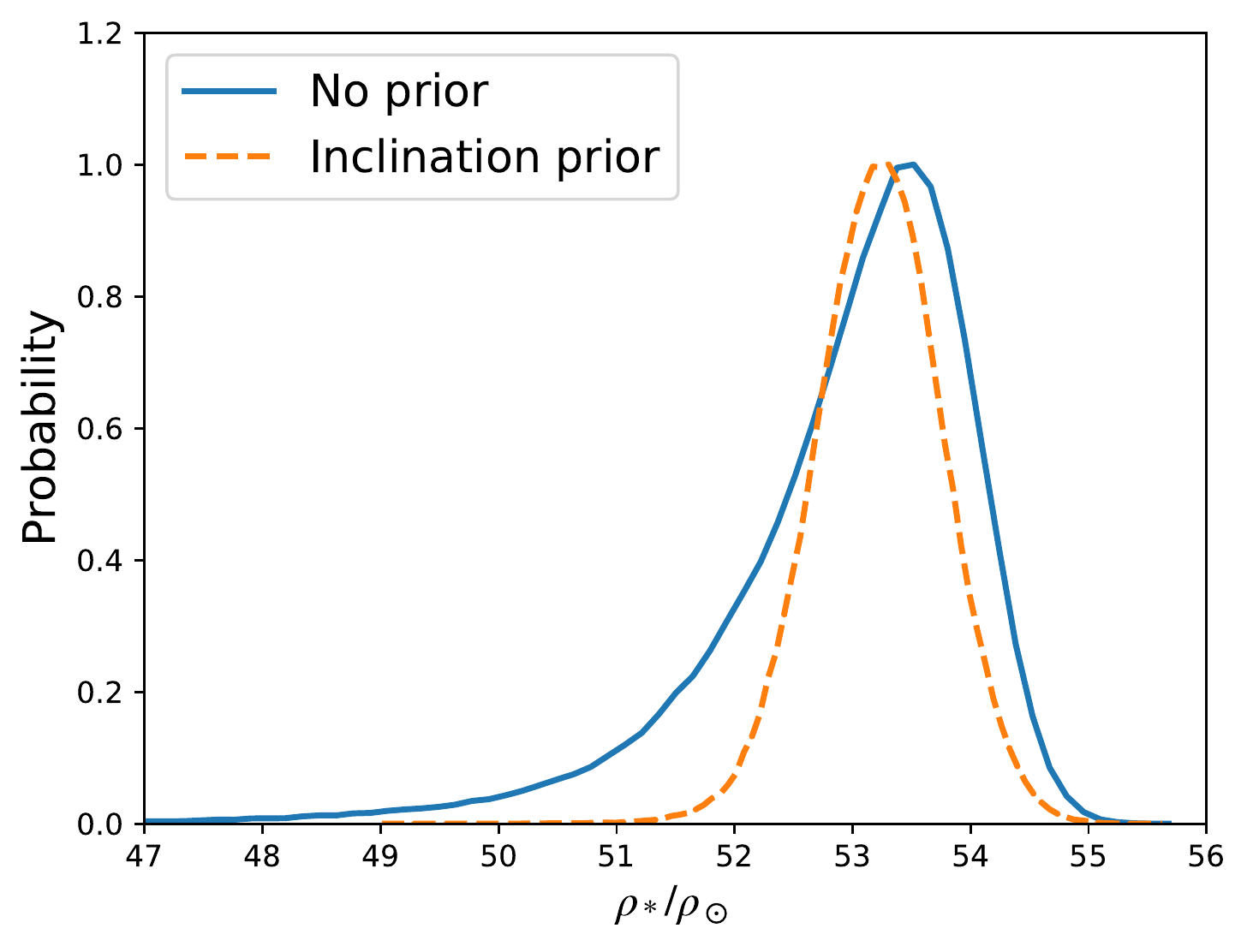}
    \oscaption{julia/photodynamic_table_noprior.jl}%
    {Stellar density derived from the photodynamic model relative to the solar density, with no prior (blue solid line) and with relative inclination prior (orange dashed line).}
    \label{fig:stellar_density}
\end{figure}

\begin{figure}
    \centering
    \includegraphics[width=\hsize]{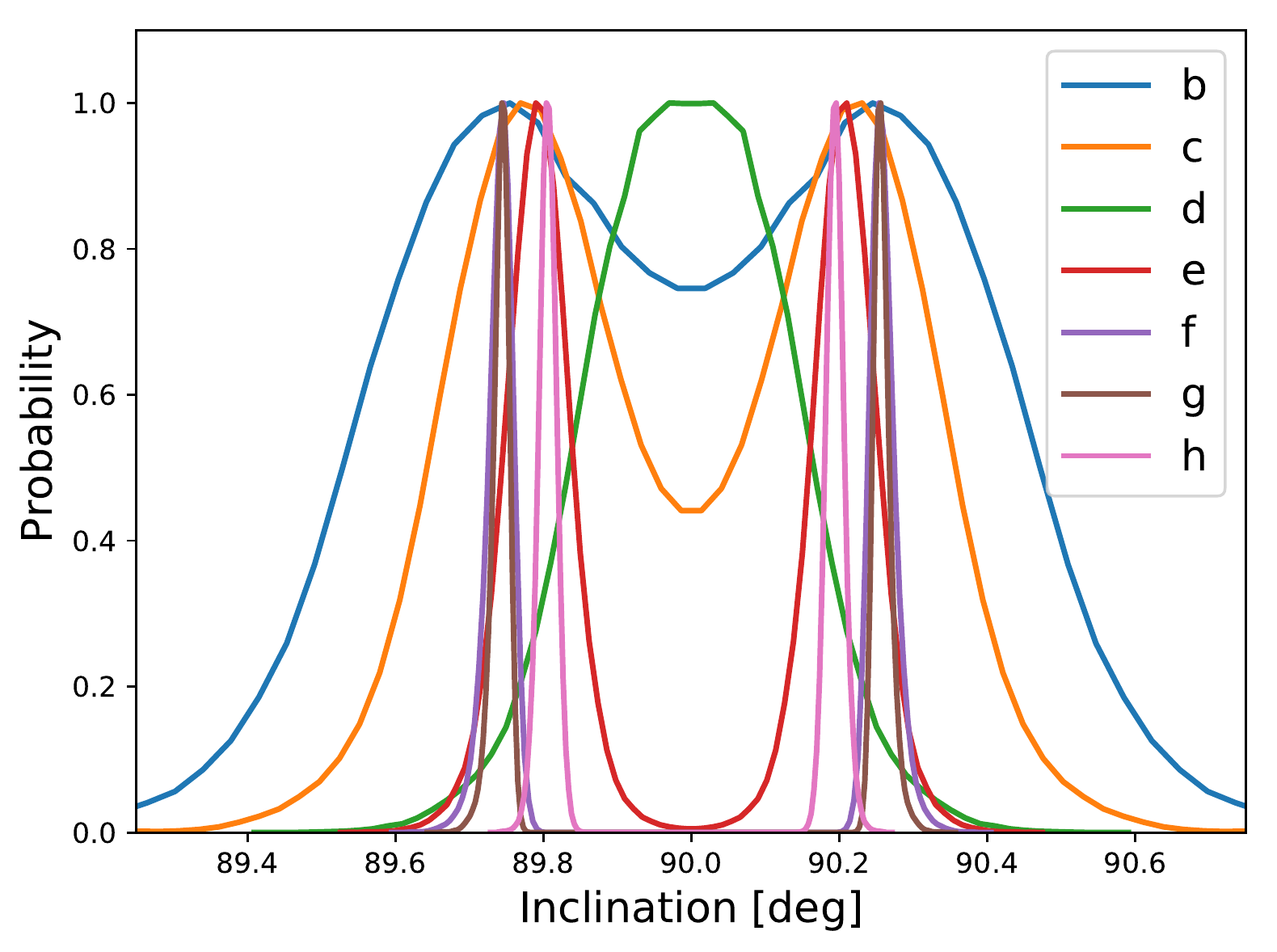}
    \oscaption{julia/photodynamic_table_noprior.jl}%
    {Posterior distribution of inclination angles of the planets given
        the photodynamical model.}
    \label{fig:inclination_distribution}
\end{figure}

\begin{figure}
    \centering
    \includegraphics[width=\hsize]{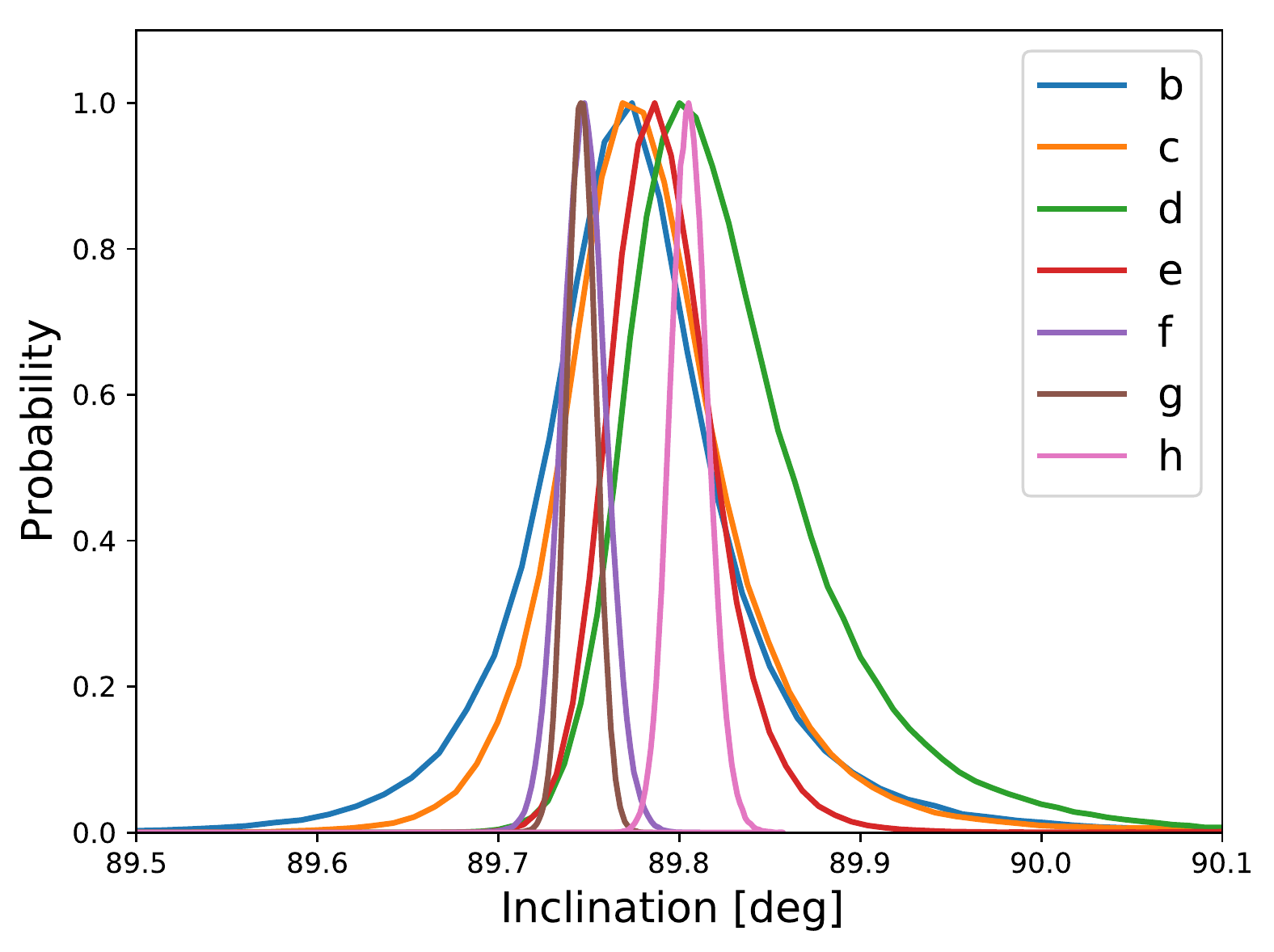}
    \oscaption{julia/photodynamic_table_nouprior_incprior.jl}%
    {Posterior distribution of inclination angles of the planets from the photodynamical model assuming
        a prior on the mutual inclinations of $\Pi_i (2\pi\sigma_\theta^2)^{-1/2} e^{(I_i-\langle I\rangle)^2(2\sigma_\theta^2)^{-1}}$.}
    \label{fig:inclination_prior}
\end{figure}

The inclination prior also enables a more precise and symmetric
estimate of the density of the star, $\rho_*/\rho_\odot {=} 53.22{\pm} 0.53$.  Why is this?  Well, the
inclination prior tightens the distribution of the impact parameters of planets b
and c (as can be seen by comparing Figures \ref{fig:inclination_distribution} and
\ref{fig:inclination_prior}).  These inner two planets have deep and frequent
transits and the sharpest ingress and egress, and hence they provide the tightest
constraint upon the density of the star of all seven planets \citep{Ducrot2020}.
Thus, given that the inclination prior tightens the distributions of inclinations
of these two planets, the stellar density posterior is correspondingly tighter, and
the low stellar density tail of the posterior is eliminated (see Figure \ref{fig:stellar_density}).  Despite this tighter
constraint upon the stellar density, we decide to forego its use in computing the
densities of the planets given the assumptions inherent in the inclination prior.

\begin{table*}
    \centering
    \scriptsize
    \setlength\tabcolsep{3pt}
    \begin{tabular}{l|c|c|c|c|c|c|c}
        \hline
        Parameter:      & $\rho_*/\rho_\odot$            & $q_{1,Ch 1}$                  & $q_{2,Ch 1}$                  & $ q_{1,Ch 2}$                 & $q_{2,Ch 2}$                  &                               & \cr
        Value:          & $53.17_{- 1.18}^{+0.72}$ & $0.133{\pm} 0.052 $         & $0.26{\pm} 0.19 $         & $0.059{\pm} 0.024 $         & $0.49{\pm} 0.20 $         &                               & \cr
        Parameter:      & $ \rho_*$ [g/cm$^3$]         & $u_\mathrm{1,Ch 1}$           & $u_\mathrm{2,Ch 1}$           & $u_\mathrm{1,Ch 2}$           & $u_\mathrm{2,Ch 2}$           &                               & \cr
        Value:          & $75.05_{- 1.66}^{+1.02}$ & $0.161{\pm} 0.093 $         & $0.20{\pm} 0.15 $         & $0.218{\pm} 0.056 $         & $0.021{\pm} 0.098 $         &                               & \cr
        \hline
        Planet:         & b                              & c                             & d                             & e                             & f                             & g                             & h\cr
        $R_p/R_*$ &  $0.08590 \pm 0.00037$ & $0.08440 \pm 0.00038$ & $0.06063 \pm 0.00052$ & $0.07079 \pm 0.00055$ & $0.08040 \pm 0.00047$ & $0.08692 \pm 0.00053$ & $0.05809 \pm 0.00087$ \cr
        Depth [\%]      & $0.7378{\pm} 0.0064$           & $0.7123{\pm} 0.0064$          & $0.3676{\pm} 0.0063$          & $0.5012{\pm} 0.0078$          & $0.6465{\pm} 0.0076$          & $0.7555{\pm} 0.0092$          & $0.3375{\pm} 0.0101$ \cr
	T [min] &  $ 36.06 \pm   0.11$ & $ 42.03 \pm   0.13$ & $ 48.87 \pm   0.24$ & $ 55.76 \pm   0.26$ & $ 62.85 \pm   0.25$ & $ 68.24 \pm   0.28$ & $ 76.16 \pm   0.56$ \cr
	$\tau$ [min] &  $ 2.889 \pm  0.046$ & $ 3.320 \pm  0.054$ & $ 2.816 \pm  0.044$ & $ 3.825 \pm  0.071$ & $ 5.158 \pm  0.089$ & $ 6.310 \pm  0.109$ & $ 4.846 \pm  0.113$ \cr
	$b/R_*$ &  $ 0.095_{-  0.061}^{+ 0.065}$ & $ 0.109_{-  0.061}^{+ 0.059}$ & $ 0.063_{-  0.043}^{+ 0.063}$ & $ 0.191_{-  0.041}^{+ 0.041}$ & $ 0.312_{-  0.018}^{+ 0.023}$ & $ 0.379_{-  0.014}^{+ 0.018}$ & $ 0.378_{-  0.023}^{+ 0.024}$ \cr
        $a/R_*$         & $20.843_{- 0.155}^{+0.094}$    & $28.549_{- 0.212}^{+0.129}$   & $40.216_{- 0.299}^{+0.182}$   & $52.855_{- 0.392}^{+0.239}$   & $69.543_{- 0.516}^{+0.314}$   & $84.591_{- 0.628}^{+0.382}$   & $111.817_{- 0.830}^{+0.505}$ \cr
        $ I $[$^\circ$] & $89.728{\pm}  0.165$           & $89.778{\pm}  0.118$          & $89.896{\pm}  0.077$          & $89.793{\pm}  0.048$          & $89.740{\pm}  0.019$          & $89.742{\pm}  0.012$          & $89.805{\pm}  0.013$ \cr
        \hline
    \end{tabular}
    \caption{Parameters derived from the photodynamic model.  Top: Stellar density (in units of solar density), limb-darkening parameters ($q_1, q_2$) in Spitzer Channel 1 and 2, and stellar density in cgs units and limb-darkening parameters $u_1$ and $u_2$. Bottom: {Planet-to-star radius ratio}, $R_p/R_*$; transit depth, $(R_p/R_*)^2$;  transit duration, $T$ (from first to fourth contact); ingress/{egress} duration, $\tau$ (from first to second contact {or} third to fourth contact); impact parameter in units of stellar radius, $b_0$ (assumed to be positive); ratio of semi-major axis to stellar radius,
        $a/R_*${; and inclination $I$ in degrees (for $b_0>0$).}}
    \label{tab:photodynamic_params}
\end{table*}

The coplanarity of the planets may be used to constrain the presence of a
more distant, inclined planet given the scatter in their mutual inclinations
induced by gravitational perturbations \citep{JontofHutter2018}.  Such an
analysis should be carried out, but we leave this to future work.

\section{Planet densities and mass-radius relation} \label{sec:mass_radius_relation}

With the completion of the transit-timing analysis and photodynamic analysis, we are now ready to revisit the mass-radius relation of the TRAPPIST-1 planets.

The only component missing is a constraint upon the mass of the host star.  We use the recent analysis by \citet{Mann2019}, who have constructed a sample of nearby M-dwarf binaries to calibrate the mass-luminosity ($M_*-M_{K_S}$) relation of M-dwarfs down to a mass of 0.075 $M_\odot$.\footnote{Note that ``M" is being used in three ways here: spectral category (M-dwarf), stellar mass ($M_*$), and
    absolute magnitude in the $K_S$ band, $M_{K_S}$.}   Given the precise parallax measurement available for TRAPPIST-1 thanks to GAIA {\citep{Lindegren2018}}, the relation yields an estimated mass of $M_* = 0.0898{\pm}0.0023 M_\odot$.

To derive the masses of the planets, we draw {planet-to-star mass ratios} from the posterior distribution of the transit-timing analysis (\S \ref{sec:transit_timing}), which we multiply by the mass of the star drawn from a normal distribution with $M_* = 0.0898{\pm}0.0023 M_\odot$. We then draw the {planet-to-star radius ratios} and stellar density from the posterior distribution from the photodynamic analysis (\S \ref{sec:photodynamics}).
With the same mass draw, we compute the stellar radius as
\begin{equation}
    R_* = \left(\frac{M_*}{M_\odot}\frac{\rho_\odot}{\rho_*}\right)^{-1/3} R_\odot,
\end{equation}
which we multiply by each of the radius ratios drawn from the same sample to obtain the planet radii.   We carry this out for a large number of samples to derive the probability distribution of the masses and radii of the entire posterior probability sample of the planets.

\begin{figure*}
    \centering
    \includegraphics[width=\hsize]{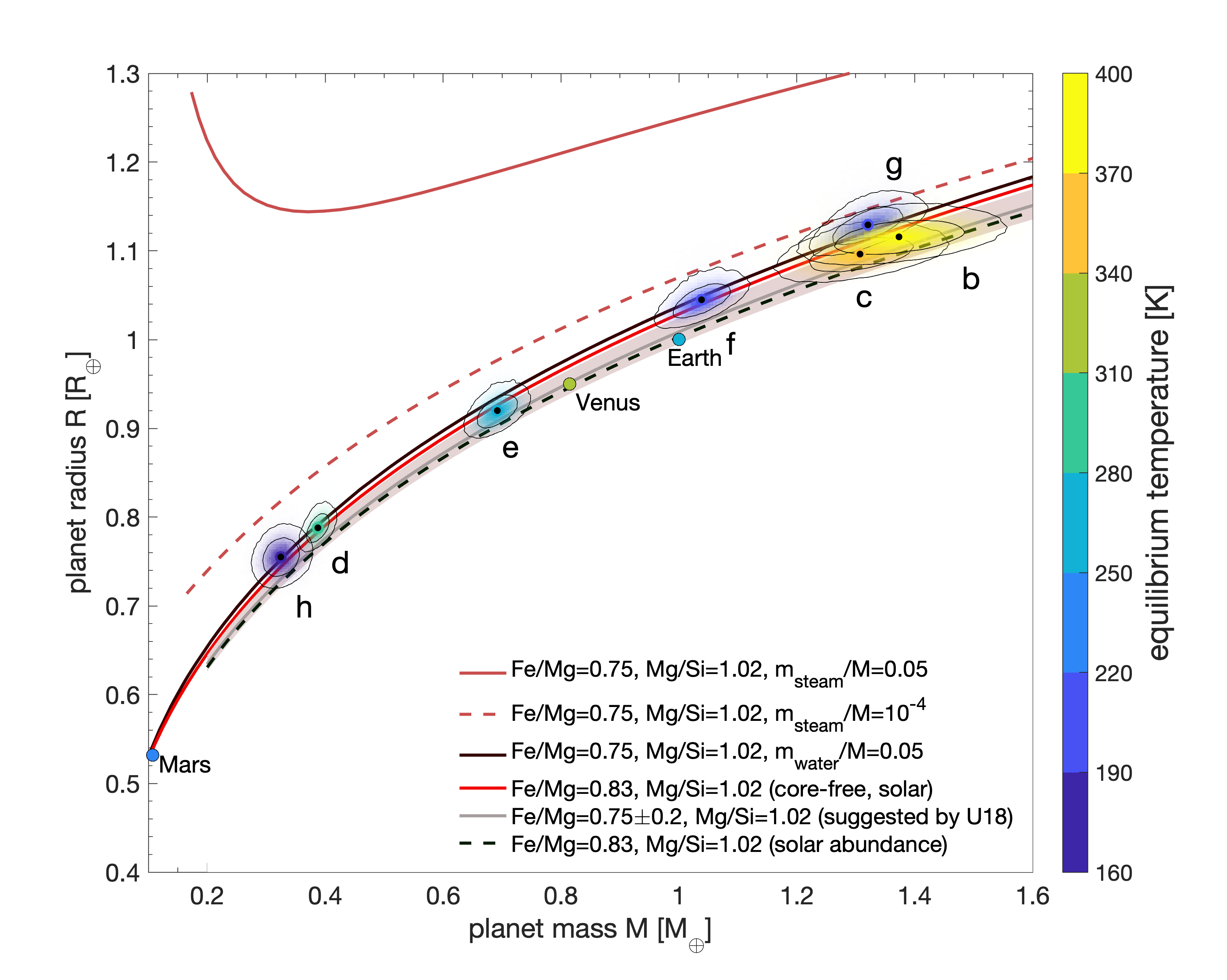}
    \oscaption{matlab/Plot_Eric.m}%
    {Mass-radius relation for the seven TRAPPIST-1 planets based on
        our transit-timing and photodynamic analysis.
        Each planet's posterior probability is colored by the equilibrium temperature
        (see colorbar), with the intensity proportional to probability, while the 1 and 2$\sigma$
        confidence levels from the Markov chain posterior are plotted with solid lines.
        Theoretical mass-radius relations are overplotted using the model in \citet{Dorn2016} for an Earth-like {molar} Fe/Mg=0.83 ratio with a core ({black} dashed) and core-free ({red}), and a range of cored models with {molar} Fe/Mg = $0.75{\pm}0.2$ ({grey}).  {{\it U18}} refers to \citet{Unterborn2018a} (see text).
        The solid black line was calculated for a 5$\%$ water composition, for irradiation low enough (i.e. for planets e, f, g and h) that water is condensed on the surface  (assuming a surface pressure of 1 bar and a surface temperature of 300 K). The {umber} dashed and solid lines were calculated for a 0.01$\%$ and a 5$\%$ water composition, respectively, for irradiation high enough (i.e. for planets b, c and d) that water has fully evaporated in the atmosphere, with the {U18} interior model with Fe/Mg = 0.83 and Mg/Si = 1.02 \citep{Turbet2020}.
        The Earth, Venus and Mars are plotted as single points, also colored by their equilibrium temperatures.}
    \label{fig:mass_radius_relation}
\end{figure*}

The probability distribution for the masses and radii of the seven planets are shown in Figure \ref{fig:mass_radius_relation}.  The maximum likelihood values and the posterior distributions (for 1- and 2-$\sigma$ confidence) are both plotted in this figure.  We postpone to \S\ref{sec:theoretical_interpretation} a detailed analysis of the densities and resulting constraints on the bulk compositions of the planets.

In addition to masses and radii, we also derive other planetary properties,
given in Table \ref{tab:uber_table}.  Each of the planets has a density
intermediate between Mars ($\rho_{\mars}$ = 3.9335 g/cm$^3$ = 0.713 $\rho_\oplus$) and
Earth ($\rho_\oplus = 5.514$ g/cm$^3$).  The surface gravities span a range
from 57\% of Earth (planet h) to 110\% of Earth (planet b).

\begin{table*}
    \centering
    \setlength\tabcolsep{3pt}
    \begin{tabular}{l|c|c|c|c|c|c|c|}
        Planet:                                                    & b                              & c                              & d                              & e                              & f                              & g                              & h \cr
        \hline
        $ R$ [$R_\oplus$]                                          & $  1.116_{- 0.012}^{+ 0.014} $ & $  1.097_{- 0.012}^{+ 0.014} $ & $  0.788_{- 0.010}^{+ 0.011} $ & $  0.920_{- 0.012}^{+ 0.013} $ & $  1.045_{- 0.012}^{+ 0.013} $ & $  1.129_{- 0.013}^{+ 0.015} $ & $  0.755_{- 0.014}^{+ 0.014} $  \cr
        $ M$ [$M_\oplus$]                                          & $  1.374{\pm}  0.069 $         & $  1.308{\pm}  0.056 $         & $  0.388{\pm}  0.012 $         & $  0.692{\pm}  0.022 $         & $  1.039{\pm}  0.031 $         & $  1.321{\pm}  0.038 $         & $  0.326{\pm}  0.020 $  \cr
        $ \rho$ [$\rho_\oplus$]                                    & $  0.987_{- 0.050}^{+ 0.048} $ & $  0.991_{- 0.043}^{+ 0.040} $ & $  0.792_{- 0.030}^{+ 0.028} $ & $  0.889_{- 0.033}^{+ 0.030} $ & $  0.911_{- 0.029}^{+ 0.025} $ & $  0.917_{- 0.029}^{+ 0.025} $ & $  0.755_{- 0.055}^{+ 0.059} $  \cr
        $ g $ [$g_\oplus$]                                         & $  1.102{\pm}  0.052 $         & $  1.086{\pm}  0.043 $         & $  0.624{\pm}  0.019 $         & $  0.817{\pm}  0.024 $         & $  0.951{\pm}  0.024 $         & $  1.035{\pm}  0.026 $         & $  0.570{\pm}  0.038 $  \cr
        $ v_\mathrm{esc} $ [$v_\mathrm{esc,\oplus}$]               & $  1.109 \pm  0.026 $          & $  1.092 \pm  0.022 $          & $  0.701 \pm  0.010 $          & $  0.867 \pm  0.012 $          & $  0.997 \pm  0.012 $          & $  1.081 \pm  0.013 $          & $  0.656 \pm  0.020 $  \cr
        $ S $ [$ S_\oplus $]                                       & $  4.153_{- 0.159}^{+ 0.161} $ & $  2.214_{- 0.085}^{+ 0.086} $ & $  1.115_{- 0.043}^{+ 0.043} $ & $  0.646_{- 0.025}^{+ 0.025} $ & $  0.373_{- 0.014}^{+ 0.015} $ & $  0.252_{- 0.010}^{+ 0.010} $ & $  0.144_{- 0.006}^{+ 0.006} $  \cr
        $ a $ [$ 10^{-2} \mathrm{AU} $]                            & $  1.154{\pm}  0.010 $         & $  1.580{\pm}  0.013 $         & $  2.227{\pm}  0.019 $         & $  2.925{\pm}  0.025 $         & $  3.849{\pm}  0.033 $         & $  4.683{\pm}  0.040 $         & $  6.189{\pm}  0.053 $  \cr
        \hline
        $ R$ [$10^8$ cm]                                           & $  7.119_{- 0.077}^{+ 0.087} $ & $  6.995_{- 0.077}^{+ 0.086} $ & $  5.026_{- 0.066}^{+ 0.071} $ & $  5.868_{- 0.075}^{+ 0.082} $ & $  6.664_{- 0.077}^{+ 0.085} $ & $  7.204_{- 0.085}^{+ 0.094} $ & $  4.817_{- 0.088}^{+ 0.091} $  \cr
        $ M$ [$10^{27}$ g]                                         & $  8.211{\pm}  0.412 $         & $  7.814{\pm}  0.335 $         & $  2.316{\pm}  0.074 $         & $  4.132{\pm}  0.130 $         & $  6.205{\pm}  0.184 $         & $  7.890{\pm}  0.226 $         & $  1.945{\pm}  0.122 $  \cr
        $ \rho $ [ g cm$^{-3}$]                                    & $  5.425_{- 0.272}^{+ 0.265} $ & $  5.447_{- 0.235}^{+ 0.222} $ & $  4.354_{- 0.163}^{+ 0.156} $ & $  4.885_{- 0.182}^{+ 0.168} $ & $  5.009_{- 0.158}^{+ 0.138} $ & $  5.042_{- 0.158}^{+ 0.136} $ & $  4.147_{- 0.302}^{+ 0.322} $  \cr
        $ g$ [$10$ m s$^{-2}$]                                     & $  1.080{\pm}  0.051 $         & $  1.065{\pm}  0.042 $         & $  0.611{\pm}  0.019 $         & $  0.801{\pm}  0.024 $         & $  0.932{\pm}  0.024 $         & $  1.015{\pm}  0.025 $         & $  0.558{\pm}  0.037 $  \cr
        $ v_\mathrm{esc}$ [$\frac{\mathrm{km}}{\mathrm{s}}$]       & $ 12.400 \pm  0.292 $          & $ 12.205 \pm  0.241 $          & $  7.839 \pm  0.110 $          & $  9.694 \pm  0.133 $          & $ 11.145 \pm  0.137 $          & $ 12.087 \pm  0.142 $          & $  7.335 \pm  0.227 $  \cr
        $ S $ [$10^6\frac{\mathrm{erg}}{\mathrm{cm}^2\mathrm{s}}$] & $  5.652_{- 0.216}^{+ 0.220} $ & $  3.013_{- 0.115}^{+ 0.117} $ & $  1.518_{- 0.058}^{+ 0.059} $ & $  0.879_{- 0.034}^{+ 0.034} $ & $  0.508_{- 0.019}^{+ 0.020} $ & $  0.343_{- 0.013}^{+ 0.013} $ & $  0.196_{- 0.008}^{+ 0.008} $  \cr
        $ a $ [$10^{11}$ cm]                                       & $  1.726{\pm}  0.015 $         & $  2.364{\pm}  0.020 $         & $  3.331{\pm}  0.028 $         & $  4.376{\pm}  0.037 $         & $  5.758{\pm}  0.049 $         & $  7.006{\pm}  0.060 $         & $  9.259{\pm}  0.079 $  \cr
    \end{tabular}
    \caption{Planetary parameters from combining the transit-timing
        and photodynamic analysis.  The units are given with respect to Earth first, and cgs second.}
    \label{tab:uber_table}
\end{table*}

\section{Stellar parameters} \label{sec:stellar_params}

A byproduct of our analysis is a revision of the properties of the host
star.  {The} empirically-based mass estimate {for} the star
based on \citet{Mann2019} is consistent with the mass derived
by \citet{vanGrootel2018}, who first proposed that the mass of the TRAPPIST-1 star
is $\approx 0.09 M_\odot$ based upon stellar evolution models and
a ground-based parallax measurement.  \citet{Ducrot2020} find a luminosity
for the star of $L = (5.53{\pm}0.19){\times}10^{-4} L_\odot$, which, when
compared with stellar evolution models, yields a mass of $M= 0.09016{\pm}0.0010 
    M_\odot$, which is also consistent with the \citet{Mann2019} value.
\citet{Burgasser2017} found an older age for the host
star, $7.6{\pm}2.2$ Gyr, which implies an inflated radius for the star compared
with evolutionary models.

\begin{table}
    \centering
    \begin{tabular}{l|c|l}
        Parameter                             & Value                         & Ref \cr
        \hline
        $ M $[$ M_\odot $]                    & $ 0.0898{\pm} 0.0023 $        & \citet{Mann2019}\cr
        $ R $[$ R_\odot $]                    & $ 0.1192{\pm} 0.0013$         & This paper \cr
        $ L $ [$ L_\odot $]                   & $ 0.000553{\pm} 0.000019 $    & \citet{Ducrot2020}\cr
        $ T_{eff} $ [K]                       & $2566{\pm}   26$              & This paper \cr
        $ \log_{10} (g \mathrm{[cm / s^2]}) $ & $5.2396_{-0.0073}^{+0.0056} $ & This paper \cr
    \end{tabular}
    \caption{Updated stellar parameters based on the combined analysis.}
    \label{tab:stellar_parameters}
\end{table}

Our analysis differs slightly from our prior Spitzer analyses \citep{Delrez2018a,Ducrot2020} in that we do not place a prior upon the {quadratic} limb-darkening coefficients of the TRAPPIST-1 host star.  This is motivated by the fact that late M dwarf atmospheres are very complex to model and have yet to match observed spectra precisely \citep{Allard2011,Allard2012,Juncher2017}, and thus it is possible that limb-darkening predictions may not be reliable.  {We investigated using a higher-order
quartic limb-darkening law, and found that this was disfavored by the Bayesian Information Criterion,
and that the best-fit model differed negligibly in the model parameters.  We also simulated
more realistic limb-darkening models based on 3D stellar atmospheres \citep{Claret2018} and 
found that a quadratic law was sufficient to recover the correct model parameters with negligible 
systematic errors.}

\begin{figure}
    \centering
    \includegraphics[width=\hsize]{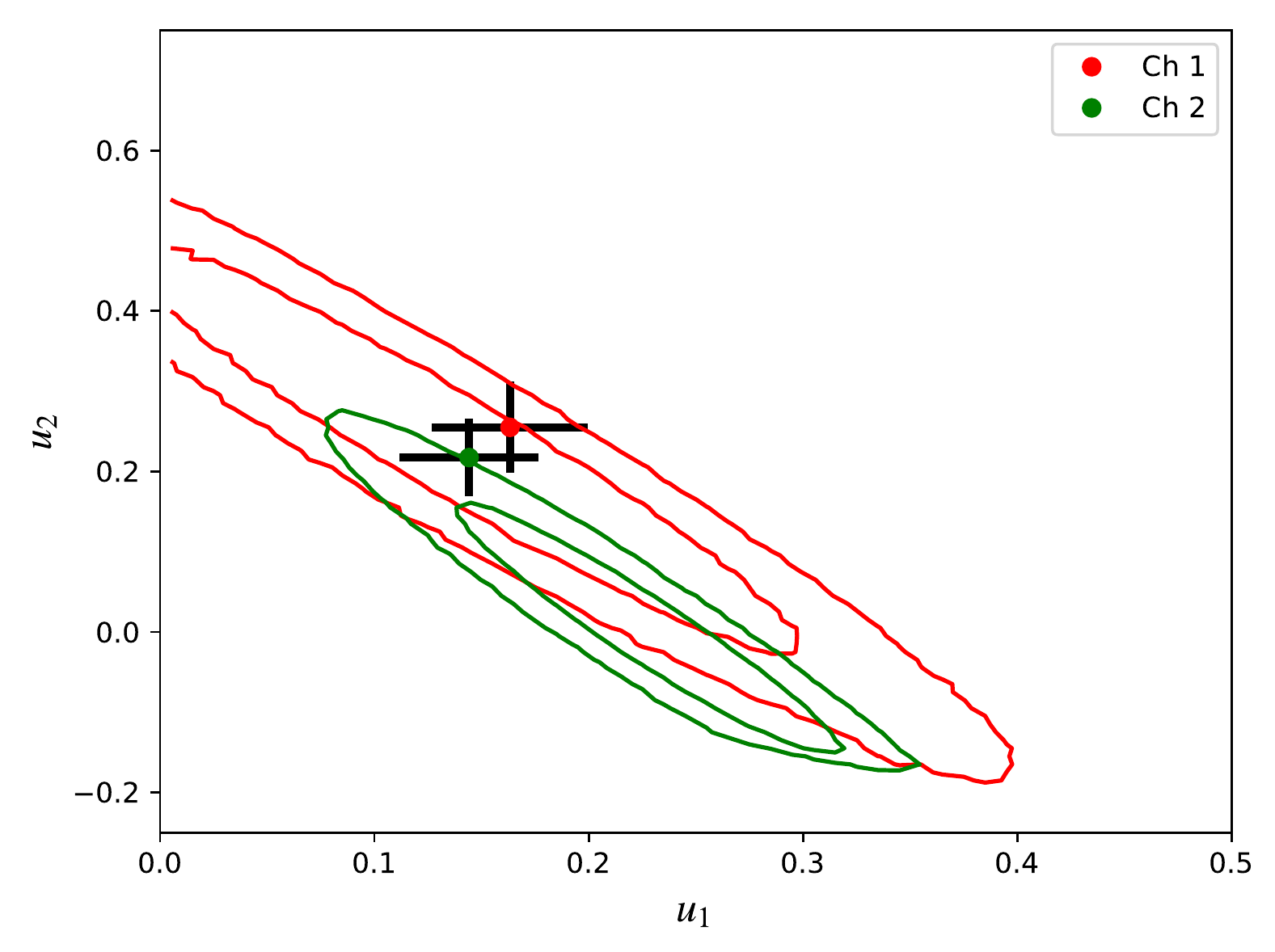}
    \oscaption{julia/photodynamic_table_noprior.jl}%
    {Limb-darkening constraints, 1 and 2$\sigma$ confidence contours.  Red is Spitzer IRAC Channel 1 (3.6$\mu$m), while green is Channel 2 (4.5$\mu$m).  Error bars indicate the limb-darkening parameters and uncertainties used as priors in \citet{Ducrot2020}}.
    \label{fig:limb_darkening}
\end{figure}

The TRAPPIST-1 system has the advantage that the planets sample different chords of the stellar disk (Figure \ref{fig:impact_parameter}{; also see \citealt{Delrez2018a}}), and given the large number of transiting planets, we are afforded multiple constraints upon the stellar limb-darkening parameters.  Figure \ref{fig:limb_darkening} shows our posterior constraints upon the limb-darkening parameters of the star based on our photodynamical model, which are reported in Table \ref{tab:photodynamic_params}.

Based on the updated stellar density, we have updated the physical
parameters of the star.  We adopt the luminosity from \citet{Ducrot2020}
and the mass from \citet{Mann2019} given the complete and careful analysis
from both of those papers.  With our updated constraint upon the density
of the star, we re-derive the other parameters of the star, which are summarized in Table \ref{tab:stellar_parameters}.
{In this table the stellar effective temperature was computed from
the stellar luminosity and radius, with errors computed via Monte Carlo.}

\section{Search for an eighth planet} \label{sec:detection}

With the detection of multiple transits of the six inner planets in TRAPPIST-1, and
a single transit of planet h, a clue as to the orbital period of planet h was the
series of GLRs found between adjacent triplets of planets
\citep{Papaloizou2014}.  This
relation was then used to predict candidate periods of planet h, based on different
integer pairs for its commensurability with planets f and g, and a search through
the prior data eliminated all but one possibility at 18.766 days.  A subsequent
observation of the TRAPPIST-1 system with the K2 spacecraft revealed four more
transits of planet h occurring at precisely the period that was predicted
\citep{Luger2017a}.  The existence of the GLRs amongst the known seven planets has been used to forecast
the possible existence of an eighth planet interior \citep{Pletser2017} and exterior
\citep{Kipping2018} to the seven known transiting planets.  There is yet to be a definitive detection of an eighth transiting planet based upon the currently available data \citep{Ducrot2020}.

It may be possible to detect an exterior eighth planet via transit-timing variations
induced on the inner seven planets.  Planet h should experience the strongest
perturbations by an exterior eighth planet due to the fact that transit-timing
variations are a very strong function of the proximity of planets to one another,
and also to resonance.  Table \ref{tab:laplace_p_i} shows predictions for the
period of planet ``i", $P_\mathrm{i}$, assuming a GLR configuration with planets g and
h given by
\begin{equation}
    P_\mathrm{i} = q/(-p P_\mathrm{g}^{-1} + (p+q) P_\mathrm{h}^{-1})
\end{equation}
for a range of $1 \le p,q \le 3$, which is the same range
of integers for the GLRs amongst the inner seven planets.
Interestingly these cases are all close to a $j{:}j{+}1$ period ratio with planet
h, and thus should strongly perturb planet h due to forcing at this frequency.

\begin{table}
    \centering
    \begin{tabular}{c|c|c|c|c}
        p & q & $P_\mathrm{i}$ [day] & $P_\mathrm{i}/P_\mathrm{h}$ & j     \\
        \hline
        1 & 1 & 39.029               & 2.08                        & 1 \cr
        1 & 2 & 25.347               & 1.35                        & 3 \cr
        1 & 3 & 22.695               & 1.21                        & 4 \cr
        2 & 3 & 28.701               & 1.53                        & 2\cr
    \end{tabular}
    \caption{Predictions for a GLR of planets g and h with
        an eighth planet, planet i, with period $P_\mathrm{i}$.  The ratio with the
        period of planet h is given, as well
        as the value of $j$ for which $P_\mathrm{i}/P_\mathrm{h} \approx (j+1)/j$.}
    \label{tab:laplace_p_i}
\end{table}

We carried out a transit-timing search for an eighth planet by placing
planets with mass ratios between $2{\times}10^{-6}{-}5{\times}10^{-5}$ at
these four trial orbital periods
in a coplanar configuration with the other seven planets drawn from
a random orbital phase at the initial time, and with eccentricity
vector elements drawn from a random normal of width 0.005.   We
placed a Gaussian prior on the eccentricity vector elements of the eighth planet
with a standard deviation of $0.14$ to avoid unstable configurations.
We then optimized the likelihood with the eight-planet model, carrying
out 11,200 optimizations {on 112 CPUs}
with 100 optimizations per {CPU}, lasting seven days each for about 20,000 CPU hours.

We then carried out a search for evidence of perturbations by planet i by
determining if the optimized likelihood of the transiting planets was improved
by adding an eighth planet to the transit timing model, using
the Bayesian Information Criterion (BIC) to penalize the additional degrees
of freedom of the eight-planet model \citep{Wit2012}.  We searched for a change
to BIC for the eight-planet model over the seven-planet model with a difference
of better than $5\log{N_\mathrm{trans}} = 30.5$.  Given that the inner seven
planets show orbital eccentricities with values $\la 0.01$, we only considered
an eighth planet candidate plausible if it shows an eccentricity less than
this cutoff.

\begin{figure}
    \centering
    \includegraphics[width=\hsize]{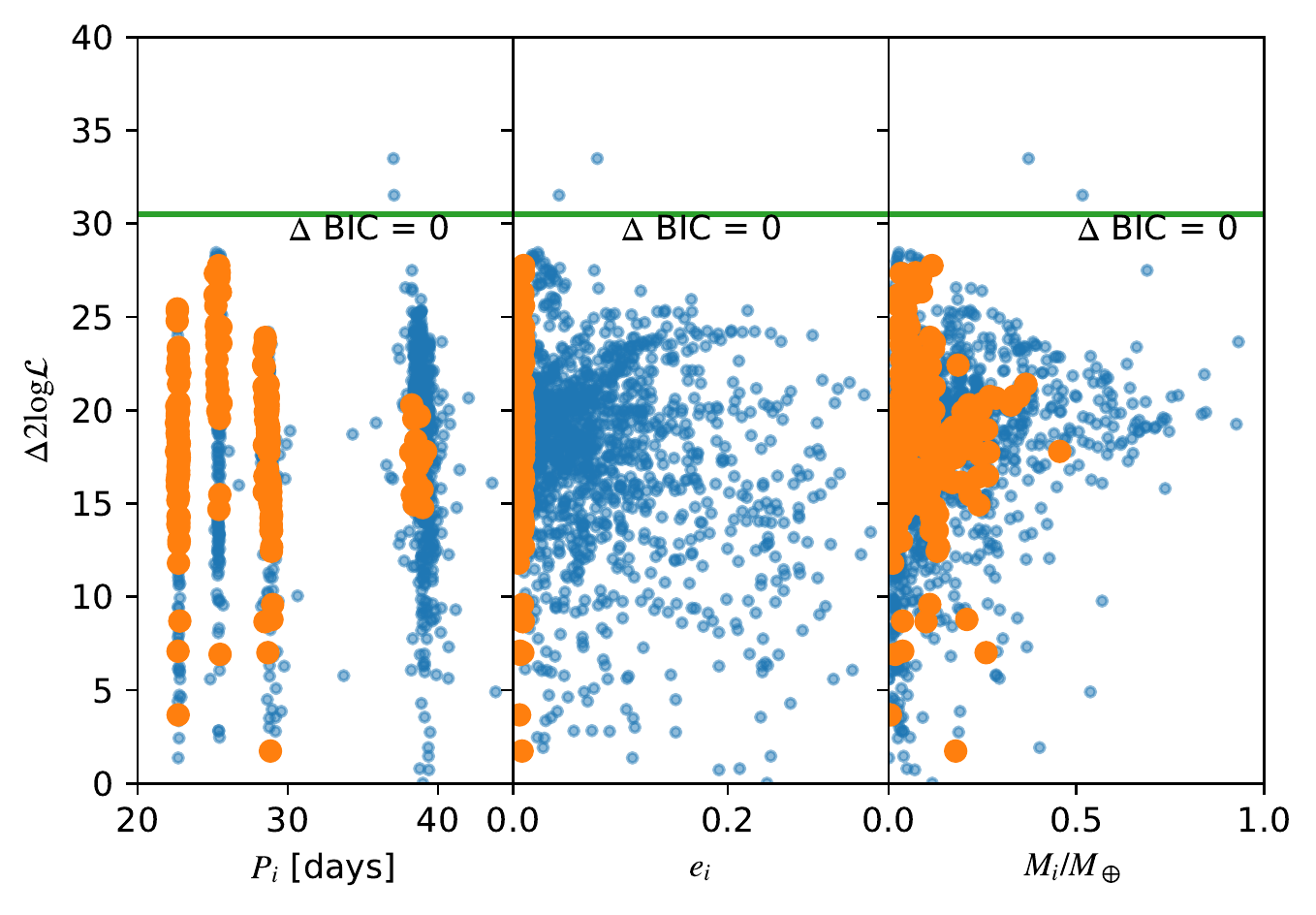}
    \oscaption{julia/TTV_search_8planet_noprior_rand.jl}%
    {Limits on an eighth planet, ``i", for a search near
        the periods in table \ref{tab:laplace_p_i}.  The 8-planet
        models are only plotted if they led to an improvement in log likelihood.
        Only two of the optimized
        likelihoods reach the difference in BIC ${>}0$ indicated on the plots; however, these two cases have an eighth planet with a relatively large eccentricity and are distant from a GLR with g and h.  Orange points have eccentricities smaller than 0.01;  light blue points have larger eccentricities.}
    \label{fig:planet_i}
\end{figure}

In all 11,200 trial optimization cases we found that only two of the eight-planet models
did exceed the BIC criterion, but both significantly exceed an eccentricity
of 0.01.  Figure  \ref{fig:planet_i} shows the change in BIC versus
orbital period and mass for planet ``i", assuming a mass of the star
of $M_* = 0.09 M_\odot$.   These two cases with $\Delta BIC {>} 0$
do not appear to be plausible
planet candidates:  they only just exceed the BIC criterion;  they
both have large eccentricities; and they are not in close proximity to a
GLR with planets g and h (even though the initial parameters of the optimization were started near a GLR).

We also carried out a search for an eighth planet interior to planet
b, and found even smaller improvements in the log likelihood than
in the exterior case.

We have not carried out an exhaustive search for eight-planet models
at other orbital periods due to the significant volume of parameter
space to search.   However, it is still possible that an exterior
eighth planet is perturbing planet h, and may modify its transit
times to a point that affects the posterior masses we infer from
our seven planet model.   In principle one could include the effect
of an eighth planet on the mass inference by adding it to the Markov
chain modeling;  in practice this would be a challenging model to
sample due to the multi-modal nature of the parameter space.
We defer such analysis to future work.

\section{Interior compositions}
\label{sec:theoretical_interpretation}

In this section we present theoretical interpretation of the planets'
interior properties based upon the mass-radius relation we inferred in \S \ref{sec:mass_radius_relation}.  As there is significant degeneracy in the possible interior compositions, we present a menu of different possibilities in \S \ref{sec:interior_compositions}.  However, we start with an approach which is less dependent upon the assumption of interior composition, which we term the ``normalized density."

\subsection{Initial analysis of planet densities across the system}

The probability distribution for the masses and radii of the seven planets are shown in Figure \ref{fig:mass_radius_relation} alongside several theoretical mass-radius relationships added for comparison.
We have added three rocky mass-radius relationships with different bulk Fe/Mg compositions:  (1) {molar}  Fe/Mg$ =0.75{\pm} 0.2$ as suggested by \citet{Unterborn2018a} to represent the rocky interior of all TRAPPIST-1 planets with a 1$\sigma$ range of Fe/Mg ratios consistent with local stellar abundances;  (2) the {Sun}-like value of {molar} Fe/Mg$ = 0.83$ \citep{Lodders2009}; and (3) a core-free model with Earth-like refractory ratios, but in which all of the iron is oxidized in the mantle \citep{ElkinsTanton2008}. Rocky interiors are calculated {similar to the models of Dorn et al. (2016) with two adaptations: we are using the equation of state of \citet{Hakim2018} for pure iron and \citet{Sotin2007} for silicates.} We have also added the theoretical mass-radius relationships for planets endowed with a water layer, both for planets which are irradiated less ({black} line; water) and more ({umber} lines; steam) than the runaway greenhouse irradiation threshold \citep{Turbet2020}.

The comparison of measured masses and radii with theoretical mass-radius relationships reveals several striking results. First, all seven TRAPPIST-1 planets appear to be consistent with a line of interior isocomposition {at the $1\sigma$ level}. There are multiple theoretical mass-radius curves that overlap with all seven planets' mass-radius probability distributions (Fig.\ \ref{fig:mass_radius_relation}), which may be a good indication that the composition varies little from planet to planet. Secondly, all of the TRAPPIST-1 planets have lower uncompressed densities than Solar System terrestrial planets. This likely means that the TRAPPIST-1 planets either have a lighter interior (e.g. lower iron content) or are enriched with volatiles (e.g. water).

\begin{figure}
    \centering
    \includegraphics[width=\hsize]{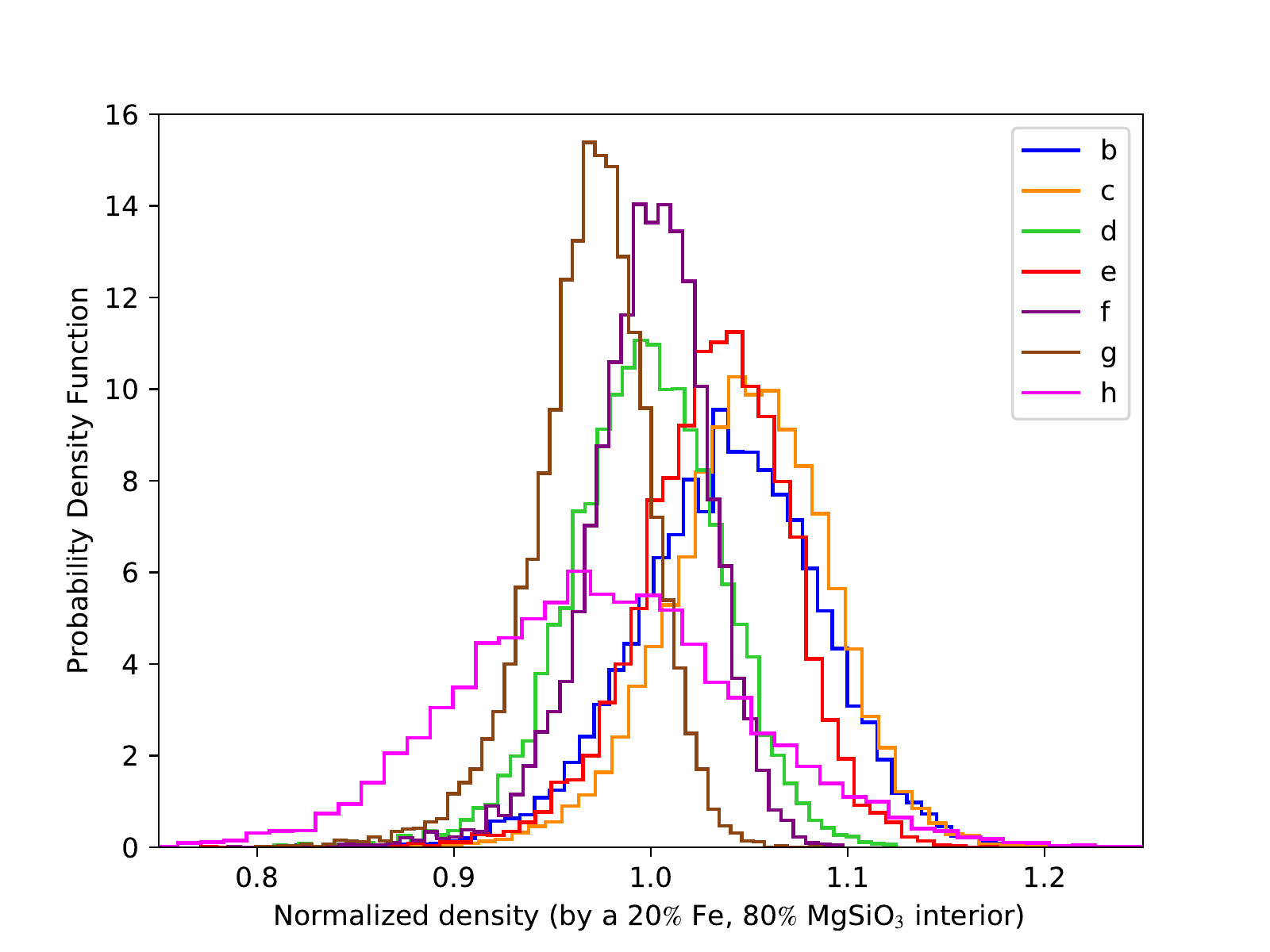}
    \oscaption{python/plot_histogram_normalized_density_agol_et_al.py}%
    {Probability density function of the normalized density of all seven planets in the system.
}
    \label{fig:norm_density_histo}
\end{figure}

\begin{figure}
    \centering
    \includegraphics[width=\hsize]{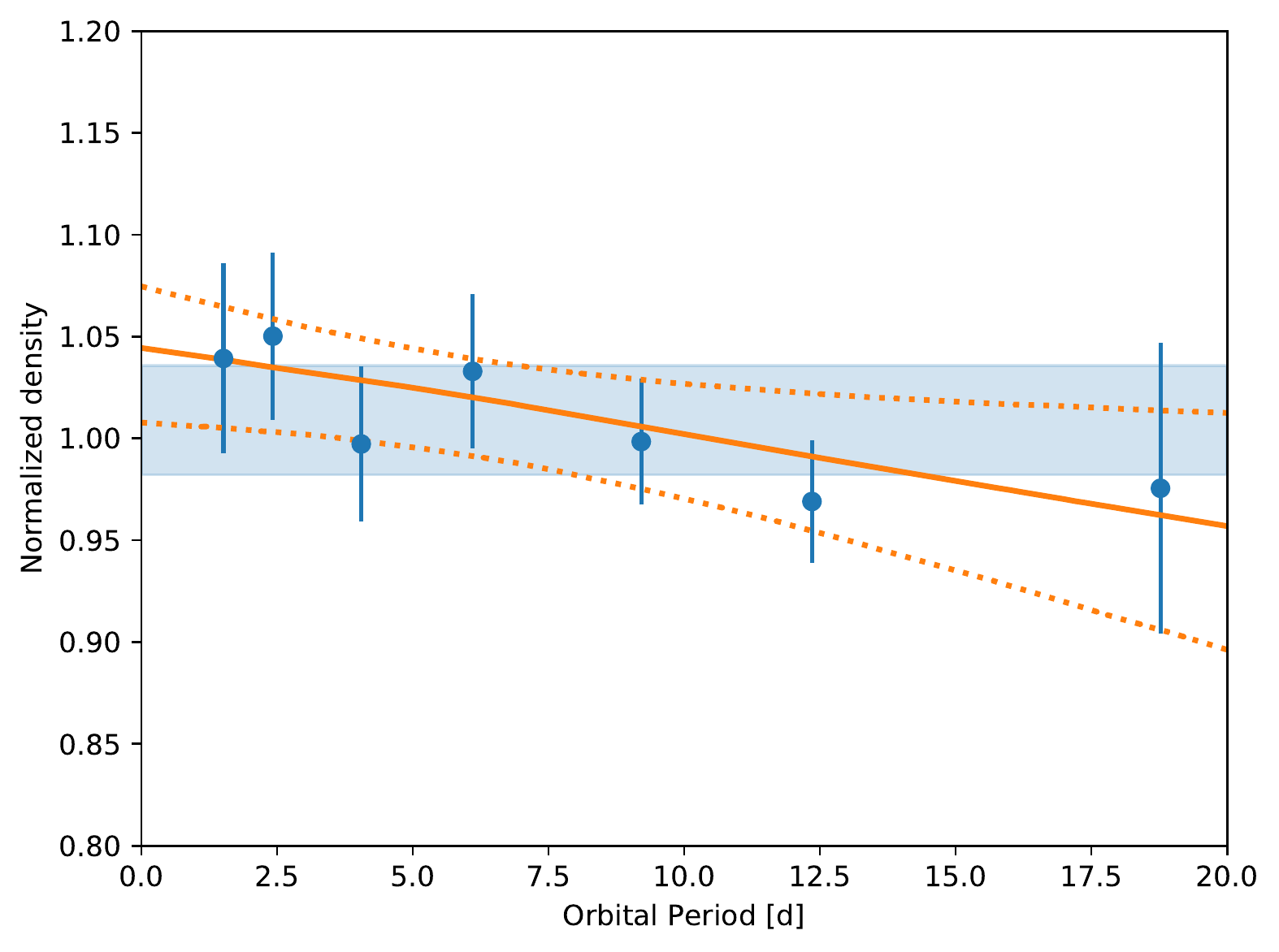}
    \oscaption{julia/plot_norm_density.jl}%
    {Normalized planet densities (with 1$\sigma$ error-bars) versus planet orbital periods. The light blue band is the 68\% confidence interval of the weighted mean normalized density of all seven planets.  The orange lines show the 68.3\% confidence intervals of linear fits to the normalized densities computed from $10^4$
        draws from the posterior.  The mean fit to the normalized density versus period is {$y = a P + b$ where ${a= 1.042{\pm}0.034}$ and ${b=-0.0043{\pm}0.0036}$.}}
    \label{fig:norm_density_vs_period}
\end{figure}

We next searched for variations of density across the planets. For this, {we took} each planetary density calculated from $10^4$ samples {and divided by the density of the closest pair of mass and radius} of a fully differentiated 20 wt\% iron, 80 wt\% silicate (MgSiO$_3$) interior planet with no surface layers, which is less iron rich than Earth.  {A planet with a normalized density of 1 has exactly the same density as the reference model, while a normalized density ${>}1$ (${<}1$) is denser (lighter), than the reference model, respectively.}
Fig.~\ref{fig:norm_density_histo} shows the resulting histograms of the posterior probability of the normalized TRAPPIST-1 planet densities. We then {plot} in Fig.~\ref{fig:norm_density_vs_period} the normalized densities (along with their 1$\sigma$ uncertainty) as a function of the orbital periods of the planets.  The normalized planet density appears very uniform across the seven planets, with perhaps a slight decrease with the increase of the orbital period (or the distance to the host star).  We fit a line to the normalized density, $y$, versus orbital period, $P$, for $10^4$ posterior samples, and found a relation of {$y=(1.042{\pm}0.034){-}(0.0043{\pm}0.0036)P$} where the
coefficients are the 68.3\% confidence interval. There is only weak evidence for a declining trend of normalized density with orbital period: {88\% of the fits to the $10^4$ posterior samples have slopes with a negative value, while 12\% of the slopes fit have a \emph{positive} value}.  If in the future more precise data strengthen this trend, then
this {may indicate} that either (i) the outer planets are depleted in heavy elements (e.g. iron) compared to the inner ones, or (ii) the outer planets are enriched in volatiles (e.g. water) compared to the inner ones.  However, based on the current data we suggest that the planets' compositions could be rather uniform in nature.

The interpretation of these observations in terms of internal compositions is discussed in more detail next.

\subsection{Range of possible interior compositions and volatile contents}  \label{sec:interior_compositions}

In this subsection, we discuss {a} range of possible compositions of the planets based on their measured densities,
starting with a volatile-poor model in which the densities are fit by varying the core-mass fraction (\S \ref{sec:subsection_CMF}),
and followed by an analysis in which the solid planets are taken to have an Earth-like composition, to which is added a water fraction needed to create the observed densities (\S \ref{sec:volatiles}).
Alternatively, the planets might be explained with an {enhanced oxygen content by} which all of the iron is oxidized making the planets core-free (\S \ref{sec:core_free}).

\subsubsection{Core Mass Fraction}
\label{sec:subsection_CMF}

If we assume that the planets' atmospheres contribute a negligible
amount to their total radius, and that the planets are fully differentiated, composed of rocky mantles (MgSiO$_3$) and iron cores only, then the densities may be used to constrain the portion of the planets' mass which is contained within their cores.

We evaluated the core mass fractions {(CMF)} of the TRAPPIST-1 planets as follows. For each mass/radius pair in our posterior distribution we have estimated the core-mass fraction by linearly interpolating {between pre-calculated mass-radius relationships with our employed interior model.} We arbitrarily set each mass/radius pair lighter than a pure silicate (MgSiO$_3$) planet to a CMF of 0. Alternatively, we repeated the same procedure but discarding all CMF values lower or equal to 0. However, we found that the estimate of the core mass fraction is only marginally changed (and only for planets g and h).

\begin{figure}
    \centering
    \includegraphics[width=\hsize]{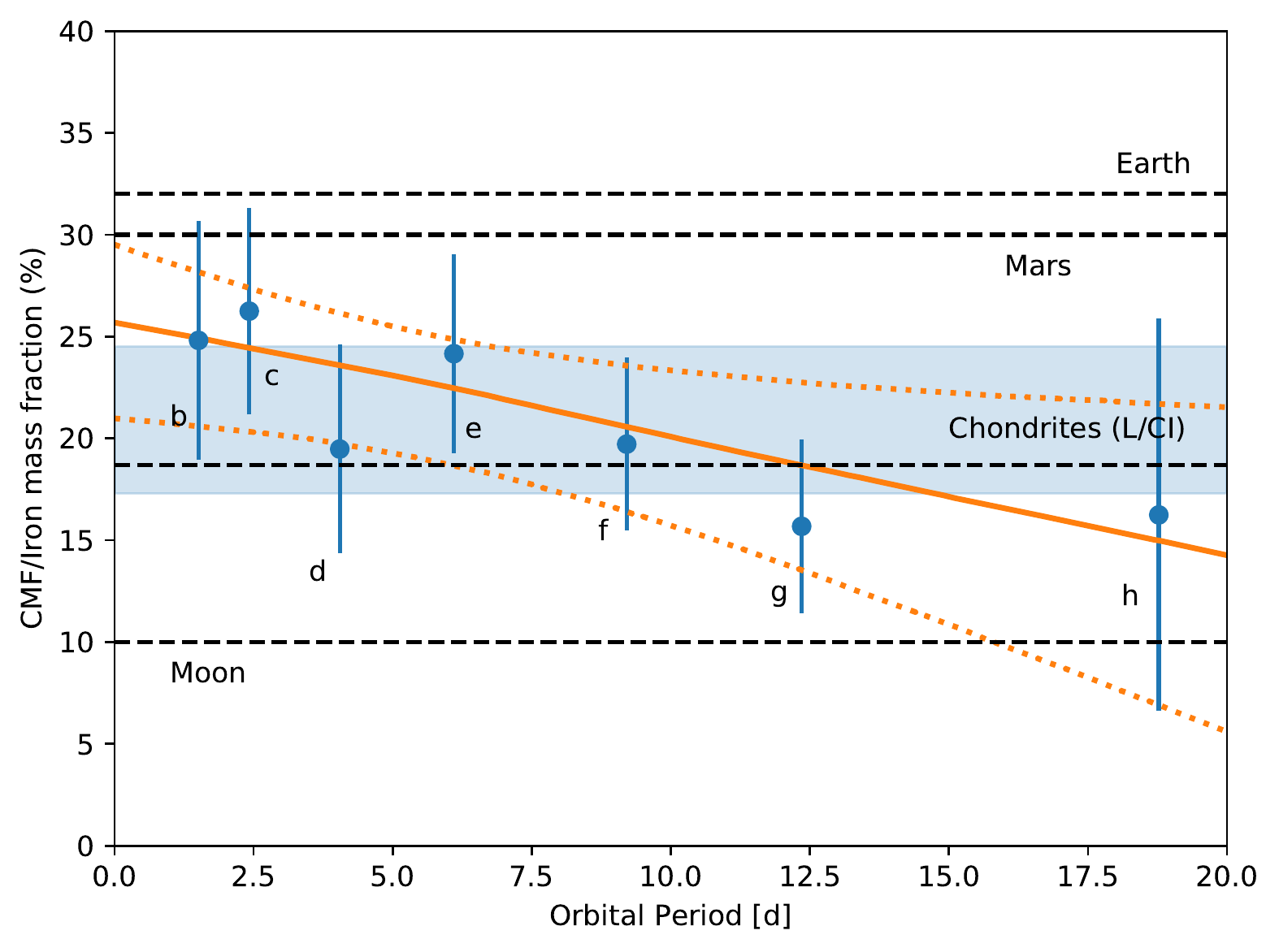}
    \oscaption{julia/plot_cmf.jl}%
    {Iron core mass fraction versus the planetary orbital periods {for a fully-differentiated model with molar Mg/Si = 1.02 and no surface layer}.  The approximate values for Earth \citep{McDonough2014}, Mars \citep{Khan2018}, the Moon \citep{Barr2016}, and common chondrites \citep{Palme2014} are indicated, as well as the 1$\sigma$ confidence intervals of the TRAPPIST-1 planets.
        The light blue box is the 68.3\% confidence region of the weighted mean of all seven planets. The orange lines show the median and 68.3\% confidence interval for linear fits to the $10^4$ posterior values for all seven planets.}
    \label{fig:cmf}
\end{figure}

Our core mass fraction estimates are provided in Fig~\ref{fig:cmf} and Table~\ref{tab:cmf}.  Estimates range from {${16.1_{-4.2}^{+3.5}}$ wt\% for planet g up to ${26.6_{-5.1}^{+4.6}}$ wt\% for planet c}, which,  despite the different central values, have considerable overlapping probability distributions.
Fig~\ref{fig:cmf} shows that within the uncertainties, the CMF/Iron fraction of the planets are very consistent with one another, with the mean of all planets of {${21{\pm}4}$} wt\% (taking into account the correlations between the planets' core-mass fractions).

There may be a slight trend of the inferred CMF, which decreases with increasing orbital period. The trend is qualitatively similar to that reported on the normalized density (see Fig.~\ref{fig:norm_density_vs_period}), with similarly weak support: {only 88\% of the linear fits to the $10^4$ posterior CMF values have a slope with orbital period which is negative, whilst 12\% are positive.}

{\begin{table*}
    \centering
    \begin{tabular}{l|c|c|c|c|c|c|c|c|}
        Planet:    & b                      & c                    & d                    & e                    & f                    & g                    & h                    & Avg b-h              \cr
        \hline
CMF [wt\%] &  $25.2_{- 6.0}^{+ 5.3}$ & $26.6_{- 5.1}^{+ 4.6}$ & $19.7_{- 5.1}^{+ 4.7}$ & $24.6_{- 4.9}^{+ 4.3}$ & $20.1_{- 4.2}^{+ 3.5}$ & $16.1_{- 4.2}^{+ 3.5}$ & $16.5_{-10.0}^{+ 9.3}$ & $20.9{\pm}3.6$ \cr
Fe/Mg molar ratio &  $0.60_{-0.18}^{+0.18}$ & $0.64_{-0.16}^{+0.16}$ & $0.44_{-0.13}^{+0.14}$ & $0.58_{-0.14}^{+0.14}$ & $0.45_{-0.11}^{+0.10}$ & $0.34_{-0.10}^{+0.09}$ & $0.35_{-0.23}^{+0.27}$ & $0.47\pm0.07$ \cr
       \hline
        H$_2$O [wt \%] for: & & & & & & & & \cr
        CMF=18\% & ${<}10^{-3}$ &  ${<}10^{-3}$ &  ${<}10^{-3}$ & $0.0_{-0.0}^{+0.0}$ & $0.0_{-0.0}^{+0.0}$ & $0.72_{-0.72}^{+1.3}$ & $0.6_{-0.6}^{+3.4}$ &  \cr
        CMF=25\% & ${<}10^{-3}$ &  ${<}10^{-3}$ &  ${<}10^{-3}$ & $0.3_{-0.3}^{+1.8}$ & $1.9_{-1.3}^{+1.5}$ & $3.5_{-1.3}^{+1.6}$ & $3.0_{-3.0}^{+3.8}$ &  \cr
        CMF=32.5\% & ${<}10^{-3}$ & ${<}10^{-3}$ & ${<}10^{-3}$ & $2.9_{-1.5}^{+1.7}$ & $4.5_{-1.2}^{+1.8}$ & $6.4_{-1.6}^{+2.0}$ & $5.5_{-3.1}^{+4.5}$ &  \cr
        CMF=50\% &  $0.05_{-0.03}^{+0.08}$ & $0.03_{-0.02}^{+0.05} $ & $0.002_{-0.0009}^{+0.002}$ & $9.4_{-1.8}^{+2.2}$ & $12_{-1.7}^{+2.0}$ & $14_{-1.7}^{+2.0}$ & $12_{-3.9}^{+4.4}$ & \cr
    \end{tabular}
    \caption{Core mass {fractions and molar Fe/Mg ratio (for a fully-differentiated model), and water mass fractions} inferred for each TRAPPIST-1 planet, as well as the weighted means.}
    \label{tab:cmf}
\end{table*}}

\subsubsection{Surface water content} \label{sec:volatiles}

\begin{figure}
    \centering
    \includegraphics[width=\hsize]{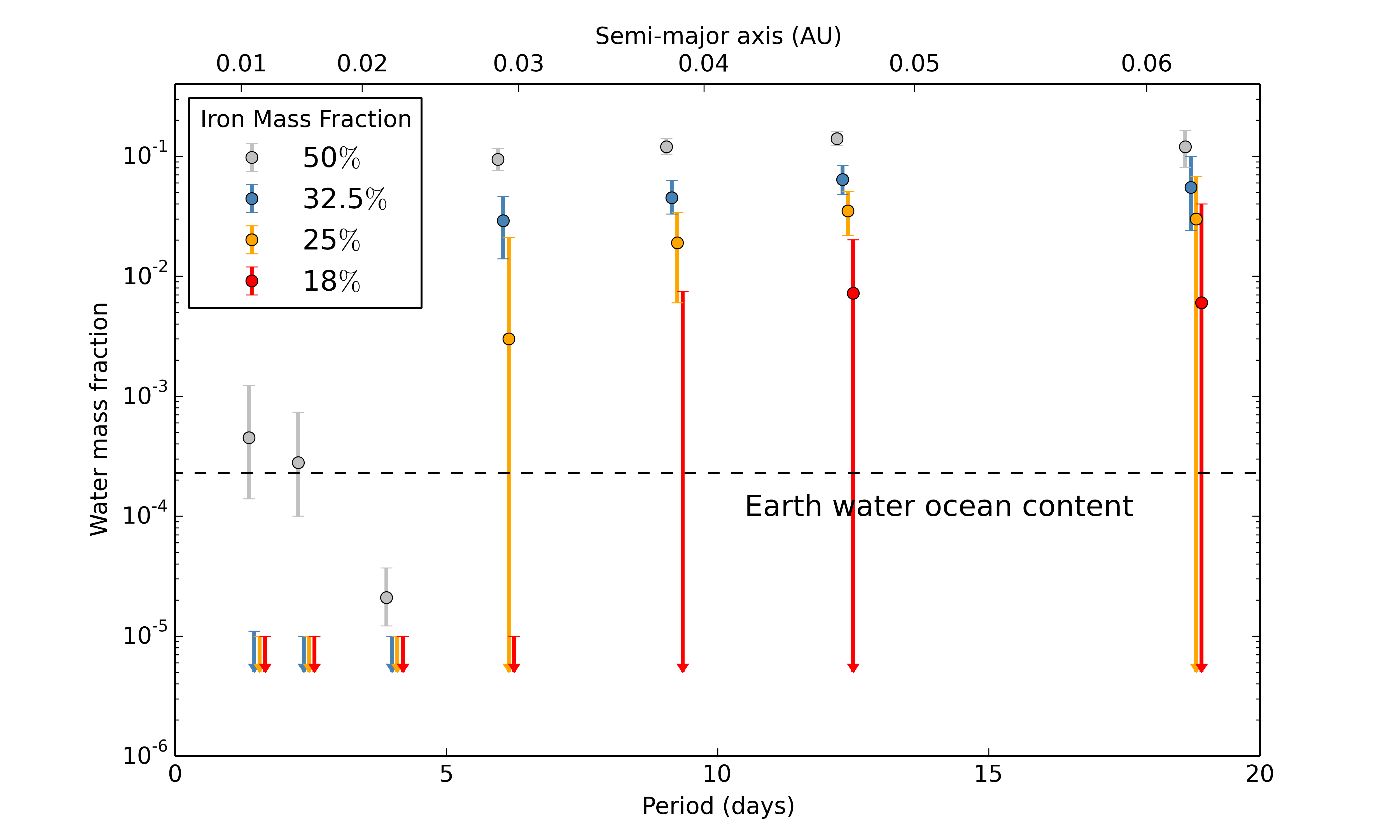}
    \oscaption{julia/copy_figure18.jl}%
    {Theoretical water content {estimates} (along with 1$\sigma$ error bar) versus planetary orbital periods. {Colors depict different compositions for the rocky interior
(18, 25, 32.5 and 50 wt\% CMF). For high CMF, estimated water contents are larger in order
to fit the total mass and radius.}}
    \label{fig:water_vs_period}
\end{figure}

The observed (weak) variation in the planet densities among all seven planets may instead be due to their differing volatile (e.g. water) inventories.

If we assume a rocky Earth-like interior {(CMF=32.5\%, fully-differentiated)} and only allow an additional {\it condensed}\footnote{Note that it is likely unwarranted to assume condensed surface water for the inner three planets given their location within the runaway greenhouse zone \citep{Turbet2020}.} water layer to contribute to the total radius, we can estimate the water mass fractions of the seven planets (b: ${2.8^{+2.1}_{-1.9}}$ wt\%, c: ${2.3^{+1.8}_{-1.7}}$ wt\%, d: ${4.4^{+2.0}_{-1.5}}$ wt\%, e: ${2.9^{+1.7}_{-1.5}}$ wt\%, f: ${4.5^{+1.8}_{-1.2}}$ wt\%, g: ${6.4^{+2.0}_{-1.6}}$  wt\%, h: ${5.5^{+4.5}_{-3.1}}$ wt\%).
The lower densities of planets d, f, g, and h can allow for two to three times as much water than for planets b, c, and e. For this simple estimate we assumed a water layer with a surface temperature of 300 K at 1 bar.

Actual surface conditions {and assumed iron content} can, however, lead to much larger differences in the estimated water budgets between the inner three and outer four planets. This stems from the fact that the inner three planets are more irradiated than the runaway greenhouse irradiation limit \citep{Kopparapu2013,Wolf2017,Turbet2018} for which all water is vaporized, forming a thick H$_2$O-dominated steam atmosphere. Taking into account the expectation that water should be vaporized for the three inner TRAPPIST-1 planets \citep{Turbet2019,Turbet2020}, their water mass fractions drop drastically to less than $0.01$ wt\%, i.e. more than several times lower than the water ocean mass fraction of the Earth.

Figure~\ref{fig:water_vs_period} shows the expected water mass fractions for each of the TRAPPIST-1 planets, and for four distinct interior compositions (18, 25, 32.5 and 50 wt\% iron content). It shows that the same {qualitative} trend of water versus orbital period is relatively robust across a large range of assumptions on the interior composition thanks to the transition from runaway greenhouse for planets b-d to surface liquid water for planets e-h. 

Higher estimated water budgets for the outer three or four planets could be a clue that they formed beyond the water condensation line at $\approx$0.025 AU \citep{Unterborn2018a}. This could also be due to the significant differences in water loss (through atmospheric escape) arising from variations of irradiation and gravity among the TRAPPIST-1 planets \citep{Lissauer2007, Bolmont2017,Bourrier2017}.  However, again, we caution again that trends in the planetary volatile content are only weakly supported by the current data.

\subsubsection{Core-free planets} \label{sec:core_free}

Given that the data may be consistent with an iso-composition mass-radius relation, we next consider another intriguing possibility:
that the {interiors of the planets are fully oxidized.} If, instead of forming a core, all of the iron is oxidized and remains in the mantle, the size of a planet may increase by a few percent \citep{ElkinsTanton2008}.  This turns out to be about the amount of radius inflation necessary to match the TRAPPIST-1 planets when compared with our Solar system planets.

If we assume that the refractory ratios match a Solar composition, and that all seven planets lack an atmosphere, then it turns out that all seven planets are consistent with a core-free, oxidized composition (Fig.\ \ref{fig:mass_radius_relation}; {red} line).  For this model the bulk mass abundance ratios for Fe/Si/Mg/O are 29.2/17.3/15.3/38.2 wt\% {with a magnesium number of 0.55 (Mg/(Mg+Fe)) {mol fraction}; this model has } a significant increase in oxygen compared to the bulk Earth with 29.7 wt\% \citep{McDonough2014}.
Such a scenario would likely require formation of the planets at large distances from the star in a highly oxidizing environment \citep{ElkinsTanton2008} and a lower  devolatization temperature intermediate between that of Earth and chondrites \citep{Wang2019}.  Hence, although this hypothesis efficiently explains the TRAPPIST-1 data, it remains to be seen whether a geochemical model can be constructed which results in high oxidation of iron throughout the processes of planet formation and evolution {\citep{Kite2020}}.

\section{Discussion} \label{sec:discussion}

Here we discuss some of the implications of the results in the foregoing sections.

\subsection{Timing uncertainties}

As reported in \S \ref{sec:outliers}, the transit timing measurements we have made show an excess of outliers with respect to the measurement uncertainties of each transit.  We were unable to identify a culprit (or culprits) for these discrepancies, but wish to speculate on what may be the origin of these outliers.  The cumulative distribution of these outliers (Fig.\ \ref{fig:timing_residuals}) indicates that about 10\% of transits
are affected at some level.  It is also interesting to note that the core of the distribution has a slightly smaller width of about 87\% of the measurement errors, indicating that for about 90\% of the transits, the uncertainties may be \emph{overestimated}.  This may be a consequence of inflating the uncertainties to account for correlated noise rather than modeling the data with, for example, a Gaussian process;  further re-analysis of the data will be needed to check this hypothesis.

Could the timing outliers be due to stellar flares?  In \citet{Vida_2017} and \citet{Ducrot2020}, the frequency distribution of stellar flares is shown to be rising towards smaller flare energies.  This could mean that the more frequent, but lower energy flares, occur at a level that is swamped by the photon noise, and thus not visible to an observer.  We used the spectrum and energy calibration of Spitzer flares measured by \citet{Ducrot2020} to extrapolate the frequency of lower energy flares (which are not detected in Spitzer due to photon noise).  As an example, for planet h the transit time can be affected by a flare which occurs at ingress or egress (duration $2\tau \approx 10$ min). 
We estimate that a flare of energy $10^{31}$ erg could cause a 1.5$\sigma$
timing outlier if it occurs during ingress or egress.  This has a probability of only $\approx 0.3$\% to occur during the 10 minutes of ingress or egress,
and thus cannot be responsible for 10\% of outliers for planet h.  We carried out a similar estimate for
the other planets, and we conclude that low-level flaring activity cannot be the cause of the timing outliers.

Other possible causes of the timing outliers are correlated stellar variability, star spot crossings, or instrumental systematics.  We don't yet have an estimate of the magnitudes of these effects, and so cannot reach a conclusion about where the origin of the timing outliers lies.

\subsection{Possible systematic errors}

In this section we consider possible factors which might affect our
inference of the densities of the planets.  Simulated planetary
densities predict core-mass-fractions which are similar to
Earth, with a very small scatter \citep{Scora2020}.  Hence, the
fact that the TRAPPIST-1 planets have inferred planetary densities
which are less than this could be due to systematic uncertainties
which are not captured by our modeling.

The transit depths
determine the {planet-to-star radius ratios}, but these
measurements are affected by the {non-}uniform surface brightness
of the star.  Fortunately the multiple impact parameters of the
planets yield a constraint upon the infrared limb-darkening, which
is fairly weak compared with optical bands.  However, star-spots
can also affect the inferred transit depths {\citep{Czesla2009,Oshagh2013, 
Oshagh2014,McCullough2014, Rackham2018, Kipping2012b}}.  If spots are present 
on an active latitude
which is not on the same hemisphere as the planetary transit chords,
this can cause all of the planet radii to be mis-inferred by a similar
factor.

TRAPPIST-1 may have complex surface inhomogeneities, including regions brighter or darker than the mean photosphere  {\citep{Morris2018d,Zhang2018,Wakeford2019}}. It is possible that bright or dark regions could bias the apparent transit depths towards larger or smaller measurements, depending on which type of inhomogeneity dominates. 
{Time-variable contamination should average out with many observations, while 
time-steady inhomogeneity will not, such as active latitudes, polar spots, or even
hemispheric asymmetry \citep{Yadav2015,Brown2020}.  We modeled the transit-transmission 
in the K2, SPECULOOS, LT, near-infrared, and Spitzer bands from \citet{Ducrot2020} for all
seven planets using the contamination formula from \citet{Rackham2018} with a time-steady, 
three-temperature model with the temperatures of the three components ranging from 2000-2980 K 
and the covering fraction varying from 0 to 1.  The mean effective temperature is constrained 
by our stellar model parameters (Table \ref{tab:stellar_parameters}). We assumed that all seven 
planets transit the region with the larger covering fraction, and that their transit depths are
achromatic.  We ran a Markov chain fit to the transmission spectra, interpolating the fluxes in 
the bands between the effective temperature grid points which were spaced by 20 K; we find
that the posterior parameters with maximum likelihood are temperatures of $(2980,2331,2071)$ K 
with covering fractions of  $(0.8,82.1,17.1)$\%.  We then computed 
the expected impact on the transit depths in the two IRAC channels.  The constraints are tight: we 
find that the observed radii should only change by a factor of $1.0072{\pm}0.0097$ in Channel 1 
and $1.0071{\pm}0.0108$ in Channel 2 (these are the ratios of the observed radii to the actual
radii).  These factors are consistent with unity at better than $1\sigma$, and have uncertainties 
which are comparable to or smaller than the uncertainties on the absolute planetary radii. We 
conclude that this form of self-contamination does not greatly influence our results, but should 
lead to caution in the interpretation.  This constraint is much stronger than the analysis of 
\citet{Morris2018c}.}

Our mass precisions are predicated on a complete model of the dynamics of the system.  {We neglect tides and general relativity, which are too small in amplitude to affect our results at the current survey duration and timing precision \citep{Bolmont2020}.} Should an eighth planet be lurking at longer orbital periods, which has yet to reveal itself via significant transit-timing variations or transits, this may modify our timing solution and shift the masses slightly.  In our timing search for an additional planet, however, we found that such a planet might only cause shifts at the ${\approx} 1\sigma$ level.  This possibility begs for caution in interpreting the potential variation of iron fraction with orbital period:  should an eighth planet be present beyond planet h, its timing impact would likely affect the masses of the exterior planets more significantly than the interior planets.  Drawing stronger conclusions about the variation of planet iron/core mass fractions will likely require longer-term monitoring, especially of planet h, and/or higher precision timing measurements such as are expected with JWST, to place tighter constraints on an eighth planet.

\subsection{Planet masses and radii in context}

In our current analysis of the transit-timing data
for TRAPPIST-1, we have found larger mass ratios for
all planets save planet e compared with our most
recent analysis in \citet{Grimm2018}.  Even though
most of the planets have shifted by 1$\sigma$ or more,
this does not indicate that the prior analysis was in
error.  In fact, the masses of all of the planets
are strongly correlated, and thus when one planet shifts
in the transit-timing solution, they all shift.  With
the more extensive dataset analyzed here, we provide
a better constraint over the transit-timing timescale,
and can also better account for outliers thanks to some
redundancy in our measurements. Given the high precision of
the Spitzer timing measurements, we expect that our current
analysis may remain the most reliable constraint upon the
masses of the planets until the transit times can be measured
with JWST.

\begin{figure*}
    \centering
    \includegraphics[width =0.8\hsize]{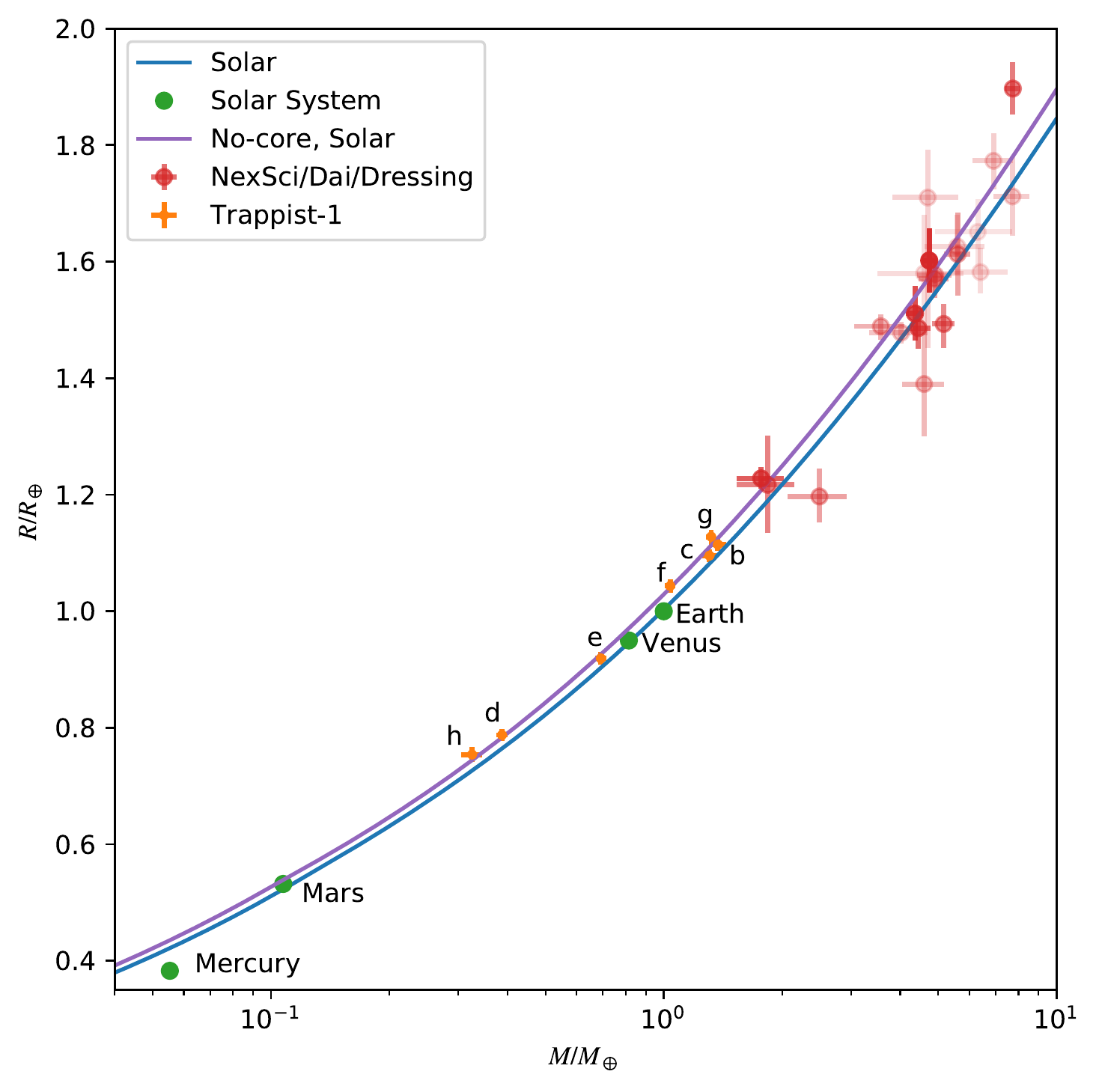}
    \oscaption{julia/plot_planets.jl}{Radius versus mass for Solar-System terrestrial planets (green dots),
        TRAPPIST-1 (orange error bars), and other potentially rocky exoplanets from the NExSci database, \citet{Dressing2015},
        and \citet{Dai2019}
        (red error bars).  Planets with smaller mass uncertainty are shown in a darker red color.  Also
        plotted is a mass-radius relation with a core-mass fraction compatible with Earth (blue), and a core-free model in which the refractory elements retain the Solar abundance ratios (purple).}
    \label{fig:mass_radius_comparison}
\end{figure*}

In Figure \ref{fig:mass_radius_comparison} we compare our measurements for
the seven TRAPPIST-1 planets with our Solar System planets and with exoplanets
with radii ${<}1.7 R_\oplus$ and masses measured to ${>}5\sigma$ retrieved from
the NexSci database on 26 Feb 2020 \citep{Akeson2013,Christiansen2018}, as well
as planet parameters reported in \citet{Dai2019} and Kepler-93b from \citet{Dressing2015}.\footnote{Note: {we corrected} Kepler-105b {with the} GAIA DR2 revised radius of
    the host star \citep{Berger2018,Fulton2018}.}   The uncertainties
on the other planets' masses are the best available to date from radial-velocity measurements, and yet they are much larger than the uncertainties for the TRAPPIST-1 planets, whether considered in
a relative or absolute sense.  The larger uncertainties of the RV planets makes the core-mass fractions difficult to constrain
for these more massive planets - core-free and cored models are consistent with most of these planets' parameters at the 1$\sigma$ level (Fig.\ \ref{fig:mass_radius_comparison}).   Nevertheless, it is notable that the
rocky planets for which we currently have data {seem to be similar in composition to the
Earth \citep{Dressing2015}; however, the actual range of bulk rock compositions of rocky exoplanets relative to their host stars is currently debated.}
This also appears consistent with the observation
that the evaporation valley requires rocky planets and their gaseous brethren to have
a composition which is a mix of silicates and iron \citep{Owen2017}.

\subsection{Comparison with radial velocities}

Given the measurements of the masses we have made with transit-timing,
this brings up the question:  what radial-velocity {uncertainties} would be
required to make mass measurements {of similar precision}?

\begin{figure*}
    \centering
    \includegraphics[width=\hsize]{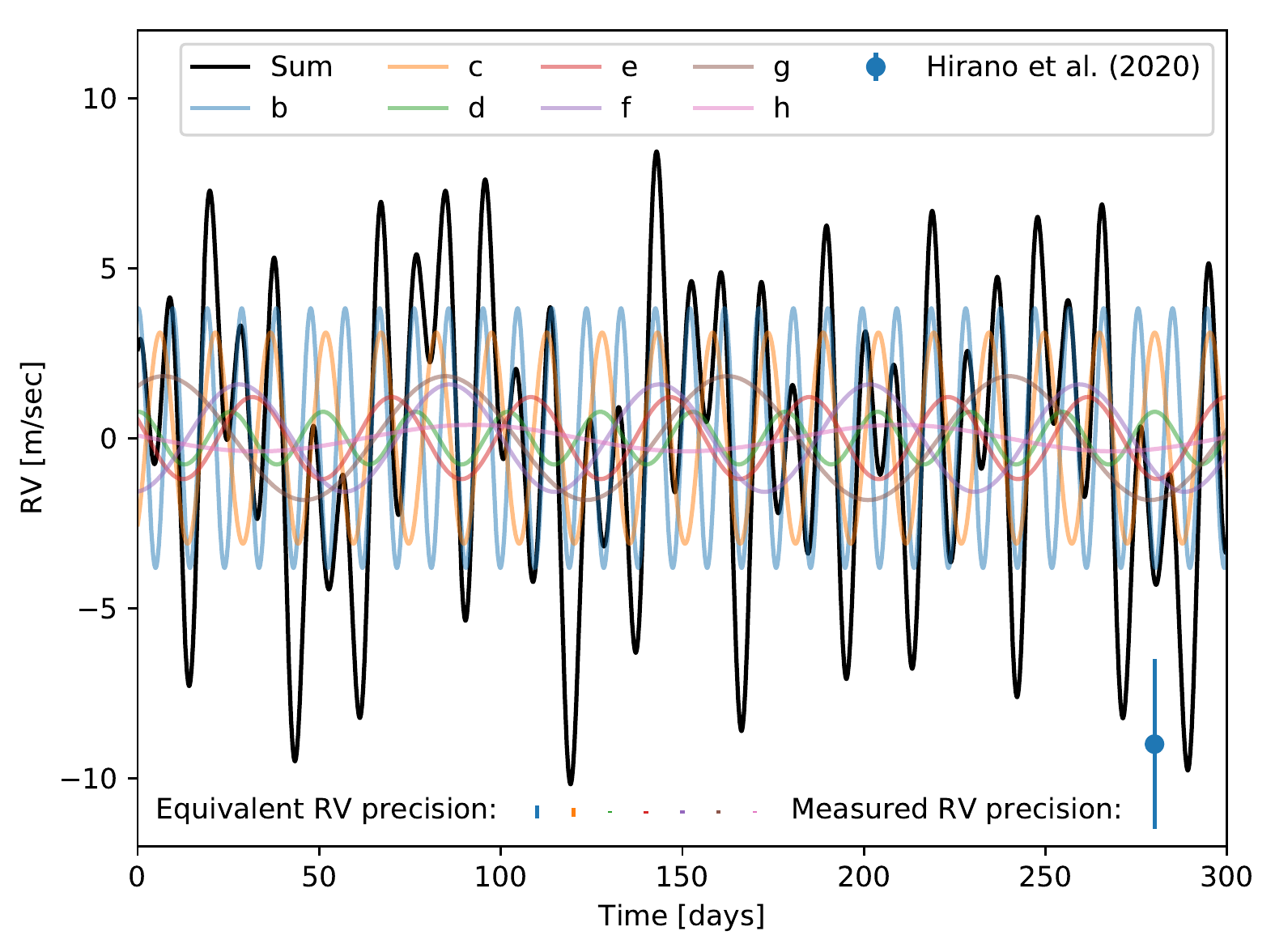}
    \oscaption{julia/T1_mass_radius_update.jl}{Predicted radial-velocity variation of the TRAPPIST-1 host star
        induced by its seven known transiting planets, as well as the current
        measurement error bar reported by \citet{Hirano2020}, which they interpret
        as an upper limit, thanks to stellar variability.  Also plotted are
        the equivalent semi-amplitudes for the seven planets which would be required
        to achieve the same mass precision as measured with TTVs.}
    \label{fig:rv_forecast}
\end{figure*}

The precision of the mass measurements may be placed in
context by comparing with current radial-velocity capabilities.
The predicted semi-amplitudes for the seven planets are given
in Table \ref{tab:rv_equivalent}.
The predicted radial-velocity variation of the star induced by the TRAPPIST-1
planets is plotted in Figure \ref{fig:rv_forecast}, also based upon our
mass measurements from transit timing.  The sums of the semi-amplitudes
of the planets equals ${\approx}12.7$ m/sec, which is close to the
peak amplitude when the planets are all orbiting on the same side of
the star (near 218 days in the plotted figure).  How does this compare
with current RV measurements?

Recently \citet{Hirano2020} were able to make high precision
measurements of the radial velocity (RV) of the TRAPPIST-1 host star,
achieving a constraint on the linear variation of the star to
a precision of 2.5 m/sec which they ascribe
to stellar variability.  To compare this with our
transit-timing results, the semi-amplitude precision
which would be needed to achieve the same mass error bars that
we have achieved with transit-timing ranges from 2.4-19 cm/sec,
up to 100 times more precise than the radial-velocity measurements.
Future observations may be able to achieve higher precision radial
velocity measurements of TRAPPIST-1, but will continue to contend with stellar
variability \citep{Klein2019}.

Were these planets orbiting a Sun-like star, the semi-amplitude RV error {would}
need to be even smaller to achieve
the same mass precision we have achieved with transit timing.
Table \ref{tab:rv_equivalent} lists what semi-amplitudes precisions
would be required if each one of these planets was placed around a
Solar twin at one astronomical unit.  The required precision {ranges}
from 1-6 millimeters/second.  This is nearly two orders of magnitude
more precise than the highest precision RV measurements
for short-period exoplanets reported to date, such as Tau Ceti g,
which has a reported RV semi-amplitude precision of 11 cm/sec \citep{Feng2017}.
We conclude that the mass precisions of Earth-sized, Earth-insolation
planets based on radial velocity must be improved by two orders
of magnitude to match our TTV precision for the TRAPPIST-1 system.

\begin{table*}
    \centering
    \begin{tabular}{l|c|c|c|c|c|c|c}
        Planet                                                        & b     & c     & d    & e     & f     & g     & h        \\
        \hline
        $K_p$ [cm/sec]                                                & 382.0 & 310.7 & 77.6 & 120.7 & 158.1 & 182.2 & 39.1 \cr
        RV equivalent precision for TRAPPIST-1 host [cm/sec]          & 19    & 13    & 2.5  & 3.8   & 4.7   & 5.2   & 2.4\cr
        RV equivalent precision for 1 $M_\odot$ host at 1~AU [cm/sec] & 0.62  & 0.50  & 0.11 & 0.20  & 0.28  & 0.34  & 0.18\cr
    \end{tabular}
    \caption{RV semi-amplitudes, $K_p$, for the TRAPPIST-1 planets predicted from our
        measured masses.  Equivalent RV precision required to measure the masses to the
        same precision as measured with TTVs around TRAPPIST-1.  Also, equivalent RV precision required {\it if}
        each planet were placed around a Solar twin at one astronomical unit.}
    \label{tab:rv_equivalent}
\end{table*}

\subsection{Planetary dynamics}

In this section we discuss some of the dynamical aspects of the planetary system: the  eccentricities, the longitudes of periastron, and the GLR angles.

\subsubsection{Eccentricities}

\begin{figure}
    \centering
    \includegraphics[width=\hsize]{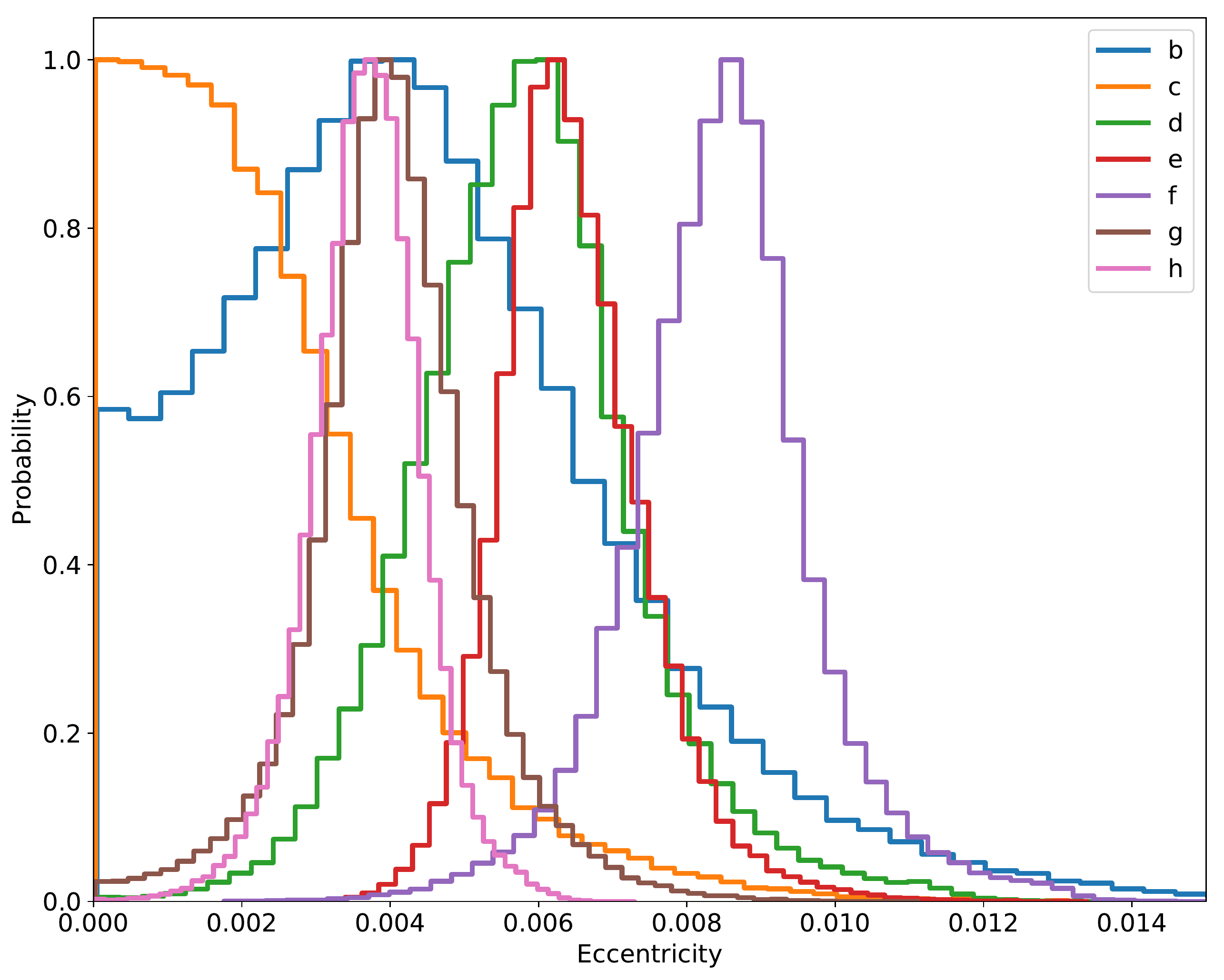}
    \oscaption{julia/plot_likelihood_profile.jl}{Probability distribution of the eccentricities of the planets at the initial time
        based upon the transit-timing model.}
    \label{fig:eccentricity_posterior}
\end{figure}

The posterior distribution of {the initial} eccentricities of the planets 
is shown in Figure \ref{fig:eccentricity_posterior}.  In prior analyses of
the transit-timing variations of the TRAPPIST-1 system we found that
the inner two planets, b and c, had significant eccentricities \citep{Grimm2018}.
In {contrast, with} the current analysis we find that the eccentricity probability
distributions of these two planets are significant {near} zero eccentricity.
This is consistent with N-body models which include tidal damping of
the orbits, which predict that the planets {b and c} should have low eccentricities,
$\la 10^{-3}$ \citep{Luger2017a,Turbet2018}.  The other planets are
all consistent with the predictions of the tidal evolution model
\citep{Luger2017a}.

\begin{figure}
    \centering
    \includegraphics[width=\hsize]{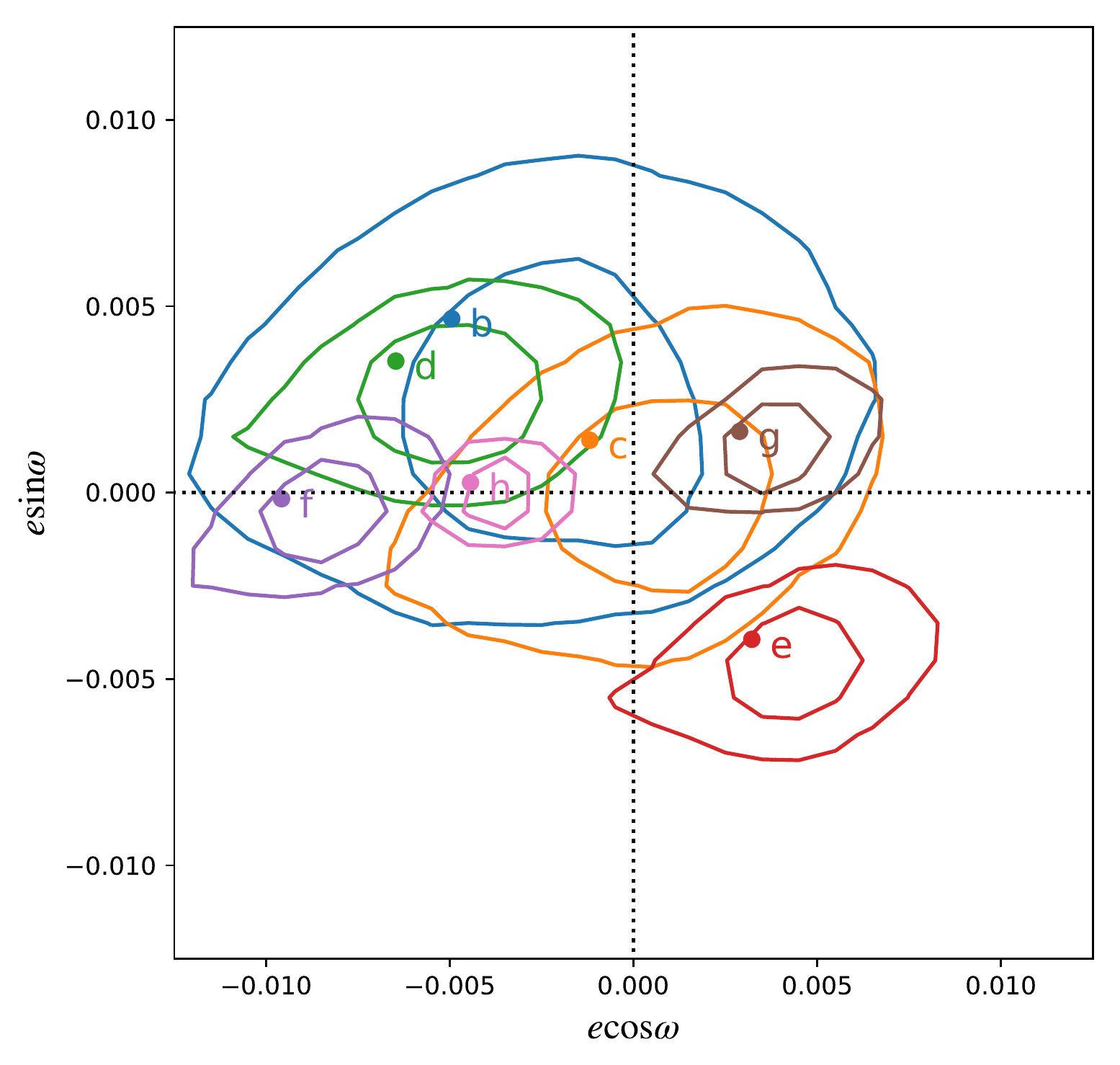}
    \oscaption{julia/plot_evectors.jl}{Posterior probability distribution for the eccentricity vectors  at the
        initial time for each of the planets.  Contours are 1 and 2$\sigma$ confidence limits.
        The maximum likelihood parameters are shown as solid points.}
    \label{fig:eccentricity_vectors}
\end{figure}

Figure \ref{fig:eccentricity_vectors} shows the posterior probability distribution
for the eccentricity vectors of each planet.  The only two planets consistent with
zero eccentricity at 1$\sigma$ confidence are planets b and
c (blue and orange contours).  The other five planets have non-zero eccentricities.

\begin{figure}
    \centering
    \includegraphics[width=\hsize]{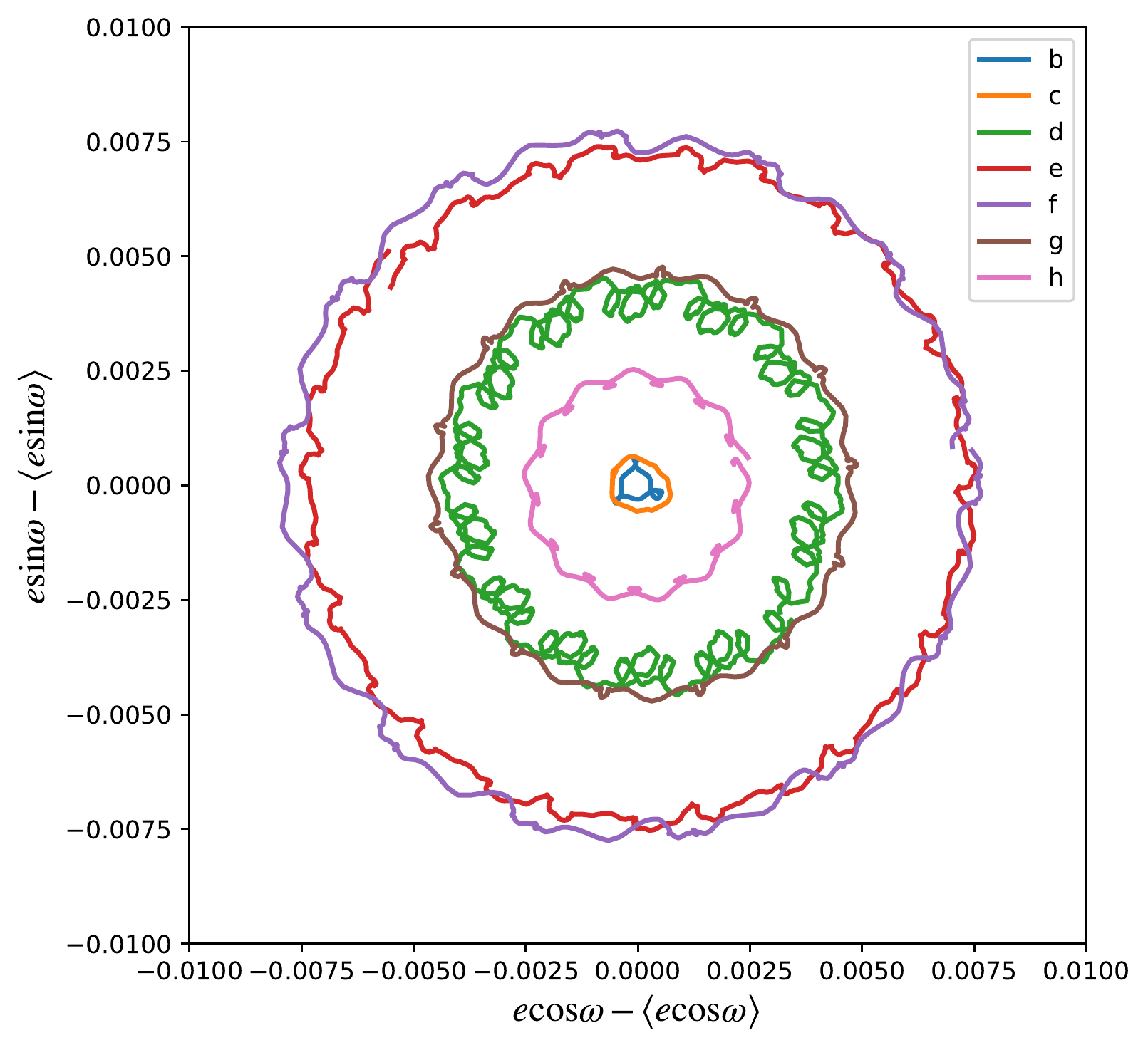}
    \oscaption{julia/plot_eccentricities.jl}{Variable component of the osculating eccentricity vectors plotted from a simulation over 12 days for planets b and c, and over $\approx 490$
        days for planets d-h, with initial parameters drawn from the posterior distribution.}
    \label{fig:forced_eccentricity}
\end{figure}

Now, the eccentricity vectors plotted in Figure \ref{fig:eccentricity_vectors} show the
values at the initial time. However, over time, the eccentricity vector of each planet can be decomposed into two components: the mean eccentricity vector (over some timescale) and the variable component (which is time variable, with multiple oscillation timescales driven by the mutual planetary perturbations).   Figure \ref{fig:forced_eccentricity}
shows the eccentricity over a single oscillation for all seven planets.  The outer five
planets are close to first-order resonances with adjacent planets, and the super-period
for each of these planets is close to ${P_{TTV}{\approx}}490$ days thanks to the near-GLR commensurability
for all triplets of planets.  This leads to a nearly circular oscillation over this timescale
due to circulation of the first-order resonances driving oscillations in the eccentricity
vectors of each of these planets.  The inner two planets are close to second and third
order resonances with adjacent planets (b and c are close to 8:5, which is third order,
while c and d are close to 5:3, which is second order).  Since the strength of these
interactions scales as a higher power of eccentricity, these planets show much smaller variation
in the time-variable components of their eccentricity vectors.  Since planets b and c are
close to a third order resonance,
their eccentricity vectors show a three-fold symmetry.  On longer timescales these patterns
precess, filling a circular pattern over time.  The time-variable eccentricity vector patterns are
very similar over the range of posterior values, indicating that it is primarily this
component which is constrained by the transit timing variations of the planets.

\begin{figure}
    \centering
    \includegraphics[width=\hsize]{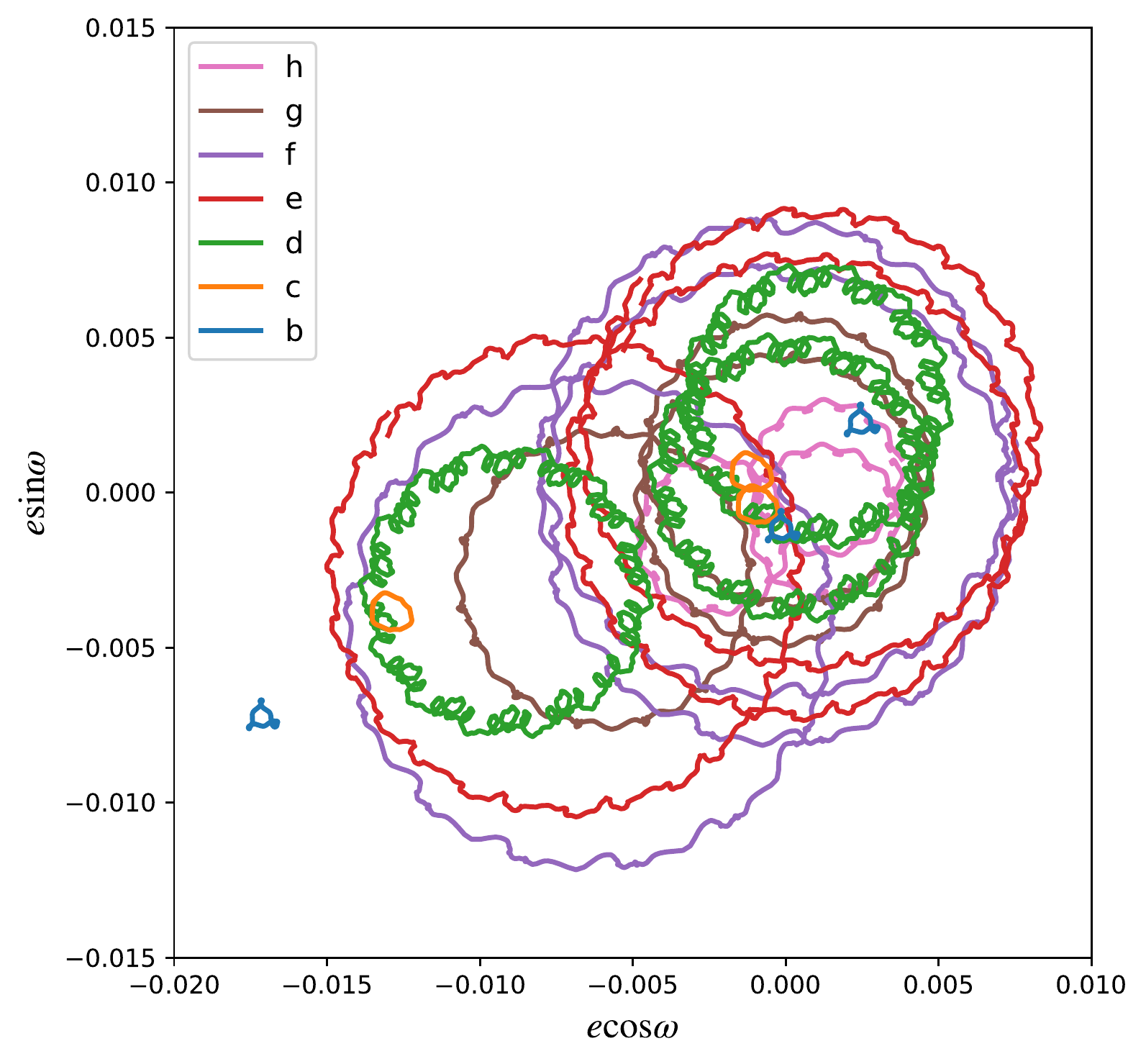}
    \oscaption{julia/plot_eccentricities.jl}{Osculating eccentricity vectors computed from a simulation for all seven planets shown for three different draws from
        the posterior:  the first with eccentricities nearest the median of the posterior distribution;
        the second with eccentricties furthest from the median, and a third drawn randomly from
        the posterior.  As with Figure \ref{fig:forced_eccentricity}, planets b and c are plotted
        over 12 days, while planets d-h are plotted over $\approx 490$ days.}
    \label{fig:total_eccentricity}
\end{figure}

The total eccentricity vectors show a wider range of behavior, thanks to a wider variation
of the mean eccentricity, as shown in Figure \ref{fig:total_eccentricity}.  It is clear
from this figure that each planet executes an eccentricity-vector oscillation about a
mean value (which was subtracted off for figure \ref{fig:forced_eccentricity}).  Unfortunately the mean eccentricity is {less} constrained by the transit-timing
variations {\citep{Linial2018}}, and so there is a much wider range of eccentricity vectors which is allowed
{which manifests as strong correlations amongst the eccentricity vectors of pairs of
planets (Figure \ref{fig:corner_ttv}).}

\subsubsection{Laplace angles}

A remarkable property of the TRAPPIST-1 system is the near-commensurability
of adjacent triplets of planets \citep{Luger2017a}, akin to Laplace resonances, with
GLR angles given by
\begin{equation}
    \phi_{i,i+1,i+2} = p \lambda_i - (p+q) \lambda_{i+1} + q \lambda_{i+2},
\end{equation}
where $\lambda_i$ is the mean longitude of the $i$th planet, and $p$ and
$q$ are small integers.  In
the case of an isolated triplet of planets, a stable configuration takes
on $\phi = 180^\circ$, but when planets are captured into a series of GLR
commensurabilities, their mutual torques displace the stable configuration
\citep{Delisle2017}.

Long-term
dynamical simulations show that these GLR angles can take on stable values
for extended durations, and sometimes can quickly jump in value, flipping symmetrically about
180 degrees \citep{Mah2018,Brasser2019},
resulting in two possible angles for each triplet of stars, $\phi$ and $360-\phi$.
Based on the prior measured {planet-to-star mass ratios}, \citet{Mah2018}
predicted the value of the three-body resonance angles resulting from the values at the
end of the simulation.

\begin{figure}
    \centering
    \includegraphics[width =\hsize]{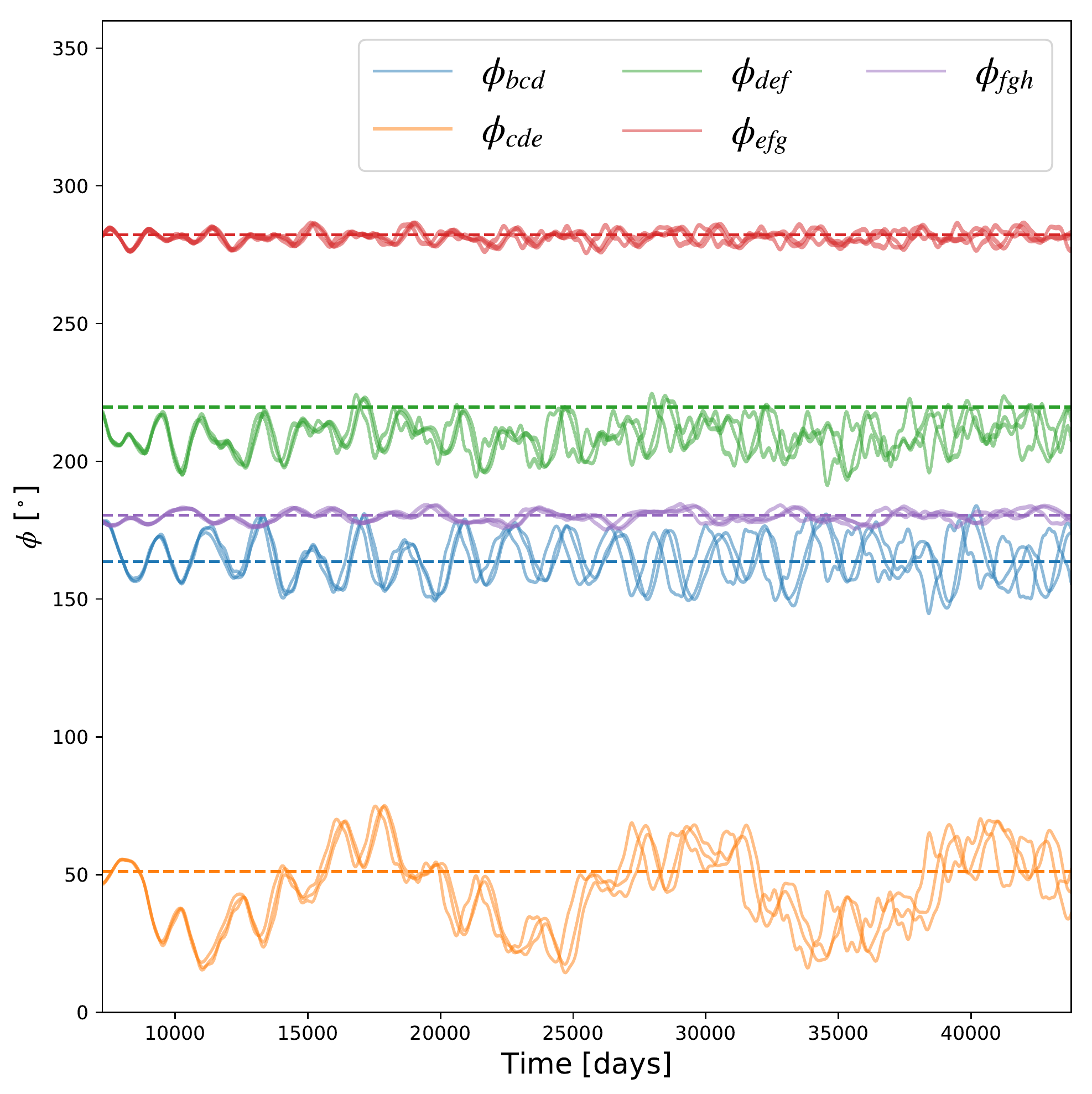}
    \oscaption{julia/plot_elements.jl}{GLR angles plotted over 100 years for
        three draws from the posterior distribution:  one with low eccentricities,
        one with high, and one randomly chosen.  These are
        compared with the predictions from \citet{Mah2018}, shown by
        dashed horizontal lines, with
        the values for $\phi_\mathrm{bcd}$ and $\phi_\mathrm{cde}$ flipped about
        180$^\circ$ (i.e. changed from $\phi$ to $360^\circ - \phi$).}
    \label{fig:laplace_angles}
\end{figure}

In Figure \ref{fig:laplace_angles} we show the GLR
angles for the following triples:
\begin{eqnarray}
    \phi_\mathrm{bcd} &=& 2 \lambda_\mathrm{b} -5\lambda_\mathrm{c} + 3\lambda_\mathrm{d},\cr
    \phi_\mathrm{cde} &=&  \lambda_\mathrm{c} -3\lambda_\mathrm{d} + 2\lambda_\mathrm{e},\cr
    \phi_\mathrm{def} &=& 2 \lambda_\mathrm{d} -5\lambda_\mathrm{e} + 3\lambda_\mathrm{f},\cr
    \phi_\mathrm{efg} &=&  \lambda_\mathrm{e} -3\lambda_\mathrm{f} + 2\lambda_\mathrm{g},\cr
    \phi_\mathrm{fgh} &=&  \lambda_\mathrm{f} -2\lambda_\mathrm{g} + \lambda_\mathrm{h}.
\end{eqnarray}
Differences between the predicted and observed angles
agree within 0.5-10 degrees, where the predicted values for $\phi$ are
taken from \citet{Mah2018},
but allowing $\phi_\mathrm{bcd}$ and $\phi_\mathrm{cde}$ to be flipped about 180 degrees.
It is possible with the updated mass-ratios from
our analysis that the predictions will be more accurate, which awaits further
simulation.

\subsubsection{Long-term stability}

Prior studies of the TRAPPIST-1 system by
\citet{Tamayo2017} found long-lived configurations for
systems which had formed via migration.  \citet{Quarles2017} examined the stability
of the TRAPPIST-1 system,  refining the large uncertainties from prior
measurement \citep{Gillon2017} to further constrain the masses of
the system.  Given the much tighter constraints we have placed upon
the masses of the planets and the orbital eccentricities, here we
re-examine the long-term stability of our posterior distribution.

We have used the GPU N-body integrator GENGA \citep{Grimm2014} to carry out long-term simulations of a set of $10^4$ posterior samples from the timing analysis.  These simulations were carried out for $10^7$ years, which corresponds to 2.4 billion orbital periods of planet b, and 195 million orbital periods of planet h. We used a time step of 0.06 days, which gives a total number of $6.1 \cdot 10^{10}$ integration steps. We find that 100\% of these posterior samples are stable over this entire timescale. 
To check the stability of the samples, we analyzed the evolution of the semi-major axis, $a$, and eccentricity, $e$, of all samples and planets. We compared the average values over the first Myr and the last Myr. Table \ref{tab:averageAE} gives the average over all samples, and the maximum differences between the first and the last Myr.  In all cases the variations are small, ${\le}0.002$.  These results suggest that the simulations could be stable even on a much longer time scale. In addition, we have carried out long-term (50 Myr) integrations with tidal damping for two posterior samples, one with low and one with high values of the eccentricity of planet b. Using a range of values of tidal damping (from 1/10 to 100 times Earth's), we find in all cases that the system remained stable \citep[using Posidonius;][]{Bolmont2020}.

\begin{figure}
    \centering
    \includegraphics[width =\hsize]{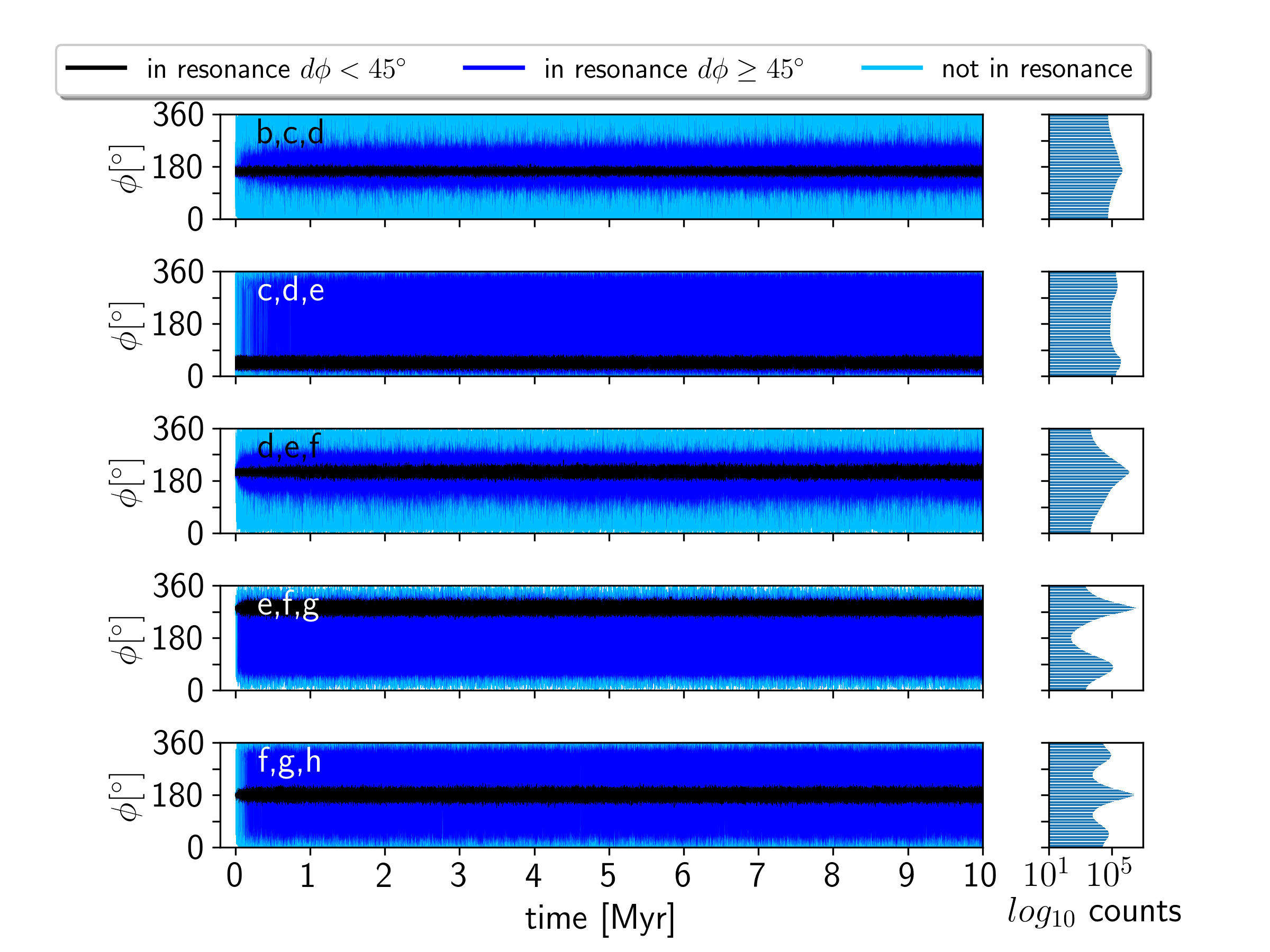}
    \oscaption{python/plotphiM2.py}{Evolution of the GLR angles $\phi$ for 10000 samples over 10 Myr. The simulations can be split into three categories: I remaining in resonance (black), II remaining in resonance but jump between states (dark blue), and III not remaining in resonance (light blue).}
    \label{fig:tlM}
\end{figure}

More interesting is the evolution of the five GLR resonant angles, shown in Figure \ref{fig:tlM}. In order to describe the evolution of the GLR angles, we define three categories:
\begin{itemize}
    \item Category I: remaining in GLR for 10 Myr, with a maximum difference to the initial value of less than $45^\circ$
    \item Category II: remaining in GLR for 10 Myr, with a maximum difference to the initial value of more than $45^\circ$. In this category, the GLR angles can jump between different states.
    \item Category III: not remaining in GLR for 10Myr.
\end{itemize}
The threshold of $45^\circ$ is chosen arbitrarily, but is found to be practical to distinguish simulations where the GLR angles jump between different states (Category II), or remain in the same state (Category I). Figure \ref{fig:tlM} shows the three different categories in different colors, as well as a histogram of all 10,000 simulations over 10~Myr for all five GLR angles.  The exact number of simulations in the three categories are given in Table \ref{tab:cat}. The GLR angles from Planets b,c and d as well as Planets d,e,f show a unique resonant state. Planets c,d,e and Planets e,f,g have a dominant state and a subdominant state, while Planets f,g,h have a dominant state and two symmetric subdominant states.
Our new samples show a better conservation of the GLR angles than was found in \citet{Grimm2018}, where the longest resonance time was found to be 2 Myr.

\begin{table}
    \centering
    \begin{tabular}{|c|c c c c c}
        planet & $\Delta \bar{a}$ & $\Delta \bar{e}$ & $\max (\Delta a)$ & $\max(\Delta e)$ & \\
        b      & -6.52e-09        & 1.73e-04         & 7.12e-07          & 0.0020             \\
        c      & -1.44e-08        & 1.62e-04         & 2.51e-06          & 0.0018             \\
        d      & -1.06e-08        & 1.45e-05         & 4.07e-06          & 0.0009             \\
        e      & 2.05e-08         & 4.44e-05         & 8.53e-06          & 0.0008             \\
        f      & 2.45e-07         & 5.13e-05         & 3.00e-05          & 0.0012             \\
        g      & 8.24e-08         & 5.01e-05         & 2.19e-05          & 0.0011             \\
        h      & -1.23e-07        & 2.11e-04         & 3.00e-05          & 0.0035
    \end{tabular}
    \caption{Evolution of the semi-major axes, $a$, and eccentricities, $e$, from $10^4$ samples over 10Myr. For each sample and planet, the difference of the average of $a$ and $e$ over the first and last Myr are compute as $\Delta \bar{a}$ and  $\Delta \bar{e}$; we report the maximum over all samples. These numbers show that all samples remain stable over 10Myr.}
    \label{tab:averageAE}
\end{table}

\begin{table}
    \centering
    \begin{tabular}{|c|c c c c c}
                     & bcd  & cde  & def  & efg  & fgh  \\
        Category I   & 130  & 178  & 755  & 7943 & 6462 \\
        Category II  & 1571 & 874  & 7653 & 1855 & 1578 \\
        Category III & 8299 & 8948 & 1592 & 202  & 1960
    \end{tabular}
    \caption{Number of posterior samples falling into the three resonant categories for the five GLR angles. The total number of posterior samples is 10000.}
    \label{tab:cat}
\end{table}

\subsection{Forecasts for JWST}

\subsubsection{Forecast transit times}

With our transit timing model we can forecast the probabilities of future transit times,
and hence better help to plan transit observations with JWST.  This is
important for both optimizing the efficient use of the telescope, and for
determining concurrent transits (i.e.\ two or more planets crossing the face
of the star at the same time).  This is especially important for transit
transmission spectroscopy as the signal will be small, and hence many
transits may need to be observed.  With observation of initial transits
with JWST the ephemerides can be refined/updated;  however, our current
forecasts provide the starting point for planning JWST observations.

Table \ref{tab:transit_times_forecast} gives our forecast for upcoming times of transit through {October, 2023 to cover the first 2 years of the JWST mission (six months after the end of Cycle 1, given the present launch date of October 2021).}

\subsubsection{Simulated JWST TTV analysis}

Based on the measured properties of TRAPPIST-1, we have carried out a preliminary analysis forecasting future transit observations with the James Webb Space Telescope. Already there are several JWST Guaranteed Time Observation (GTO) programs which plan to observe the TRAPPIST-1 planetary system, primarily for the purposes of spectroscopic characterization (GTO programs 1177, 1201, 1279 and 1331).\footnote{For specifications of these programs, see \texttt{https://www.stsci.edu/jwst/observing-programs/ approved-gto-programs}.}  It is very likely that additional observations will be scheduled during guest observing  time throughout the duration of the JWST mission as the detection of spectroscopic features requires observations of multiple transits for each of the planets \citep{Morley2017,Barstow2016,Lustig-Yaeger2019,Fauchez2020b}.   An effort to coordinate these observations amongst the exoplanet and planetary science communities is underway via the TRAPPIST-1 JWST Community Initiative \citep{Gillon2020}.  All to say, long-term studies of TRAPPIST-1 for spectroscopy will also yield transit times for each transit observed, enabling a transit-timing analysis of the results.

To estimate the {\it maximum} possible precision of observations with JWST, we have simulated a five-year program in which {\it every} transit of every planet in TRAPPIST-1 is observed with NIRSPEC \citep{Birkmann2016}.   The NIRSPEC instrument was chosen as its prism mode covers 0.5-5 microns, covering the peak of the SED of the star, and thus maximizing the number of photons detected, which is about two orders of magnitude per transit greater than collected by Spitzer. Although such a complete set of transits will be impossible to collect (thanks to limits due to scheduling and time-allocation), this analysis yields an estimate of the most optimistic results we might expect from JWST.

We have carried out simulations of transits of each of the planets as observed by NIRSPEC.  We include realistic estimates of photon noise and correlated stellar variability based on the pattern of variations detected with the Spitzer Space Telescope, using a Gaussian Process model created with \texttt{celerite} \citep{ForemanMackey2017}.  We do not include instrumental systematics under the assumption that over the timescales of ingress/egress, which are what limit the timing precision, that the noise contribution will be dominated by photon noise and stellar variations.
From these simulations, we
found that the posterior timing precision ranges from 0.6-1.7 second per transit, much more precise than the measurements reported in the present paper.

\begin{figure*}
    \centering
    \includegraphics[width=\hsize]{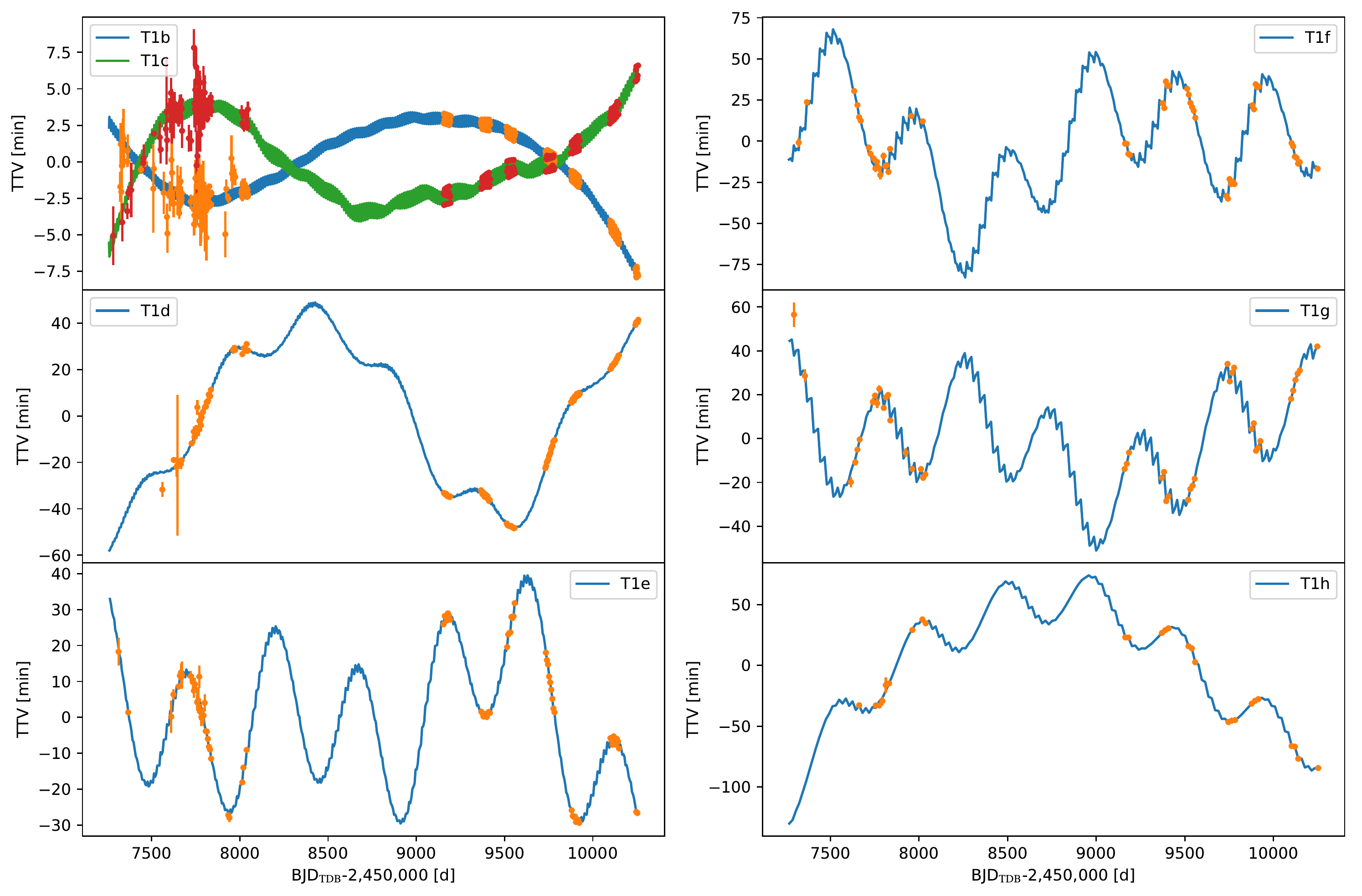}
    \oscaption{julia/plot_TTV_JWST_T1.jl}%
    {Simulated observations of all of the transits of TRAPPIST-1
        detectable with JWST.  Each transit has an
        uncertainty of ${\approx} 0.6-1.7   $ seconds, assumed to be observed with
        NIRSPEC (which maximizes the number of photons collected of any JWST
        instrument).  From retrieval, we obtain
        ${\la}$0.1\% mass precision.}
    \label{fig:JWST_all_the_transits}
\end{figure*}

Next, we created a simulated set of transit-timing observations at the two windows each year when the TRAPPIST-1 system is observable with JWST (Figure \ref{fig:JWST_all_the_transits}).   For each transit time, we drew the time from the distribution of uncertainties from the posteriors of the simulated transit data. 

\begin{figure}
    \centering
    \includegraphics[width=\hsize]{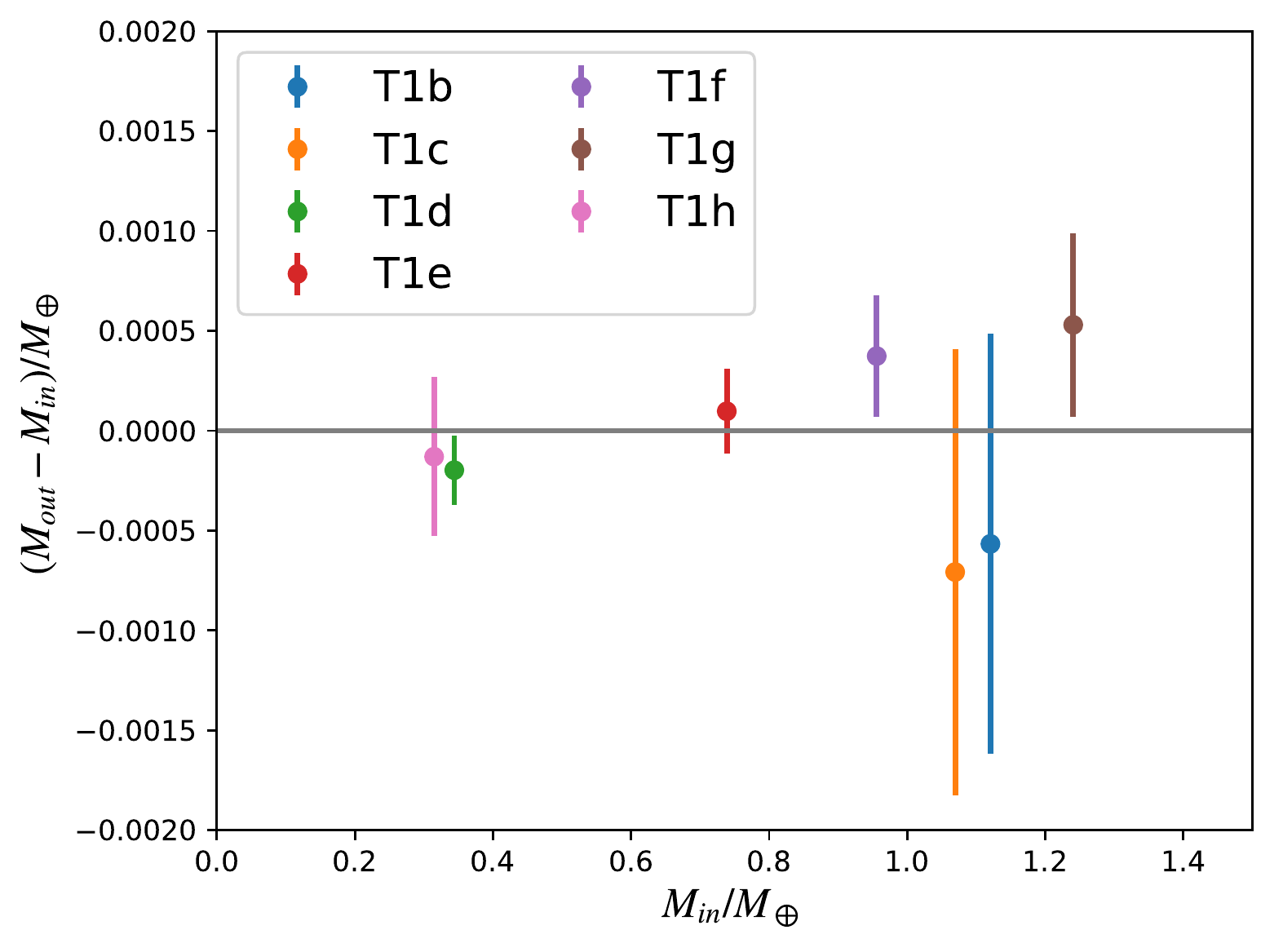}
    \oscaption{julia/plot_mass_recovery.jl}{Simulated planet massses based on 5 years of JWST observations
        of \emph{every} TRAPPIST-1 transit with NIRSPEC.  The recovered mass ($M_{out}$) minus
        the input mass versus the input mass ($M_{in}$).  The masses {relative to the star} can be recovered to
        better than 0.1\% precision.}
    \label{fig:JWST_all_mass}
\end{figure}

Finally, we utilized our code for transit-timing analysis  to optimize a plane-parallel model with seven planets.  At the maximum likelihood of the fit, we computed the Hessian to estimate the uncertainties on the model parameters.  Figure \ref{fig:JWST_all_the_transits} shows the simulated transit-timing observations with JWST.   This includes about 600 transits observed with the telescope (again, the maximum possible over the nominal 5-year JWST mission).  Figure \ref{fig:JWST_all_mass} shows the results of the mass measurements in the simulations.  We find that the masses can be recovered to better than 0.02\% for planets d-h, and to {0.1}\% for planets b and c.

Of course, it will be impossible to arrange such a large number of transit observations of this system.  But, even if the number of observations is an order of magnitude smaller, we expect that the {signal-to-noise} should scale with the square root of the number of measurements made, and thus {the outer} planets will still have mass measurements precise to the order of a part-per-thousand.

\subsection{Stellar parameters}

The stellar density we derive using the photodynamic model, $\rho_* =
    53.17_{- 1.18}^{+0.72} \rho_\odot$, is {in $1\sigma$
agreement with} prior analyses. 
Most recently, \citet{Delrez2018a}
found a density of $\rho_* = (52.3{\pm}2.2)\rho_\odot$,
twice as uncertain as our analysis.   {Our} approach yields a density of superior
precision due to several factors.
The transit times in the Spitzer data are constrained by
    {\it all} of the measured transits in the photodynamic model
so that fewer degrees of freedom
are needed to fit the times (37 free parameters in the N-body
model versus 447 transit times fit to each transit).

The stellar mass we take from the analysis by \citet{Mann2019},
$M_* = 0.0898{\pm}0.0023 M_\odot$.\footnote{\texttt{https://github.com/awmann/M\_-M\_K-}}
This mass has a precision of 2.6\%, which limits the mass precision
for several of the planets.  We are at the point that to improve the mass measurements
of the planets we will need to improve the measurements of the star.

We used the luminosity estimate from \citet{Ducrot2020}, which is slightly lower than
that estimated by  \citet{Gonzales2019} due to a difference in the measured bolometric
flux.  We are consistent with \citet{Gonzales2019}
for the reported value of $R^2T_{eff}^4$ at 1$\sigma$, while our $T_{eff}$ is more precise
(28 K vs 42 K), $R$ is 2.5 times more precise, and our $\log{g}$ is more
precise by an order of magnitude.

\section{Conclusions} \label{sec:conclusions}

The Spitzer discovery of seven transiting planets orbiting the TRAPPIST-1 star by \citet{Gillon2017} promised the determination of the interior compositions of these planets via dynamical analysis.
We have now analyzed the complete set of transit time measurements
of the TRAPPIST-1 planets from  Spitzer, augmented by
additional transits from the ground, K2, and HST.  Our primary conclusions
are:

\begin{enumerate}
    \item  We have measured the masses, radii and densities to high {fractional} precision, {1-8\%}, based on an N-body model {and a photodynamical model} with seven planets.  This improves upon RV current precision by {up to} two orders of magnitude.
    \item  The pattern of masses and radii {may be} consistent with a uniform planetary composition for all seven planets
          which have lower uncompressed densities than the Earth, Mars or Venus, with weaker evidence for a declining normalized density with orbital period ({88\%} confidence). The planet properties may either be consistent with
          a core mass fraction of ${21{\pm}4}$ {wt\%}, or an Earth-like core and mantle with a {surface water}
          content which varies from {${<}0.001$\%} for the inner three planets to {${\approx} 5$\%}
          for the outer four, or core-free planets with highly oxidized iron in the mantle which elevates the interior {light element} content.  These are not unique explanations.
    \item  The planets appear to be dynamically-cold, with eccentricities {less} than
          ${\approx }1$\%, and inclinations which {may be} coplanar to a {few hundredths} of a degree.
    \item The system is stable on long timescales, and shows a pattern of generalized Laplace resonances with angles which match predictions from migration simulations of \citet{Mah2018}.
    \item We provide a forecast of the future times of transit for the planets (Table \ref{tab:transit_times_forecast}) to help in planning observations with JWST, which may yield more precise constraints upon the planets' masses.
    \item We have yet to find strong evidence for an eighth planet.
\end{enumerate}

Based upon these properties, we next speculate on some possible scenarios for the formation and evolution of the
system.

\subsection{Expectations for the compositions of the TRAPPIST-1 planets from formation scenarios} \label{sec:formation}

As mentioned, our analysis suggests that the TRAPPIST-1 planets have somewhat lower uncompressed bulk densities than Earth (see Table \ref{tab:uber_table} and Fig. \ref{fig:mass_radius_relation}).  It is possible that these lower densities result from a deficit of high-density material (e.g., less iron) relative to Earth, or an excess of low-density material (e.g., having more water), or both;  in this section we speculate about formation scenarios which may be consistent with these planets' bulk densities.

In general, planets which formed within the same proto-planetary disk {are expected to} have similar budgets in relative refractory elements \citep{bond2010making,elser2012origin} but can have
very different volatile element budgets \citep{oberg2016excess}. Similar relative refractory elements (Fe, Mg, Si) implies
similar core mass fractions for all seven planets, assuming full differentiation. As suggested by \citet{Dorn2018}, the refractory composition may best be described by studying the densest planet of the system, planet c with {22-31\%} CMF. Thus, with this assumption, all of the planets may likely have a {22-31\%} CMF but different {light element} mass fractions (that may increase slightly with orbital period, Fig. 19).

Is an overall CMF of {22-31\%} realistic for terrestrial planet interiors? This range of CMF implies lower Fe/Mg and Fe/Si values compared to Earth (and the Sun). Elemental abundances of rocky interiors are expected to be reflected in the photospheric abundance of the host star as argued by \citet{Unterborn2018a} and \citet{Dorn2018}. Unfortunately, measuring the photospheric abundances of this cool and active host star remains very challenging. However, \citet{Unterborn2018a} estimated the stellar {molar} Fe/Mg number ratio to be $0.75{\pm}0.2$ by analysing Sun-like stars of similar metallicity to TRAPPIST-1, which may be slightly lower than the Solar value. {This corresponds to a CMF of 24-35\% for a fully-differentiated model.} The corresponding mass-radius curve for a rocky interior of this range of Fe/Mg value is plotted in Figure \ref{fig:mass_radius_relation} (gray curve and shaded region). It overlaps well with the densest planets c and b. This means that the expected range of stellar abundances supports a possible overall CMF value of {22-31\%}, assuming full differentiation.

Could there be a variation of Fe/Mg ratios among the planets?
Rocky planet accretion should preserve the integrated iron/rock ratio. Consider a population of planetary embryos and planetesimals that accrete into a system of rocky planets. Giant collisions between growing planetary embryos can change the iron/rock ratios of individual objects by preferentially stripping the outer, rock-dominated layers from differentiated embryos~\citep[e.g.][]{benz88,marcus10,asphaug14}. But from a system-wide perspective, it is a zero-sum game unless rock or iron is preferentially lost from all of the planets. Rock is the major component of loosely-bound impact debris and more likely to be lost either by differential aerodynamic drag~\citep{weidenschilling77} or solar wind drag~\citep{spalding20}, and so the integrated iron/rock ratio should only increase.
Hypothetical variations in Fe/Mg can otherwise be caused if large portions of planetary building blocks condense at different high temperatures (${>} 1200$ K). During planet formation, such temperatures are only reached in a tiny region very close to the ultracool dwarf star. Consequently, both \citet{Unterborn2018a} and \citet{Dorn2018} have assumed that all seven planets have similar refractory element ratios (i.e., Fe/Si, Fe/Mg). {Whether rocky planets can have a wider compositional distribution than that of stars remains to be seen \citep{Plotnykov2020}.}

Alternatively, the lower measured bulk densities of the TRAPPIST-1 planets relative to Earth-like composition might be explained by core-free interiors \citep{ElkinsTanton2008} in which the oxygen content is high enough such that all iron is oxidized. If the refractory elements (Mg, Fe, Si) follow Solar abundances, a fully oxidized interior would contain about 38.2 wt\% of oxygen, which lies between the value for Earth (29.7 wt\%) and CI chondrites (45.9 wt\%). Such an interior scenario can easily describe the observed bulk densities (red line in Fig. \ref{fig:mass_radius_relation}).  And this may bolster the long-range migration scenario in which the planets formed in a highly oxidizing environment which enabled the iron to remain in the mantle even after migration.  {Based on the elemental composition, these models have an oxygen fugacity of $\Delta \mathrm{IW} = -0.91,$\footnote{{Oxygen fugacity is stated relative to the Iron-W\"ustite equilibrium reaction ${\mathrm{Fe}{+}0.5\mathrm{O}_2{=}\mathrm{FeO}}$ (W\"ustite) such that ${\Delta \mathrm{IW} = \log(f_{\mathrm{O}_2})_\mathrm{rock} - \log(f_{\mathrm{O}_2})_\mathrm{IW}}$}} which is more oxidized than Earth or even Mars, but is comparable to the oxidation state of small bodies, both in our solar system and accreted by white dwarfs \citep{Doyle2019}.}

However, the evidence for a core-free planet may rest on knowing the refractory abundances of the TRAPPIST-1 host star, which have yet to be constrained.  Alas, our interpretation of the planets' compositions may be limited by our imprecise knowledge of the host star:  its radius, its mass, its photospheric inhomogeneity, and its refractory abundances all affect our measurement and interpretation of the masses, radii, and compositions of the TRAPPIST-1 planets.  In this paper our measurements of the relative planetary radii and masses have reached such a precision that the fault may now lie in the star.

\subsection{Future work}

We conclude by pointing out directions for building upon the work described in this paper:
\begin{enumerate}
    \item We have yet to identify the origin of timing outliers which
          show an excess relative to a normal distribution.  This may be addressed
          with higher precision measurements which may be able to identify a source
          of noise responsible for these outliers.
    \item Our analysis assumes a plane-parallel system with seven planets, and
          does not yet couple the dynamical and photometric analysis (our photodynamics
          held the dynamical model fixed).  Future analysis with a fully-coupled
          photodynamical model with 3D orbits and more than seven planets may be valuable.
    \item We need more transits measured for planets d and h, in order
          to better measure the amplitude and phase on the transit-timing variation
          timescale, as well as to better constrain the presence of planets beyond h.
    \item The interpretation of the compositions of the planets is limited by the unknown composition of the host star. A measurement of the Mg/Fe and Fe/Si ratios would help to interpret the core and mantle compositions.  Both sets of constraints would help to limit the range and break degeneracies of possible interior compositions of the planets \citep{Dorn2015,Bitsch2019b}.
    \item Without a constraint on the detailed abundance ratios of the host star, a Bayesian interpretation of the bulk densities of the planets should be warranted \citep{Dorn2016} to better quantify the range of possible compositions.
    \item More detailed spectral analysis of the stellar photosphere to ascertain the impact of an inhomogeneous stellar atmosphere on the radius ratios would be warranted.
\end{enumerate}

We anticipate that once JWST launches, we will obtain higher precision constraints upon the dynamics of the system, yielding much improved constraints upon the planets' bulk densities, which will further improve the interpretation of their interior compositions.

\section*{Acknowledgements}

This work is based in part on observations made with the Spitzer Space Telescope, which
is operated by the Jet Propulsion Laboratory, California Institute of Technology under
a contract with NASA. Support for this work was provided by NASA through an award issued
by JPL/Caltech.  EA was supported by a Guggenheim Fellowship and NSF grant AST-1615315.
This research was partially conducted during the Exostar19 program at the Kavli Institute 
for Theoretical Physics at UC Santa Barbara, which was supported in part by the National
Science Foundation under Grant No.\ NSF PHY-1748958.
We also acknowledge support from NASA's NExSS Virtual Planetary Laboratory, funded under
NASA Astrobiology Institute Cooperative Agreement Number NNA13AA93A, and the NASA Astrobiology
Program grant 80NSSC18K0829.
This work was facilitated though the use of the advanced computational, storage, and
networking infrastructure provided by the Hyak supercomputer system at the University of
Washington. TRAPPIST is a project funded by the Belgian Fonds (National) de la Recherche 
Scientifique (F.R.S.-FNRS) under grant FRFC 2.5.594.09.F. TRAPPIST-North is a project funded 
by the University of Li\'ege, in collaboration with Cadi Ayyad University of Marrakech (Morocco).
B.-O.D., C.D.\ and J.H. acknowledge support from the Swiss National Science Foundation (grants
PP00P2-163967, PZ00P2\_174028, and 200020\_19203, respectively).
Calculations were performed on UBELIX (http://www.id.unibe.ch/hpc), the HPC
cluster at the University of Bern. This work has been carried out in the framework of the PlanetS National Centre
of Competence in Research (NCCR) supported by the Swiss National Science Foundation (SNSF).
This research has made use of NASA's Astrophysics Data System and of services produced by
the NASA Exoplanet Science Institute at the California Institute of Technology.
This project has received funding from the European Union's Horizon 2020 research and
innovation program under the Marie Sklodowska-Curie Grant Agreement No. 832738/ESCAPE and European Research Council (ERC; grants agreement n$^\circ$ 679030/WHIPLASH and 803193/BEBOP). M.T.
thanks the Gruber Foundation for its generous support to this research.
The research leading to these results has received funding from the European Research Council
under the European Union's Seventh Framework Programme (FP/2007-2013) ERC Grant
Agreement n° 336480, and from the ARC grant for Concerted Research Actions, financed by the
Wallonia-Brussels Federation. V.V.G. is a F.R.S.-FNRS Research Associate. M.G. and E.J. are F.R.S.-FNRS Senior Research Associates. R.M. acknowledges support for this research from NASA (grant 80NSSC18K0397). Z.L. acknowledges support from the Washington NASA Space Grant Consortium Summer Undergraduate Research Program.

We thank Trevor Branch for discussions about HMC and automatic differentiation.
We thank Pramod Gupta, Diana Windemuth, and Tyler Gordon for help in
using Hyak.  We thank Vardan Adibekyan, Rory Barnes, Jim Davenport, Natalie Hinkel, Dave Joswiak,
and Sarah Millholland for useful discussions. {Finally, we thank
the referees for valuable feedback which improved this paper.}

{\software{Matplotlib \citep{Hunter:2007,https://doi.org/10.5281/zenodo.592536},
Julia \citep{Bezanson2017}, Limbdark.jl \citep{Agol2019}.}}



\bibliographystyle{aasjournal}
\bibliography{trappist1_references} 



\appendix

\section{Approximate Hessian matrix} \label{sec:hessian}

Here we approximate the posterior probability distribution as a multi-dimensional Gaussian, assuming a uniform prior.  
The log likelihood for each data point with indices $i$ and $j$ may be written as a function of the observed transit times and uncertainties, the modeled transit times, and the Student's t-distribution model parameters, such that
\begin{equation}
    \mathcal{L}_{ij}(\mathbf{x}) = \mathcal{L}_{ij}(t_{ij}(\mathbf{x}_{dyn});\log{\nu},V_1e^{1/2\nu};t_{ij,obs},\sigma_{ij}),
\end{equation}
where all of the dependence on the dynamical model parameters enters through $t_{ij}(\mathbf{x}_{dyn})$.  
The maximum posterior probability also corresponds to the maximum likelihood in this limit, in which case we expand the log likelihood for the $i$th planet and $j$th transit as a Taylor series:
\begin{equation}
    \begin{split}
        \log{\mathcal{L}_{ij}(\mathbf{x})} &\approx  \log \mathcal{L}_{ij}(\mathbf{x_0})\\
        &+ \frac{1}{2} \sum_{k,l} \left.\frac{\partial^2 \log\mathcal{L}_{ij}(\mathbf{x})}{\partial x_k \partial x_l}\right\vert_{\mathbf{x}_0}(x_k - x_{0,k})(x_l - x_{0,l}),
    \end{split}
\end{equation}
where we have used the fact that the gradient of the log likelihood vanishes at the maximum likelihood value of the model parameters, $\mathbf{x}_0$, and the indices $k,l = 1,...,5N_p+2$ for $x_k$ and $x_l$, where the first $5N_p$ parameters are the dynamical parameters, $\mathbf{x}_{dyn}$, and the last two parameters are the Student's t-distribution likelihood parameters, $\log{\nu}$ and $V_1e^{1/2\nu}$.
Now, the width of the Gaussian distribution at the maximum likelihood is governed by the Hessian matrix, with elements given by
\begin{equation}
    \mathcal{H}_{kl}(\mathbf{x}_0) =  -\sum_{i,j} \left.\frac{\partial^2 \log\mathcal{L}_{ij}(\mathbf{x})}{\partial x_k \partial x_l}\right\vert_{\mathbf{x}_0},
\end{equation}
which involves second derivatives of the {negative} log likelihood with respect to the model parameters.  
The derivatives of $t_{ij}$ with respect to $\mathbf{x}_{dyn}$ we compute with the \textsf{NbodyGradient} code;  however, the second derivatives of the transit times with respect to the dynamical model parameters are not computed with our N-body code.  We drop these transit time second derivative terms, which we justify as follows.

For the Hessian matrix elements which involve second derivatives with respect to both dynamical model parameters, $1 \le k,l \le 5N_p$, we can write:
\begin{equation}
    \frac{\partial^2 \log\mathcal{L}_{ij}(\mathbf{x})}{\partial x_k \partial x_l} = \frac{\partial^2 \log\mathcal{L}_{ij}(\mathbf{x})}{\partial^2 t_{ij}} \frac{\partial t_{ij}}{\partial x_k} \frac{\partial t_{ij}}{\partial x_l} + \frac{\partial \log\mathcal{L}_{ij}(\mathbf{x})}{\partial t_{ij}} \frac{\partial^2 t_{ij}}{\partial x_k \partial x_l},
\end{equation}
where $t_{ij} = t_{ij}(\mathbf{x}_{dyn})$ is implied in this and subsequent equations.

Now, at the maximum likelihood there is a balance of residuals which are both positive and negative, such that the second component of this equation has terms with positive and negative signs for different values of $i$ and $j$. This causes the second term in this equation to average to a small value compared with the first term when the sum is carried out over $i$ and $j$ (the planet and transit indices).  So, we drop the second term in this equation.

Adding in the cases of the Hessian matrix elements which involve the likelihood parameters, $(x_{5N_p+1},x_{5N_p+2}) = (\log \nu, V_1e^{1/2\nu})$, we compute the Hessian as
\begin{eqnarray}\label{eqn:hessian}
    \mathcal{H}_{k,l}(\mathbf{x})= -\sum_{i,j}
    \begin{cases}
        \frac{\partial^2 \log \mathcal{L}_{ij}(\mathbf{x})}{\partial^2 t_{ij}} \frac{\partial t_{ij}}{\partial x_k} \frac{\partial t_{ij}}{\partial x_l} & 1 \le k,l \le 5N_p \cr
        \frac{\partial^2 \log \mathcal{L}_{ij}(\mathbf{x})}{\partial t_{ij} \partial x_l} \frac{\partial t_{ij}}{\partial x_k}                           & 1 \le k \le 5N_p; l > 5N_p \cr
        \frac{\partial^2 \log \mathcal{L}_{ij}(\mathbf{x})}{\partial t_{ij} \partial x_k} \frac{\partial t_{ij}}{\partial x_l}                           & k > 5N_p; 1 \le l \le 5N_p \cr
        \frac{\partial^2 \log \mathcal{L}_{ij}(\mathbf{x})}{\partial x_k \partial x_k}                                                                   & k,l > 5N_p
    \end{cases},
\end{eqnarray}
where the partial derivatives with respect to $t_{ij}(\mathbf{x}_{dyn})$, $x_{5N_p+1}=\log \nu$, and $x_{5N_p+2} = V_1e^{1/2\nu}$
are computed with automatic differentiation.

The inverse of the Hessian matrix is used in the Levenberg-Marquardt optimization, and when evaluated at the maximum likelihood, is used to estimate the covariance matrix, from which the square root of the diagonal components are used to estimate the widths of the posterior distribution for each model parameter, $\mathbf{x} = (\mathbf{x}_{dyn},\log \nu, V_1 e^{1/2\nu})$, which are plotted in Figures \ref{fig:mass_likelihood_profile}, \ref{fig:ecc_likelihood_profile}, and \ref{fig:student_param_likelihood_profile}.  This approximated Hessian is also used to define the mass matrix for the HMC simulations.

\begin{table*}
    \centering
    \begin{tabular}{l|c|c|c|c|c}
        Parameter                 & {Bounds function $f_j$} & Lower bound & Lower transition & Upper transition & Upper bound \\
        \hline
                                  &                                  & $\xi_1$     & $\xi_2$          & $\xi_3$          & $\xi_4$ \cr
        Mass-ratio                & $\log_{10}{\mu}$                 & -8          & -7               & -3               & -2 \cr
        Eccentricity              & $e$                              & -           & -                & 0.2              & 0.3 \cr
        Period of b               & $P_\mathrm{b}$ [d]               & 1.49        & 1.50             & 1.52             & 1.53\cr
        Period of c               & $P_\mathrm{c}$ [d]               & 2.40        & 2.41             & 2.43             & 2.44\cr
        Period of d               & $P_\mathrm{d}$ [d]               & 4.03        & 4.04             & 4.06             & 4.07\cr
        Period of e               & $P_\mathrm{e}$ [d]               & 6.08        & 6.09             & 6.11             & 6.12\cr
        Period of f               & $P_\mathrm{f}$ [d]               & 9.18        & 9.19             & 9.22             & 9.23\cr
        Period of g               & $P_\mathrm{g}$ [d]               & 12.33       & 12.34            & 12.36            & 12.37\cr
        Period of h               & $P_\mathrm{h}$ [d]               & 18.75       & 18.76            & 18.78            & 18.79\cr
        Initial transit time of b & $t_{0,\mathrm{b}}-2,457,257$ [d] & 0.53        & 0.54             & 0.56             & 0.57\cr
        Initial transit time c    & $t_{0,\mathrm{c}}-2,457,257$ [d] & 0.57        & 0.58             & 0.60             & 0.61\cr
        Initial transit time d    & $t_{0,\mathrm{d}}-2,457,257$ [d] & 0.05        & 0.06             & 0.08             & 0.09\cr
        Initial transit time e    & $t_{0,\mathrm{e}}-2,457,257$ [d] & 0.81        & 0.82             & 0.84             & 0.85\cr
        Initial transit time f    & $t_{0,\mathrm{f}}-2,457,257$ [d] & 0.05        & 0.06             & 0.08             & 0.09\cr
        Initial transit time g    & $t_{0,\mathrm{g}}-2,457,257$ [d] & 0.70        & 0.71             & 0.73             & 0.74\cr
        Initial transit time h    & $t_{0,\mathrm{h}}-2,457,249$ [d] & 0.58        & 0.59             & 0.61             & 0.62\cr
        Degrees of freedom        & $\nu$                            & 0.5         & 1.0              & 50               & 100\cr
        Log variance factor       & $\log{V_1}$                      & -2          & -1               & 5                & 10
    \end{tabular}
    \caption{Prior probability boundary limits for the TRAPPIST-1 planet parameters.  The bounds are chosen so as to not affect the parameters as much as possible.}
    \label{tab:prior_parameterization}
\end{table*}

\section{Transit Timing Prior}\label{sec:prior}

We use a uniform prior for each mass and orbital element,
with {smooth} bounds on each, with the exception of the initial eccentricity vectors.
Since we sample in the eccentricity vector of each planet, $\mathbf{e}_i = (e_i\cos{\omega_i},e_i\sin{\omega_i})$, the volume
of parameter space scales $\propto e_i$, and so an $1/e_i$ prior is needed to
yield a posterior which has a uniform probability with eccentricity, $e_i$,
for the $i$th planet
\citep{Eastman2013}.

In addition to the eccentricity prior, we place {smooth bounds on the parameters}.
For each {bound
we choose upper and lower limits} for which the prior starts to transition from 1 to 0 with a cubic dependence.
For the bound on a {function of our parameters} of value $\xi$ we specify
\begin{eqnarray}
    \Pi_\mathrm{bound}(\xi) =
    \begin{cases}
        0                                            & \xi {\le} \xi_1             \\
        3y^2-2y^3; y = \frac{\xi-\xi_1}{\xi_2-\xi_1} & \xi_1 {<} \xi {<} \xi_2     \\
        1                                            & \xi_2 {\le} \xi {\le} \xi_3 \\
        3y^2-2y^3; y = \frac{\xi_4-\xi}{\xi_4-\xi_3} & \xi_3{<} \xi {<} \xi_4      \\
        0                                            & \xi {\ge} \xi_4
    \end{cases},
\end{eqnarray}
so that the total prior is given by
\begin{equation}
    \Pi(\mathbf{x}) = \prod_{i=1}^{N_p} e_i^{-1} {\times} \prod_{j=1}^{N_\mathrm{bound}} \Pi_\mathrm{bound}(f_j(\mathbf{x})),
\end{equation}
where the values of $\xi_1{-}\xi_4$ {and each transformation of
parameters,
$\mathbf{f} {=} \{f_j(\mathbf{x}); j{=} 1,...,N_\mathrm{bound}\}$,} are given in Table
\ref{tab:prior_parameterization}, where $N_\mathrm{bound} = 4 N_p + 2$.
The prior probability, then, is given by $\Pi(\mathbf{x})$, which we
multiply by the likelihood function before sampling.

\section{Corner plots}

\begin{figure}
    \centering
    \includegraphics[width=\hsize]{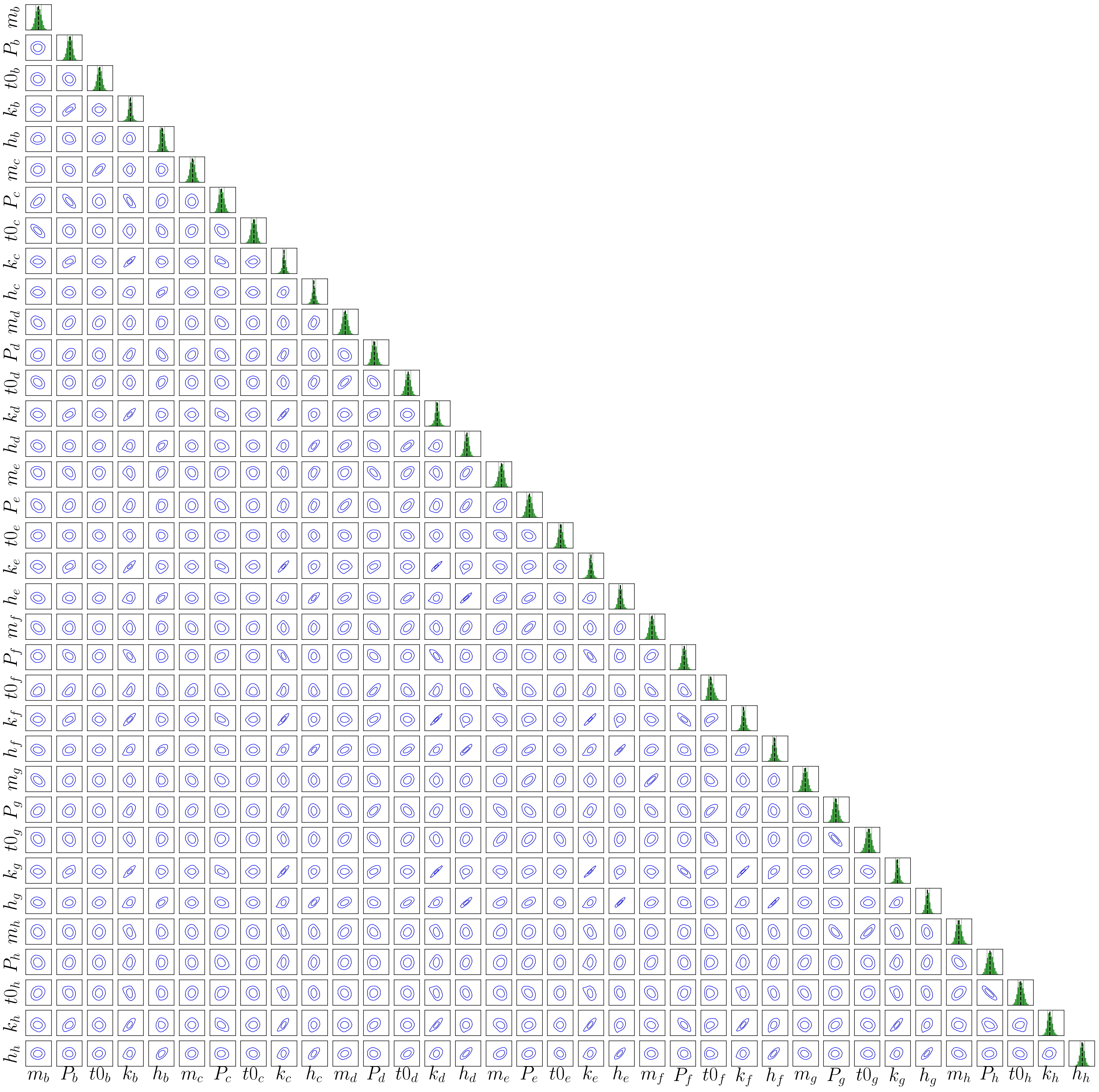}
    \oscaption{python/plotCorner_ttv.py.script}{Corner plot of variables in the transit-timing analysis with $1\sigma$
       and $2\sigma$ confidence contours.
       Lagrangian orbital elements are defined as $k_b = e_b \cos{\omega_b}$ and $h_b = e_b \sin{\omega_b}$,
       and similarly for the other planets.  Planet masses are defined relative to the star.}
    \label{fig:corner_ttv}
\end{figure}

\begin{figure}
    \centering
    \includegraphics[width=\hsize]{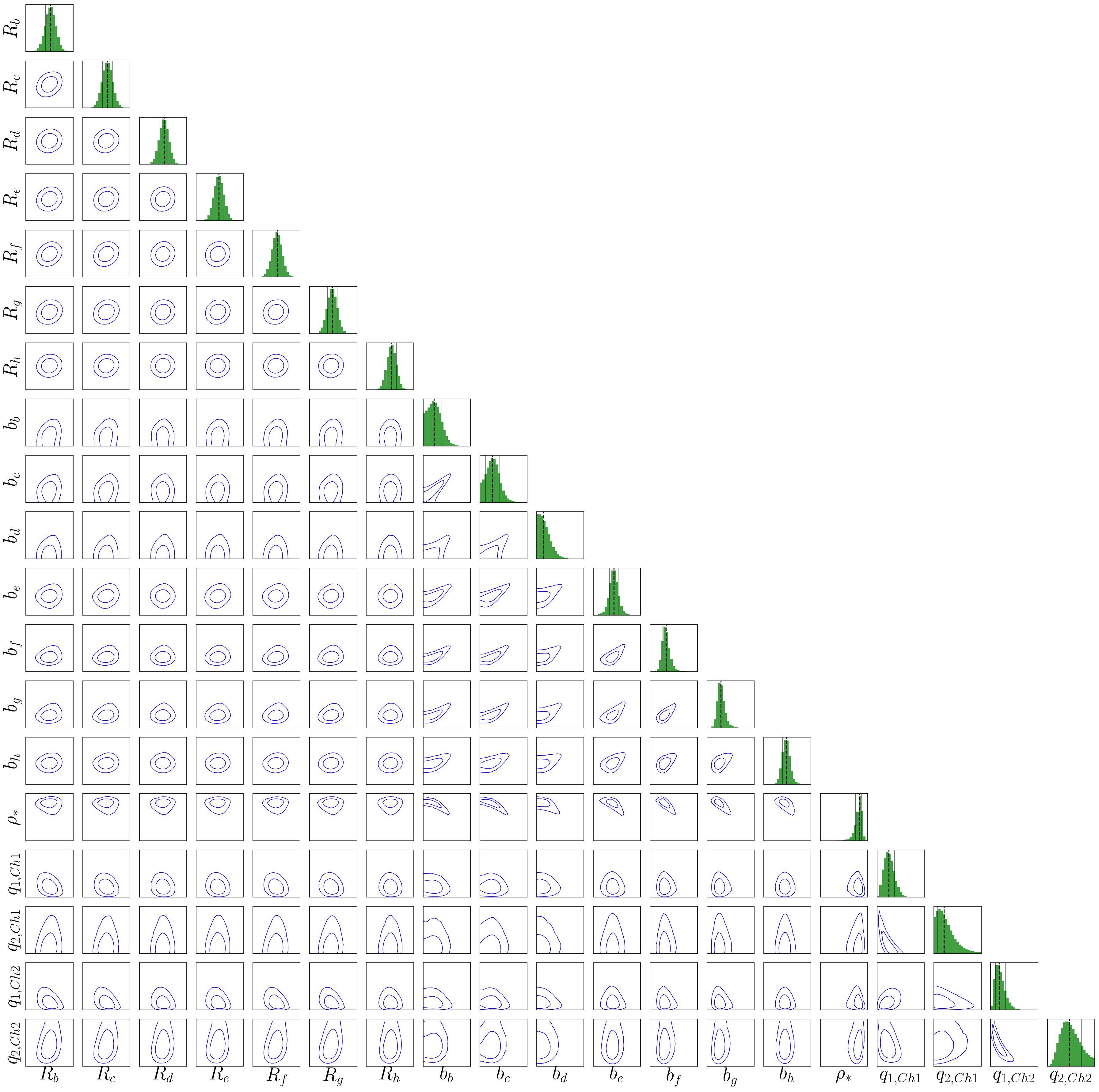}
    \oscaption{python/plotCorner_photdyn.py}{Corner plot of variables in the photodynamical analysis with $1\sigma$
       and $2\sigma$ confidence contours. Planet radii are relative to star.}
    \label{fig:corner_photdyn}
\end{figure}

Figures \ref{fig:corner_ttv} and \ref{fig:corner_photdyn} 
show corner plots of the variables
from the transit-timing and photodynamical analyses, respectively.

\section{Tables}

Tables of the
best-fit transit times (Table \ref{tab:transit_times_observed_and_calculated}), and
the  forecast times (Table \ref{tab:transit_times_forecast})
are given in this appendix.

\begin{longtable}{|c|c|c|c|c|c|c|}
    \hline
    Planet                & Epoch             & $t_\mathrm{obs}$ &
    $\sigma_\mathrm{obs}$ & $t_\mathrm{post}$ & $\sigma_{post}$  & Source \endfirsthead
    Planet                & Epoch             & $t_\mathrm{obs}$ &
    $\sigma_\mathrm{obs}$ & $t_\mathrm{post}$ & $\sigma_{post}$  & Source \endhead
    \hline
    b & 0 &  7322.515310 &  0.000710 &  7322.517901 &  0.000127 &  TS  \cr
b & 2 &  7325.539100 &  0.001000 &  7325.539171 &  0.000115 &  TS  \cr
b & 4 &  7328.558600 &  0.001300 &  7328.561188 &  0.000126 &  TS  \cr
b & 6 &  7331.581600 &  0.001000 &  7331.582577 &  0.000115 &  TS  \cr
b & 8 &  7334.604800 &  0.000170 &  7334.604786 &  0.000121 &  LT + VLT  \cr
b & 10 &  7337.626440 &  0.000920 &  7337.626038 &  0.000109 &  TS  \cr
b & 12 &  7340.648200 &  0.001400 &  7340.648105 &  0.000121 &  TS  \cr
b & 15 &  7345.180280 &  0.000800 &  7345.180807 &  0.000116 &  HCT  \cr
\end{longtable}

\pagebreak
\begin{longtable}{|c|c|c|c|}
    \hline
    Planet & Epoch & $t_\mathrm{post}$ & $\sigma_\mathrm{post}$ \endfirsthead
    Planet & Epoch & $t_\mathrm{post}$ & $\sigma_\mathrm{post}$ \endhead
    \hline
    b & 1 &  7259.061221 &  0.000152\cr
b & 2 &  7260.572662 &  0.000163\cr
b & 3 &  7262.083452 &  0.000159\cr
b & 4 &  7263.594099 &  0.000150\cr
b & 5 &  7265.104973 &  0.000151\cr
b & 6 &  7266.615845 &  0.000153\cr
b & 7 &  7268.126697 &  0.000155\cr
b & 8 &  7269.637314 &  0.000148\cr
\end{longtable}


\end{document}